\documentclass[10pt,notitlepage,letterpaper,aps,prd,tightenlines,preprintnumbers,nofootinbib,superscriptaddress]{revtex4-2}
 
\usepackage[T1]{fontenc}
\usepackage{mathptmx}
\usepackage{tgpagella}
\usepackage{bbm}
\usepackage{fullpage}
\usepackage{amsfonts}
\usepackage{amsmath} 
\usepackage{slashed}
\usepackage{amssymb}
\usepackage{graphicx}
\usepackage{array}
\usepackage{epic}
\usepackage{eepic}
\usepackage{epsfig}
\usepackage{latexsym}
\usepackage[table,xcdraw,dvipsnames]{xcolor}
\usepackage{booktabs}
\usepackage{tikz}
\usetikzlibrary{shapes.geometric}
\usepackage[export]{adjustbox}
\usepackage{float}
\usepackage{multirow}

\definecolor{NiceURL}{HTML}{A9341F}
\definecolor{NiceLinks}{HTML}{524fc0}
\definecolor{NiceCite}{HTML}{008B72}

\usepackage[caption=false]{subfig}
 
\usepackage{natbib}
\usepackage{relsize}
\usepackage[left=2.4cm,right=2.4cm,top=2.0cm,bottom=2.5cm]{geometry}
\usepackage[linktocpage]{hyperref}
\hypersetup{colorlinks=true,citecolor=NiceCite,linkcolor=NiceLinks,urlcolor=NiceURL}

\usepackage{shorthand}

\definecolor{CN}{HTML}{0071BC}

 \usepackage{orcidlink}
\begin{document}

\title{\Large\bf\textcolor{CN}{Unsupervised and lightly supervised learning in particle physics}}
\author{Jai Bardhan $\orcidlink{0000-0001-9348-3459}$}
\email{jai.bardhan90@googlemail.com}
\affiliation{Center for Computational Natural Sciences and Bioinformatics, International Institute of Information Technology, Hyderabad 500~032, India}

\author{Tanumoy Mandal $\orcidlink{0000-0001-7268-549X}$}
\email{tanumoy@iisertvm.ac.in}
\affiliation{Indian Institute of Science Education and Research Thiruvananthapuram, Vithura, Kerala, 695~551, India}

\author{Subhadip Mitra $\orcidlink{0000-0002-7107-0343}$}
\email{subhadip.mitra@iiit.ac.in}
\affiliation{Center for Computational Natural Sciences and Bioinformatics, International Institute of Information Technology, Hyderabad 500~032, India}
\affiliation{Center for Quantum Science and Technology, International Institute of Information Technology, Hyderabad 500 032, India}

\author{Cyrin Neeraj $\orcidlink{0000-0002-6748-1692}$}
\email{cyrin.neeraj@research.iiit.ac.in}
\affiliation{Center for Computational Natural Sciences and Bioinformatics, International Institute of Information Technology, Hyderabad 500~032, India}

\author{Monalisa Patra $\orcidlink{0000-0002-3242-0267}$}
\email{monalisa.p@ihub-data.iiit.ac.in}
\affiliation{iHub-Data, International Institute of Information Technology, Hyderabad 500 032, India}
\date{\today}

\begin{abstract}
\noindent {\small We review the main applications of machine learning models that are not fully supervised in particle physics, i.e., clustering, anomaly detection, detector simulation, and unfolding. Unsupervised methods are ideal for anomaly detection tasks---machine learning models can be trained on background data to identify deviations if we model the background data precisely. The learning can also be partially unsupervised when we can provide some information about the anomalies at the data level. Generative models are useful in speeding up detector simulations---they can mimic the computationally intensive task without large resources. They can also efficiently map detector-level data to parton-level data (i.e., data unfolding). In this review, we focus on interesting ideas and connections and briefly overview the underlying techniques wherever necessary.}
\end{abstract}

\maketitle
\vfill
\noindent\rule{\textwidth}{0.5pt}\vspace{-55pt}
\tableofcontents
\makeatletter
\let\toc@pre\relax
\let\toc@post\relax
\makeatother

\setcounter{page}{1}
\setcounter{footnote}{1}
\setcounter{equation}{0}

\section{Introduction}
\noindent In recent years---especially since the advent of deep learning---machine learning (ML) techniques have undergone what can be termed dramatic revolutions. Their performance and range of applicability have increased manyfold. ML tools are now standard in almost every field of study. Particle physics is no exception. These techniques are not new to particle physics---from the LEP to the LHC, high-energy data have been analysed with multivariate techniques like boosted decision trees, neural networks (NNs), etc. (see the references in, e.g., Ref.~\cite{Radovic:2018dip}). The famous discovery of the Higgs boson, which completed the experimental verification of the particle spectrum of the Standard Model (SM), relied on ML techniques. Recently, however, there has been a surge in interest in these methods. Modern and sophisticated ML techniques are being adapted to particle physics problems at an unprecedented rate (see Ref.~\cite{Feickert:2021ajf}\footnote{The Particle Data Book~ \cite{ParticleDataGroup:2022pth} now includes a review on ML.}) for various reasons. For example, precise measurements in some unexplored parts of the SM, like the Higgs self-coupling or light-quark Yukawa couplings, will be beyond the reach of the traditional analyses \cite{Amacker:2020bmn,Abdughani:2020xfo,Adhikary:2020fqf,Alasfar:2022vqw}. Also, the fact that the LHC is yet to find any signature of physics beyond the SM (BSM) implies that we need to look for processes with even smaller cross-sections or leading to minor departures from known distributions. With the planned high luminosity upgrade (HL-LHC), the LHC will gather, analyse, and record a vast amount of collision data. Modern ML techniques are ideal for such tasks---they can handle tiny signals much better than the usual cut-based analyses designed by physicists' intuitions or find (learn) patterns from a vast amount of data. 

In this review, we survey the learning methods used in particle physics that are not supervised. In this context, supervision means the external information we provide to the ML model through the labels of the training data. Depending on the degree of supervision, these methods can be either fully unsupervised (where models are trained on unlabelled data) or partially unsupervised (equivalently, weakly or semi-supervised, where models use some extra/partial information while training). Unsupervised methods are data-driven and are usually model-agnostic. They can identify patterns or deviations from known patterns in data. These methods have become popular recently as the model-specific (or top-down) searches at the LHC are yet to identify any new physics (NP) signatures. In model-specific searches, supervised classifiers are trained to discriminate simulated signal events from simulated background events generated with a Monte Carlo (MC) simulator. Training supervised classifiers requires large amounts of such simulated data. However, even with that, a supervised classifier may perform sub-optimally with the real data, as the model can learn to make predictions from approximate correlations present in the simulations (e.g., the hadronisation models, the absence of higher-order effects on simulated distributions, etc.). Even when the search results are presented in a model-independent manner, the selection cuts might be model class-specific, limiting reinterpretation in the context of a different set of models. Unsupervised methods aim to address these shortcomings.

Unsupervised tasks are not new to particle physics. Common jet clustering is essentially an unsupervised task as the algorithm (like anti-$k_T$ or Cambridge/Aachen) groups objects without any specific predefined target jet.\footnote{In practice, some tuning of the parameters may be necessary to tag boosted objects; for example, see Refs.~\cite{Bardhan:2022sif,Bhardwaj:2022wfz}, where a dijet signal from a boosted BSM singlet boson is tagged.} We look at clustering tasks in Sec.~\ref{sec:clustering}. In Sec.~\ref{sec:AnomalyDetection}, we focus on anomaly detection, another major task for nonsupervised methods. Anomaly detection without supervision is essentially finding deviations or departures (anomalies) from known patterns without any prior expectation about the nature of the anomalies (like knowledge of any NP model or interaction). If the ML model learns the distribution of the background (SM) events directly from the data, it can identify events not coming from that distribution (i.e., out-of-distribution events). In this section, we discuss data-driven background estimation, autoencoders, variational autoencoders, weakly supervised methods, topic modelling in jet space, and self-supervised learning methods. In the next section (Sec.~\ref{sec:det_sim}), we look at how ML models, especially generative adversarial networks (GANs), can play a key role in fast detector simulations and unfolding the data distributions recorded at the detectors. Finally, we conclude in Sec.~\ref{sec:conclusions}. One comment is in order before we move on to the next section. Most developments we discuss here are recent, and applications of nonsupervised techniques in particle physics are still nascent.\footnote{The HEP community has put out several challenges related to unsupervised learning in the last five years: Fast Calorimeter Challenge: \url{https://calochallenge.github.io/homepage/}, Anomaly Detection Challenge: \url{https://mpp-hep.github.io/ADC2021/}, etc.} Hence, in this review, we focus more on the methods rather than quantitative comparisons of the results. For specialised reviews on anomaly detection and generative modelling in detector simulations, see Ref.~\cite{Belis:2023mqs} and Ref.~\cite{Hashemi:2023rgo}, respectively.

\section{Clustering}\label{sec:clustering}
Clustering typically means clubbing or putting similar objects together in groups or clusters. In ML, clustering algorithms group similar data points into clusters where points within a cluster are closer (more similar) to each other than those in different clusters. Clustering is an unsupervised learning technique that works with unlabeled data without relying on predefined categories. There are different clustering techniques, depending on whether the data is directly partitioned into some clusters (like K-means) or the clusters are nested hierarchically, as in hierarchical clustering.
\medskip 

\noindent 
{\bf K-means clustering:} 
The K-means clustering algorithm (not to confuse with K-nearest neighbours, a supervised algorithm), helpful in identifying different types of interactions or physics processes, partitions a dataset like an event record based on some particle properties (like the particles' four-momenta). It works by minimising the intra-partition distances by clustering $N$ particles into a fixed number of clusters ($K$, specified by the user) containing $N^k_{cl}$ particles $(k=1, 2, \ldots, N)$. The algorithm attempts to find the $K$ centroids as far away as possible and associates each point to the nearest centroid (cluster centre). It proceeds by iteratively assigning each data point to the nearest centroid and then updating the location of the centroid as the mean of the data points assigned to the cluster. It repeats this process until the cluster assignments no longer change or a maximum number of iterations is reached. This algorithm minimises an objective function, which is a squared error function defined as:
\begin{equation}\label{eq:sumKmeans}
     SS = \sum_{j=1}^{K} \sum_{i=1}^N ||x_i^{(j)}-c_j||^2,
\end{equation}
where $c_j$ denotes the $j$th centroid. The above minimisation forces the distances between the data points within a cluster to be shorter than those from data points in other clusters. The choice of the distance metric is significant for clustering results. 

Various reconstruction algorithms for reconstructing heavy states from jets have been developed using the K-means clustering technique. Jets in high-energy experiments are collimated showers of particles originating from partons. Normally, sequential clustering algorithms like $k_t$, anti-$k_t$, or Cambridge/Aachen are used to find infrared and collinear (IRC)-safe jets (where the jet observables remain unaffected by a collinear splitting or the emission of an infinitely soft particle) in events. These sequential recombination algorithms follow a bottom-up approach of combining particles starting from the nearest ones, which is very similar to the agglomerative hierarchical clustering\footnote{Agglomerative hierarchical clustering is a bottom-up approach where each data point starts as its own cluster and pairs of clusters are successively merged based on a specified criterion until only one cluster remains. This process creates a hierarchical structure of clusters, represented as a dendrogram, which can be cut at different levels to obtain a partitioning of the data into clusters.} scheme. Jet construction using the K-means algorithm was first proposed in Ref.~\cite{Chekanov:2005cq}.

The algorithm can be understood with a simple example. Let us consider a six-jets process, $e^+e^- \rightarrow t \bar{t} \rightarrow b \bar{b} W^+ W^- \rightarrow b\bar{b} q_1 \bar{q}_2 q_3 \bar{q}_4$. Since we expect the hadrons to be clustered into six jets, we require the K-means algorithm to randomly locate six centroids in the $\eta$ (rapidity)-$\phi$ (azimuthal angle) space. The algorithm then minimises the distances from the centroids to the particles in this space. The sum $SS$ [Eq.~\eqref{eq:sumKmeans}] is evaluated after the initial iteration, and the process is repeated till the minimum value of $SS$ is obtained. The process is repeated some $m$ number of times using different initialisations for the centroids, leading to $m$ different values of $SS$. Finally, the smallest $SS$ configuration is chosen as it leads to the strongest particle collimation. Ref.~\cite{Chekanov:2005cq} uses an explicit weight factor to assign the highest priority to configurations resembling the $W$-bosons (i.e., dijet pairs with invariant masses $\sim m_W$) in an event. It uses the K-means algorithm to reconstruct heavy states, such as top quarks and W bosons, decaying into jets with improved efficiency and reduced systematic uncertainties. The paper shows that the K-means algorithm has advantages over traditional jet-finding algorithms, leading to better mass resolutions and peak positions. 

The dependence of the result on the chosen value of $K$ is one of the limitations of K-means. Similarly, choosing the relevant features (properties like energy, momentum, and charge) for clustering is also crucial. K-means also relies on distance metrics, wherein for hep data, appropriate metrics might involve specialized measures of particle similarity. Overall, K-means clustering effectiveness depends on careful consideration of data characteristics, feature selection, and meaningful interpretation of results.
\medskip

\noindent 
{\bf Fuzzy-jets clustering:} 
A new type of infrared and collinear-safe jets known as fuzzy jets is proposed in Ref.~\cite{Mackey:2015hwa}. Unlike the jets obtained by deterministic sequential-recombination algorithms, which are clearly defined by their constituents, fuzzy jets are probabilistic (mixture model). In Fuzzy-jets clustering, one constructs a probability manifold describing the event data distribution. The mixture-model density function for grouping $m$ $n$-dimensional data points into $K$ clusters can be written as,
\begin{equation}\label{eq:fuzzy1}
\rho(x_1,...,x_m|\theta) = \prod_{i=1}^m \left(\sum_{j=1}^K \pi_j \Phi(x_i|\mu_j,\Sigma_j) \right),
\end{equation}
where $x_i$ is the position vector of any point in the detector; $\Phi$ is a Gaussian density with its mean $\mu_j$ at the location of the cluster $j$ and $\Sigma_j$ describing its shape in the $n$-dimensional space; $\pi_j$ is some unknown weight of the cluster $j$ such that $\sum_j \pi_j =1$. With the number of clusters ($K$) specified by the user, the goal is to learn the parameters $\mu_j,\Sigma_j$ which
maximizes the likelihood of the observed dataset. To make Eq.~\eqref{eq:fuzzy1} infrared (IR) safe, particle $p_T$'s are directly inserted in the likelihood function:
\begin{equation}
\log \mathcal{L}({p_{T,i},\rho_i}|\mu, \Sigma)=\prod_{i=1}^m  p^\alpha_{T,i} \log \left(\sum_{j=1}^K \pi_j \Phi(\rho_i|\mu_j,\Sigma_j) \right),
\end{equation}
where $\alpha$ is a weighing factor. Setting $\alpha>0$ leads to IR-safe solutions, and $\alpha = 1$, in particular, leads to IRC-safe jets. Additionally, $\pi_j$ and $\Sigma_j$ are initialised as $1/K$ and $\sigma^2 I$, respectively, to create circular distributions. 
Here, $\sigma$ is a measure of the size of a jet core. It offers complementary information to the $n$-subjettiness variables $\tau_{21}$ and $\tau_{32}$ for tagging the flavour of the jet (like $W$-boson or top-quark jets). 

To maximise the likelihood function, the authors in Ref.~\cite{Mackey:2015hwa} use the iterative expectation-maximisation algorithm where first, the expected values $q_{ij}$ of the probabilities $\rho_i$ are calculated, given the values of the parameters, $\pi_j,\mu_j$, and $\Sigma_j$ (expectation), and then those values are used to maximise the modified complete log-likelihood over the parameters $\pi_j,\mu_j$, and $\Sigma_j$ (maximisation). Jets produced with this method are called fuzzy jets, as every particle can belong to every jet with some probability. The log-likelihood function is a multi-variate convex function, which is extremised for $\nabla \mathcal{L} = 0$, to find the optimal value for $\pi, \mu,\Sigma$. The momentum dependence in the updates of the parameters and the log-likelihood function ensures infrared and collinear safety. 
\medskip

\noindent 
{\bf Nearest-neighbours clustering:} 
Ref.~\cite{DeSimone:2018efk} proposes a model-independent non-parametric unsupervised method to look for new physics. The method tests for the degree of compatibility between the two data samples (e.g., LHC data and SM background) by building a statistical test upon a test statistic. The test statistic measures the distance between the two with a nearest-neighbours technique to estimate the ratio of the local densities of points in the samples. The authors define two samples, $\mathcal{B}$ (`benchmark' or `control' or `reference') sample, and  $\mathcal{T}$  (`trial' or `test') sample.
\begin{eqnarray}
\mathcal{T}&\equiv&\{\boldsymbol{x}_i\}_{i=1}^{N_T}
\stackrel{\textrm{iid}}{\sim}\rho_T\,,\\
\mathcal{B}&\equiv&\{\boldsymbol{x}'_i\}_{i=1}^{N_B}
\stackrel{\textrm{iid}}{\sim}\rho_B\,.
\end{eqnarray}
with  $\{\boldsymbol{x}_i | \boldsymbol{x}_i\in\mathbb{R}^D\}_{i=1}^{N_T}$
and $\{\boldsymbol{x}'_i | \boldsymbol{x}'_i\in\mathbb{R}^D \}_{i=1}^{N_B}$
being two independent and identically distributed $D$-dimensional samples drawn independently from the probability density functions (PDFs) $\rho_T$ and $\rho_B$, respectively. The $\mathcal{T}$, $\mathcal{B}$ samples consist of $N_T$, $N_B$ points, respectively. In conventional approaches, the NP signals are simulated in specific models and compared with the experimental data to test for those models. Since the theoretical landscape for NP signals is enormous, there is always a possibility that an actual signal may be missed. This method looks for deviations between the simulated SM background ($\mathcal{B}$ samples) and the true data ($\mathcal{T}$), agnostic to any NP scenarios. They check to what significance the two samples are compatible with each other (i.e., $\rho_B=\rho_T$) by performing a statistical test of the null hypothesis $\{H_0:\rho_T=\rho_B\}$ versus the alternative hypothesis $\{H_1:\rho_T\neq \rho_B\}$. The test statistic ($\textrm{TS}$) quantifies how much the sample data deviates from what is expected under the null hypothesis. The $\textrm{TS}$ over the trial sample is defined as:
\begin{equation}
\textrm{TS}(\mathcal{B}, \mathcal{T})\equiv \log\hat\lambda^{1/|\mathcal{T}|}
=\frac{1}{N_T}\sum_{j=1}^{N_T}\log\frac{\hat \rho_T(\boldsymbol{x}_j)}
{\hat  \rho_B(\boldsymbol{x}_j)}\,,
\label{eq:ts}
\end{equation}
where $|\mathcal{T}|=N_T$ is the size of the trial sample. Since the true PDFs $\rho_{B,T}$ are not known, their estimators $\hat \rho_{B,T}$ are used. $\textrm{TS}$ takes a non-zero value when the null hypothesis is false, i.e., if the estimated trial and benchmark distributions are different; \textrm{TS} is close to zero otherwise.\footnote{Eq.~\eqref{eq:ts} closely resembles the Kullback-Leibler divergence (see Appendix \ref{sec:KLD}) between the estimated distributions of trial and benchmark samples, $\mathbb{KL}(\hat \rho_T||\hat \rho_B)$, with the expectation value replaced by the empirical average.}
The ratio of the estimators $\hat \rho_{B,T}$ is obtained via a nearest-neighbours approach:
\begin{eqnarray}
    \hat \rho_B(\boldsymbol{x}_j) &=& \frac{\overline K}{N_B}\frac{1}{\omega_D r_{j,B}^D}, \quad \quad
\hat \rho_T(\boldsymbol{x}_j) = \frac{\overline K}{N_T-1}\frac{1}{\omega_D r_{j,T}^D}\,,
\end{eqnarray}
(for any $\boldsymbol{x}_j\in \mathcal{T}$) where $\omega_D=\pi^{D/2}/\Gamma(D/2+1)$ is the volume of the unit sphere in $\mathbb{R}^D$, and $\overline K$ is the number of nearest neighbours.

To evaluate the probability associated with \textrm{TS}, they use the resampling method known as the permutation test to reconstruct its probability distribution for the null hypothesis $H_0$ $\left[f(TS|H_0)\right]$. In the permutation test, first, the two samples are merged: $\mathcal{U}=\mathcal{B} \cup \mathcal{T}$ and then the elements of $\mathcal{U}$ are randomly shuffled. The first $N_B$ elements are then assigned  to $\tilde{\mathcal{B}}$, and the remaining $N_T$ elements to $\tilde{\mathcal{T}}$.  The value of $\textrm{TS}$ on $\tilde{\mathcal{T}}$ is computed, and the procedure is repeated for every possible permutation of the sample points. The $\textrm{TS}$ distribution is then reconstructed to a standard normal distribution:
\begin{equation}
    \textrm{TS} \rightarrow \textrm{TS}' \equiv \frac{\textrm{TS} - \hat\mu}{\hat \sigma}\,, 
\end{equation}
which is distributed according to 
\begin{equation}
f'(\textrm{TS}'|H_0)=\hat\sigma f(\hat\mu+\hat\sigma\textrm{TS}'|H_0)\,,
\end{equation}
with zero mean and unit variance. The two-sided $p$ value can be easily computed as  
\begin{equation}
p = 
2 \int_{|\textrm{TS}'_{\rm obs}|}^{+\infty} f'(\textrm{TS}'|H_0)d\textrm{TS}'\,.
\label{pvalue}
\end{equation}
The $p$ value associated with the null hypothesis is compared with $\alpha$ to reject the null hypothesis with statistical significance.

The authors try their technique on a dark matter signal in the monojet channel. They consider a simple model of a fermionic dark matter  $\chi$ (of mass $100$ GeV) and an $s$-channel $Z'$ mediator (of three different masses: $1200$, $2000$, and $3000$ GeV). The experimental ‘monojet’ signal consists of events with missing energy and at least one high-$p_T$ jet. The trial dataset $\mathcal{T}_i$ is constructed for each of the three $Z'$ masses. The corresponding $p$ values [Eq.~\eqref{pvalue}] are computed to determine the compatibility of the datasets ($\mathcal{T}_1$-$\mathcal{T}_B$, $\mathcal{T}_2$-$\mathcal{T}_B$, and $\mathcal{T}_3$-$\mathcal{T}_B$). Their results show that the background-only hypothesis is strongly excluded for $\mathcal{T}_1$ and $\mathcal{T}_2$, even though the traditional LHC searches do not exclude these parameter points. Theoretically, though the statistical test proposed is a powerful technique, including the experimental systematic uncertainties and the uncertainties associated with MC simulations weakens the limits. The authors argue that this technique can also be used in the context of data-driven background calculations to measure the compatibility of MC simulations with the data in the control regions.


\section{Anomaly detection}\label{sec:AnomalyDetection}
\noindent Many collider searches---especially those for NP---can be mapped to out-of-distribution learning problems. If the background distribution is known accurately, the signal can be something that is not coming from it (i.e., an anomaly). For example, NNs can generally learn the known distribution well. Models like autoencoders (AEs) can learn background distributions and recreate them efficiently from much smaller latent-space representations. The reconstruction loss can then be used to filter anomalous events. Of course, for this to work, one needs to model the background (or normal) probability distribution precisely. For fairly simple final states, one can construct precise statistical models of the background with MC simulations. In other cases, density estimations of the background distribution are sampled directly from the data.
\medskip
\begin{figure}[t]
\centering
\includegraphics[width=0.5\textwidth]{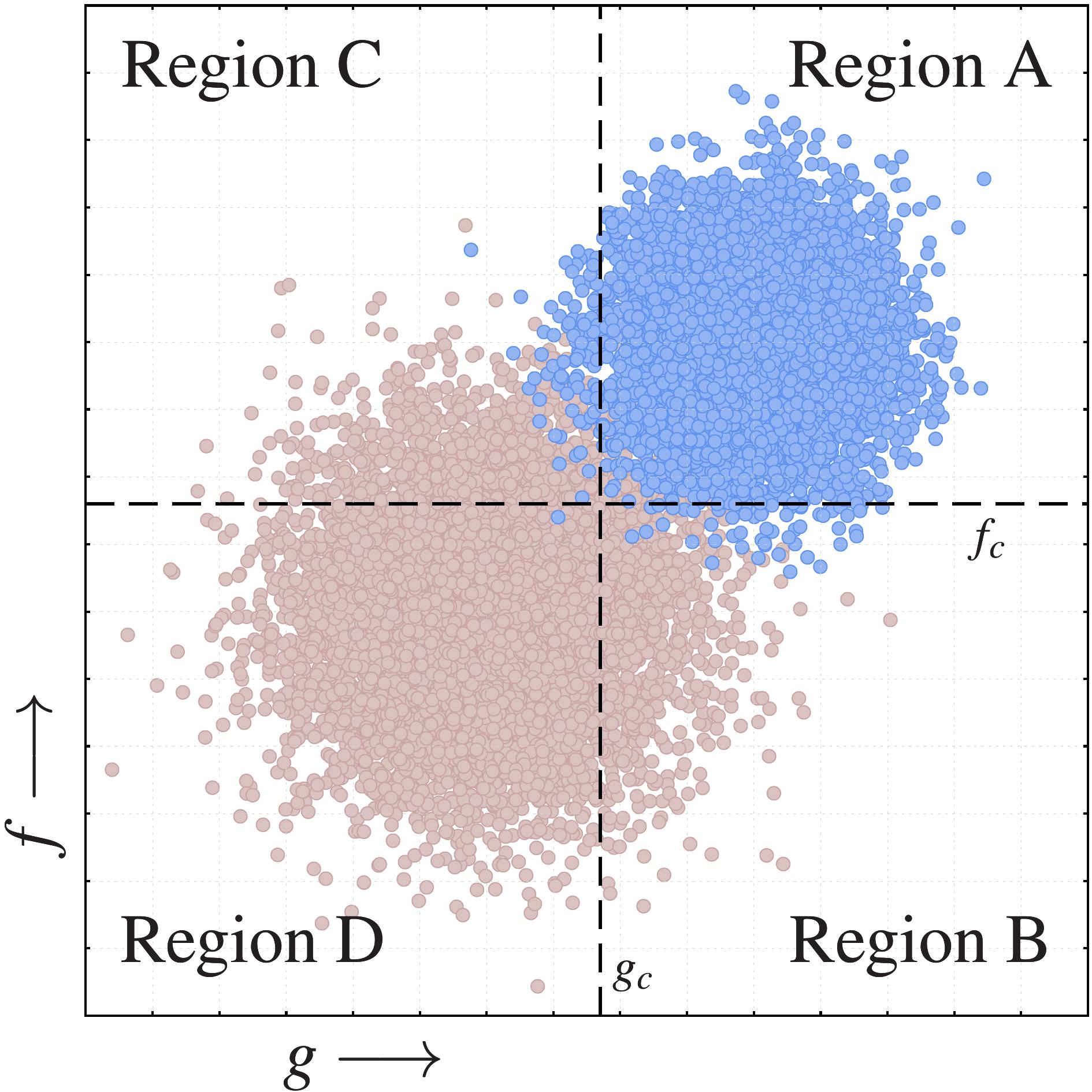}
\caption{\emph{ABCD method}: Figure adapted from Ref.~\cite{Buttinger2018BackgroundEW}. Signal (background) points are coloured blue (light brown). The independence of $f$ and $g$ implies Pr$(f\geq f_c {\rm~and~}g\geq g_c)={\rm Pr}(f\geq f_c)\times{\rm Pr}(g\geq g_c)$.\label{fig:ABCD}}
\end{figure}
\begin{figure}[h]
\centering
\includegraphics[width=0.6\textwidth]{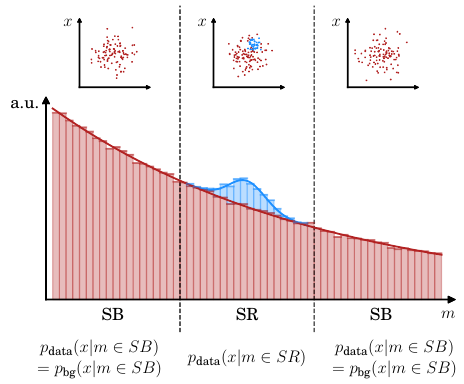}
\caption{Schematic plot from Ref.~\cite{Hallin:2021wme} of a feature of interest (i.e. sensitive to the presence of signal) in the case of a resonant anomaly showing signal regions (SR) and sideband (SB) regions.\label{fig:ResAD}}
\end{figure}

\noindent
{\bf ABCD of background modelling:} 
One well-known data-driven background estimation method used at the LHC is called the ABCD method. It requires two uncorrelated observables, $f$ and $g$. Some threshold values of these, $f_c$ and $g_c$, are used to define four regions: A, B, C, and D. The region A, where both observables are larger than the thresholds ($f > f_c$ and $g > g_c$), is called the signal region (SR). B, C, and D are the control regions (CRs) where either one or both of the observables fall under the thresholds; B has $f < f_c$ and $g>g_c$, C has $f > f_c$ and $g < g_c$ and D is the region where $f < f_c$ and $g < g_c$ (see Fig.~\ref{fig:ABCD}). A machine-learning version of the ABCD method is presented in Ref.~\cite{Kasieczka:2020pil}, where they show that 
the background in the SR can be estimated from the following relation
\begin{equation}\label{eq:abcd}
N^{bg}_A = \frac{N_B\;N_C}{N_{D}},
\end{equation}
where $N_{i}$ is number of events in region $i$. It follows from the independence of $f$ and $g$ and a condition
$$r \equiv \frac{\delta_B + \delta_C +\delta_D}{\delta_A} \ll 1.$$
where $\delta_i$ is the ratio of signal and background events in the region $i$. The ratio, $r$, called the normalised signal contamination, is the signal contamination of the CRs normalised by the signal fraction in the SR. 
\medskip

\noindent{\bf Sideband methods:}
If the signal is localised in a particular region $x_i \in \left(a_0, a_1\right)$ of feature $x_i$, a fit for the background distribution can be performed in the regions $x_i \leq a_0$ and $x_i \geq a_1$, called the sidebands and interpolated into the SR. Such methods are called sideband methods. In such a case, weakly supervised techniques can used to tag the anomaly, as we will see later. Fig.~\ref{fig:ResAD} shows the SR and sideband regions (SBRs) in the presence of a resonant anomaly. 

\medskip

\noindent
{\bf Flow-based methods:} Flow-based methods have also been employed in the recent literature to learn densities ~\cite{Kobyzev_2021, papamakarios2017masked,rezende2015variational}. Flow-based density estimation aims to find the target density distribution of the data from a known, simple probability density function through a set of finite number of invertible transformations. If $x$ and $u$ are two $D$-dimensional real vectors related by a transformation, $x = T(u)$, where $u \sim \rho(u)$; $\rho(u)$ is called the base distribution of the flow-based model. Transformation $T$ has to obey two conditions: it has to be invertible and both $T$ and its inverse have to be differentiable. These requirements are evident when we consider the change-of-variable formula to obtain the density of $x$,
\begin{align}
 \rho_x(x) = \rho_u(u) \left|\det\left(\mathcal{J}_T (u)^{-1}\right)\right|,   \label{eq:NF} 
\end{align} 
where $ u=T^{-1}(x)$ and $\mathcal{J}_T$ is the $D \times D$ jacobian
accounting for the volume change in the neighbourhood around $u$ 
due to the transformation $T$. The transformation $T$ can be modelled to a deep neural network (DNN) and $\rho(u)$ to be an appropriate simple density function, e.g. normal distribution to get the probability distribution of $x$. Normalising flows can model complex distribution starting from simple distributions~\cite{HYVARINEN1999429,NF-review}, a property commonly called expressivity. For detailed reviews on flow-based models, see Refs.~\cite{NF-review,Kobyzev_2021,kobyzev2020normalizing}.


\begin{figure*}[t!]
\captionsetup[subfigure]{labelformat=empty}
\centering
\subfloat[\quad\quad(a)]{\includegraphics[width=0.47\textwidth]{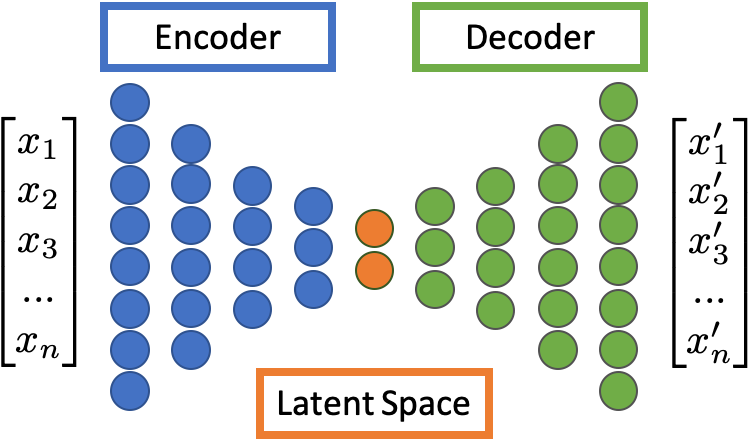}\label{fig:AE-schematic}}\hspace{1cm}
\subfloat[\quad\quad(b)]{\includegraphics[width=0.38\textwidth]{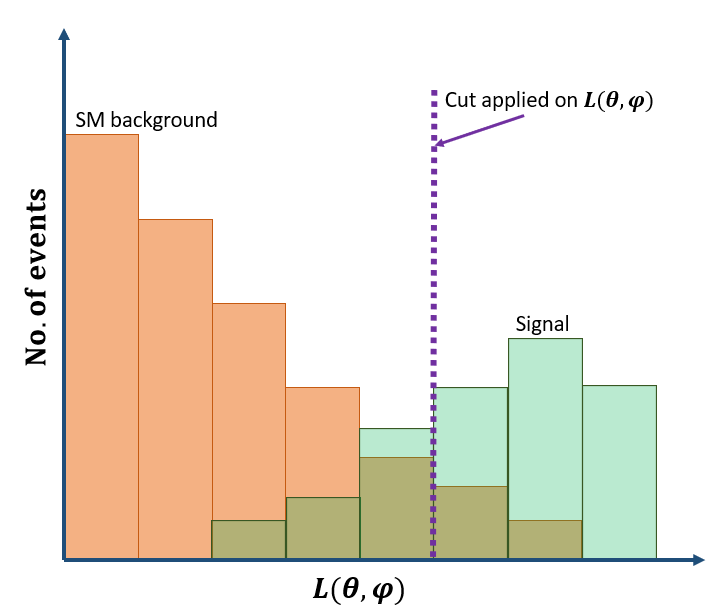}\label{fig:AE-AnomalyScore}}
\caption{Left Panel: Illustration of the AE architecture (taken from Ref.~\cite{Edelen:2021vcs}). An AE comprises two neural networks -- the encoder and the decoder. The encoder learns a transformation from the input $x$ to a compressed (bottleneck, shown in orange) latent space representation $z$, and the decoder learns a transformation from the latent space representation $z$ to reconstructed output $x'$. The AE is trained such that the reconstructed output is ideally identical to the input ($x \approx x'$). AEs learn a data-biased algorithm that is excellent at compressing and reconstructing data. The reconstruction quality of $x'$ with respect to $x$ can be modelled as an anomaly score. Right Panel: The difference between the reconstruction loss of the signal and the background. Since AEs are trained to construct the background very well, the reconstruction error for the background is lower than that for the signal. A cut on the loss can be used to screen for anomalies.}\label{fig:AE}
\end{figure*}

\subsection{Autoencoders}
\noindent AEs are generally used for dimensionality reduction and feature learning, but they can also be used as unsupervised classifiers in some scenarios. In its basic form, an AE has a symmetrical architecture and consists of an encoder and a decoder, as shown in Fig.~\ref{fig:AE-schematic}. The encoder consists of one or more layers of neurons that map the input data ($x \in \mathbb{R}^n$) into a lower-dimensional representation (latent space, $y \in \mathbb{R}^k$ with $k<n$). Most salient features of the input data captured by this compressed representation can be implemented by the function $f_\theta:\mathbb{R}^n \rightarrow 
\mathbb{R}^k$. The number of neurons in the encoder decreases with each layer until it reaches the bottleneck layer. The bottleneck layer (also known as the latent space or the coding layer) has the lowest number of neurons and represents the space with the compressed form of the input data. After that, the decoder reconstructs the input data from the compressed representation, $g_\varphi:\mathbb{R}^k \rightarrow \mathbb{R}^n$. The decoder consists of the same number of layers as the encoder, with the number of neurons in each decoder layer increasing until it reaches the output layer, which has the same number of neurons as the input layer. Each hidden layer in this network typically uses activation functions like the rectified linear unit (ReLU) to introduce non-linearity. The autoencoder training involves minimising a loss function [e.g., mean-squared error (MSE)] that measures the difference between the input data and the reconstructed output:
\begin{equation}\label{eq:ae_loss}
    L(\theta,\varphi) = \frac{1}{N}\sum_{i=1}^N(x^i-g_\varphi(f_\theta(x^i)))^2,
\end{equation}
where $N$ is the number of samples in the training dataset, with $\theta$ and $\varphi$ being the corresponding weights and biases. 

During training, the network learns to map the input data to a lower-dimensional representation and later reconstructs the input from that representation. As mentioned earlier, AEs can be used to identify events inconsistent with the expected background distribution of the SM as a sign of physics beyond the SM. 
The signal events would result in larger $L$ (Eq.~\eqref{eq:ae_loss}) than usual, i.e., a signal is more likely to manifest in the tail of the $L$ distribution, as shown in Fig.~\ref{fig:AE-AnomalyScore}.

\medskip

\begin{figure}[t]
\begin{center}
\includegraphics[width=0.6\textwidth]{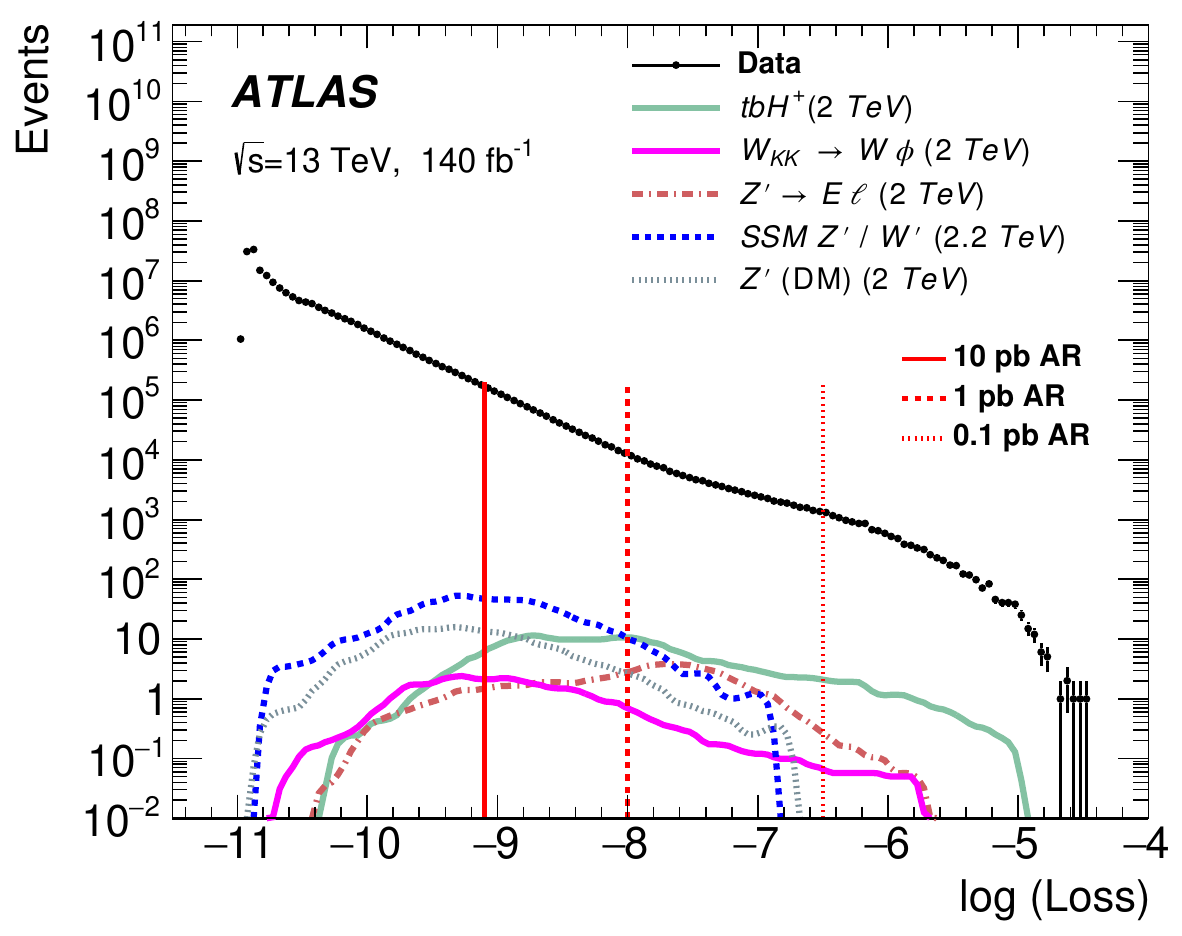}
\end{center}
\caption{The distribution of the anomaly score from the AE is presented for the data and the five benchmark models ($H^+$, $W_\text{KK}$, $Z'$, $W'$ and $Z'$, respectively). The predictions from these BSM models describe the expected number of events from 140fb$^{-1}$ of data for heavy particle masses around 2 TeV. The vertical lines represent the start of the three anomaly regions (AR). The labels associated with the three ARs state the visible cross section for the BSM processes which yields the same number of events as observed in the 140fb$^{-1}$ dataset. For further details refer to Ref.~\cite{ATLAS:2023ixc}.
}
\label{fig:atlas_ae_loss}
\end{figure}

\noindent 
{\bf Autoencoders in 
use:} 
The ATLAS collaboration has used unsupervised anomaly detection to search for new resonances~\cite{ATLAS:2023ixc}. 
The dataset considered for the analysis was a collection of randomly selected $1$\% samples of collision events passing some preselection cuts, not tuned for any specific BSM signal.
The input layer consisted of the kinematic features of the final state objects in the preselected events. The features were in the form of the rapidity-mass matrix (RMM)~\cite{Chekanov:2018nuh} of boost-invariant variables: $E^{\rm{miss}}_T$, transverse energies, transverse masses, two-particles invariant mass, and rapidity differences. An AE was trained and an anomaly score $(\log {L})$ for each event was calculated. The scores tend to be larger for various BSM models characterised by the presence of an isolated lepton in the final states than the SM. Depending on the cut on $\log{L}$, three anomalous regions were considered (Fig.~\ref{fig:atlas_ae_loss}) and nine invariant-mass spectra ($m_{jY}$, where $j$ is a light or $b$ jet and $Y$ is a second light or $b$ jet or a charged lepton ($e,\mu$) or a photon) in each anomaly region were examined to search for any localised excesses above the background. They found that compared to the previous limits obtained using traditional methods (such as peak finders or decomposition into complex functions or employing various soothing methods to describe the complex backgrounds of high-statistics data), the discovery sensitivity improved after the anomaly region selections, showing a factor of $2$–$3$ improvement.

The autoencoding technique is used in Ref.~\cite{Chekanov:2023uot} to boost sensitivity towards NP in dijet resonance searches with RMM inputs to the encoder. The AE based on dense and convolutional neural networks is also used on jets represented as images or $4$-vectors~\cite{Farina:2018fyg, Roy:2019jae}, where QCD jets are the background and boosted top/new particle jets are the signal. A factor of $6$ improvement is observed in Ref.~\cite{Farina:2018fyg}, where they use CNN AE in a jet mass bump hunt for the 400 GeV gluino signal. The reconstruction error in the CNN autoencoder is found to be decorrelated with jet mass, compared to other AE architectures considered in the analysis. The decorrelation effect is important as it results in obtaining data-driven side-band estimates of the QCD background and helps perform a bump hunt. The robustness of the autoencoder in identifying anomalous jets is explored in Ref.~\cite{Roy:2019jae}, wherein the autoencoder loss function does not depend on the initial jet mass, momentum, or orientation of the jets. They additionally show that their AE loss can be readily combined with existing discriminating jet variables (like the Nsubjettiness) or supervised learning techniques like boosted trees to improve their performance as anomalous jet taggers.

Graphs (a set of objects and pair-wise relations among them) can be used to represent data in an intuitive, permutation-invariant way. Graph representations are well-suited to problems in particle physics. For example, parton showering can be visualised as a directed graph along the direction of increasing quark/gluon multiplicity. One could also construct a graph of an event from reconstructed final states and the relationships among them. Graph neural networks (GNNs) are a class of DL models that learn functions from graphical data. Typically training GNNs involves several iterations of message-passing (where nodes exchange information with their neighbours) and learning functions that update each node's state after every iteration (based on its current status and the incoming messages). For a detailed review of general GNN architecture and applications in particle physics, see Ref.~\cite{Shlomi:2020gdn}. The prediction task for a GNN can be at the node (or vertex) level, edge level, or graph level. Ref.~\cite{Atkinson:2021nlt} uses a GNN-based AE for anomaly detection at the graph level. The AE is trained on QCD jets to tag hadronically decaying $W$ jets, top jets, and a boosted scalar (which decays to two $W$ jets to give a four-pronged signature) as anomalies. The authors introduce a novel edge-reconstruction decoder that helps the AE reconstruct the jet graph of the jet.  Ref.~\cite{Atkinson:2022uzb} shows that graph AEs can be made IRC-safe by constructing the graphs in an IRC-safe manner and by constraining the learned function to rely on energy weights for node/edge relationships within the message-passing algorithm.

Hidden-valley models can have an additional strong (confining) gauge group under which all SM particles are singlets~\cite{Strassler:2006im}. However, these hidden dynamics might form light (dark) hadrons, which could be produced at the LHC through some higher-dimensional operators or an intermediary like the dark photon that mixes with the ordinary photon or the Higgs [e.g., the Higgs decaying to dark hadrons], etc. Like the bound states in QCD, dark hadrons can produce dark showers. Since modelling the signals is difficult in these cases, they make a good case for unsupervised learning in general. Ref.~\cite{Barron:2021btf} demonstrated this by looking for soft-unclustered energy patterns (SUEPs) produced by dark showers, which manifest as isotropic and soft energy deposits. The authors analysed one of the experimentally challenging scenarios, where dark hadrons produced in the Higgs decay promptly decay to SM hadrons. Very recently, AEs have been shown to be effective for detecting dark showers or SUEPs~\cite{Anzalone:2023ugq,Chhibra:2023tyf}. 
\medskip

\noindent 
{\bf Robustness of autoencoders:}
The robustness of an AE as an anomaly tagger is tested in an interesting manner in Ref.~\cite{Finke:2021sdf}, which attempts to distinguish boosted top jets from QCD jets with two simple convolutional AEs working on jet images. They train the first AE on a pure QCD-jets sample (direct tagger), and the second one on a pure top-jets sample (inverse tagger). The inverse tagger is used to tag QCD jets as anomalies in a background sample of top jets. The direct tagger AE is found to be a powerful top tagger, agreeing with the findings in Refs.~\cite{Heimel:2018mkt, Farina:2018fyg}. The inverse tagger fails to tag an anomaly, i.e., the QCD jets, and performs worse than randomly tagging jets as anomalous. The authors find that the AE shows a strong complexity bias, i.e., the images that would intuitively be labelled as simpler are reconstructed better. The simpler images are less probable to be identified as anomalous, as AE cannot reconstruct the complex structure of top images when ignoring all but a few pixels. Limitations to the reconstruction of structures in high-energy physics have also been noted recently in Refs.~\cite{Batson:2021agz, Collins:2021nxn}. This correlation between the complexity of an image and the reconstruction loss makes AE unreliable as a model-independent anomaly tagger, as this leads to the assumption that the BSM signal should look more complex than the SM background. They discuss two modifications to the AE to learn more relevant features within the jet images: 1) an intensity remapping for the jet images and 2) the kernel MSE loss function. They find improvement in their results both in terms of the AE performance and the model independence of the tagger. Other papers like Refs.~\cite{Dillon:2021nxw, Hallin:2021wme} also discuss the challenges of using AE for anomaly detection.
\medskip

\noindent
{\bf Challenges from smearing:}
Reconstruction of the final-state objects is one of the main sources of experimental uncertainties (smearing) affecting data-driven methods. The AE learns the kinematic features of the background samples and identifies the signal using the loss function. Smearing shifts the background distribution, making anomaly detection difficult in the case where the smeared background distribution comes closer to the signal. The traditional AE can be combined with an adversarial network (AN, more about it in Sec.~\ref{sec:det_sim}) to improve the reliability and robustness of this approach, making the autoencoder's reconstruction independent of the smearing of the background, as shown in Ref.~\cite{Blance:2019ibf}. The authors consider the $3$-momenta of the jets, leptons, and the missing transverse energy to be smeared with Gaussian functions. They train their network with simulated (or pseudo) data with the background events smeared in three possible ways (upwards, downwards, and zero smearing for all objects). The AN takes the AE loss as the input and attempts to discriminate the smearing class of the background. This, in turn, forces the classifier to make its prediction independent of 
the smearing, making it hard for the AN to discriminate the background samples. Since AE training is generally insensitive to minor signal contaminations, even though their analysis is done on pseudo-data, the authors argue that this procedure can be applied analogously to experimental data by creating labelled datasets that have been systematically smeared.

\subsection{Variational autoencoders}
Variational autoencoders (VAEs) are a more recent and interesting take on autoencoding. VAEs, instead of mapping an input to a fixed vector, map the input to a distribution. 
In a VAE, the encoder models the posterior distribution of the model $\rho(z|x)$ and the decoder models the likelihood $\mathcal L=\rho(x|z)$. Two separate vectors replace the latent representation: the means and the standard deviations of the distribution. The probabilistic nature of the latent space allows for the generation of new samples by sampling from the distribution in the latent space.
There being no global representations shared by all data points, the loss function can be decomposed into terms that depend on a single data point $L_i$. The loss function $\ell_i$ for a datapoint $x_i$ is:
\begin{eqnarray}
    L_i(\theta,\varphi) = -\mathbb{E}_{z \approx \pi_\theta(z|x_i)}[\log \rho_\phi(x_i|z)] +\mathbb{KL}(\pi_\theta(z|x_i)\;||\;\rho(z)).\label{eq:VAEloss}
\end{eqnarray}
The first term is the reconstruction loss or the expected negative log-likelihood for the $i$-th data point, given the latent variable $z$. The expectation is taken with respect to the encoder’s distribution over the representations. The second term is the Kullback-Leibler divergence (see Appendix \ref{sec:KLD} for more details) between the encoder’s distribution $\pi_\theta(z|x)$ and the prior $\rho(z)$:
\begin{equation}\label{eq:KLloss}
    \mathbb{KL}(\pi_\theta(z|x)\;||\;\rho(z))=-\frac{1}{2}\sum_{j=1}^J \left(1+ \log \sigma_j^2 -\mu_j^2 -\sigma_j^2 \right),
\end{equation}
where $J$ is the dimension of the latent space, $\mu_j$ and $\sigma_j$ are the mean and the standard deviation of the approximate posterior $\pi_\theta(z|x)$ for the $j$-th dimension. The term helps in regularising the latent space and promoting a smoother structure. In the VAE, the prior $\rho(z)$ is specified as a standard normal distribution, $\rho(z)=$ Normal($0,1$). If the encoder outputs representations $z$ that are not from a standard normal distribution, it will receive a penalty in the loss. (The encoder could learn to cheat without this term and give each data point a representation in a different region of Euclidean space.) Note that the weights and biases of the VAE are optimised such that the evidence lower bound (ELBO) of the probabilistic model evaluated on the dataset is maximised:
\begin{equation}
    \text{ELBO} = \log \rho(x) + \mathbb{KL}(\pi_\theta(z|x_i)\;||\;\rho(z|x))
\end{equation}
By Jensen’s inequality, the Kullback-Leibler divergence is always nonnegative. This means that minimising the Kullback-Leibler divergence is equivalent to maximising the ELBO, as $\rho(x)$ is a constant value.

The information encoded in the latent space of the VAE has been used effectively for anomaly detection in dijet events~\cite{Bortolato:2021zic}. The paper shows that $\mathbb{KL}$ divergence could be a good indicator of anomalous jets in the dataset if the anomaly scores calculated from the VAE are used for selecting signal events with a cut on the $\mathbb{KL}$ divergence for each event. The authors further modify the decoder architecture so that the latent space also uses the invariant mass information from each event as input for the final reconstruction apart from the means and variances from the encoder network. This helps in interpreting anomalous events as localised bumps in the (invariant mass contours in the latent space) spectrum and defining signal windows. 
\medskip

\noindent
{\bf Real-time anomaly detection:} Approximately $40$ million proton-beam collisions occur every second at the LHC. However, due to limited bandwidth and storage resources, only $\sim 1000$ collision events/sec can be stored by the ATLAS and CMS experiments. The particle detectors discard most collision events with an online selection system of two stages: Level 1 Trigger (L1T) and High-Level Trigger (HLT). In L1T, the algorithms are deployed as programmable logic on custom electronic boards equipped with field-programmable gate arrays (FPGAs). These cut the event rate by $2.5$ orders of magnitude within a few microseconds~Refs.~\cite{Cerri:2018anq,Govorkova:2021utb}. The HLT consists of selection algorithms that asynchronously process the events accepted by the L1T on commercially available CPUs. Physicists, while looking for BSM scenarios, design the triggers motivated by the theoretical considerations of the model they are looking for. However, there is always a possibility that BSM scenarios without theoretical construction bypass the triggers and thus escape detection. 

Ref.~\cite{Cerri:2018anq} proposes that VAEs in the HLT system can identify recurrent anomalies otherwise escaping detection. The VAE trained on known SM processes in Ref.~\cite{Cerri:2018anq} would be able to identify BSM events as anomalies. It considers a single-lepton data stream as an example of the initial trigger. The VAE is trained on the four largest-production-cross-section SM processes with a lepton in the final states. The features used for input to VAE are all high-level, chosen mainly to represent the physics aspects of the SM processes. The authors demonstrate the robustness of their architecture against systematic uncertainties, along with its potential to provide a high-purity sample of interesting anomalous (BSM) events. They show that events produced by not-yet-excluded BSM models with cross sections in the range of $\mathcal{O}(10)$ to $\mathcal{O}(100)$ pb could be isolated in a $\sim$ 30\% pure sample of $\sim 43$ events selected per day. 

In Ref.~\cite{Govorkova:2021utb}, the idea of using AEs and VAEs for online selection in the L1T is explored. L1T mainly uses theory-motivated simple requirements like the minimum energies of reconstructed leptons, jets, etc. The authors propose that an anomaly detection algorithm deployed in the LIT will be more effective as it will collect data in a model-independent way. The recent development of the \textsc{hls4ml} open-source library has allowed the authors to translate the AD algorithms on the FPGAs firmware mounted on the L1T boards. They use a DNN AE and a CNN AE separately on FPGAs. Instead of using the standard loss function as the anomaly score, Ref.~\cite{Govorkova:2021utb} uses the $\mathbb{KL}$ divergence term and the $Z$-score of the origin $\vec{0}$ in the latent space with respect to the Gaussian distribution with standard deviation $\vec{\sigma}$ and centred at $\vec{\mu}$ ($\vec{\sigma}$ and $\vec{\mu}$ values are returned by the encoder). 
This helps in saving resources and latency by not having to evaluate the decoder in the trigger. There is also no need to do Gaussian sampling on the hardware, which will preserve the deterministic output of the trigger. They demonstrate that BSM signatures can be enhanced by three orders of magnitude while staying within the strict latency and resource constraints of a typical LHC event filtering system. The anomaly detection is shown to be performed in as little as $80$ ns using less than $3$\% of the logic resources in the Xilinx Virtex VU9P FPGA. This capability to perform real-time anomaly detection on FPGAs has generated attention within the community and has opened the way to use this idea during the next data-taking campaign at the LHC.

The CMS experiment has recently used an ML-based trigger algorithm, AXOL1TL (using a VAE, dense neural network)~\cite{Zipper:2023ybp}, which selects anomalous events in real-time. It was trained with unbiased data collected by the CMS experiment in 2023 at a centre-of-mass energy of $\sqrt{s}$=13.6 TeV and was implemented in Level-1 Global Trigger Test Crate for 2023 data taking. The Global Trigger Test Crate is a copy of the main Global Trigger system---it receives the same input data but does not trigger the readout of CMS. This provides a platform for thoroughly testing new trigger algorithms on live data without interrupting data taking. The AXOL1TL algorithm has shown an improvement of the signal efficiency by up to $46$\% for the BSM signal of a Higgs decaying to two (pseudo)scalars of mass $15$ GeV each of which decays to bottom quark pairs. Their results show that the Anomaly Detection trigger built on Xilinx Virtex-7 FPGA operates at an extremely low latency of $50$ ns and takes up only a small fraction of the resources.

\medskip
\noindent
{\bf Non-resonant anomaly detection:} Ref.~\cite{Mikuni:2021nwn} proposes a new method for unsupervised anomaly detection in high-dimensional data.  It is based on the idea that a complete anomaly detection algorithm, apart from being sensitive to anomalous events, should also be able to estimate the rate of the SM events that are labelled as anomalous, mostly false positive rate. The authors propose to train two (or more) statistically independent AEs at the same time. This enables data-driven background estimation through a regulariser based on the distance correlation (DisCo) measure of statistical dependence. The events are classified as anomalous if their reconstruction quality is poor across all autoencoders. The loss function for the two autoencoders takes the form:
\begin{eqnarray}
    L[f_1,f_2,g_1,g_2]&=& \sum_i R_1(x_i)^2 + \sum_i R_2(x_i)^2 +\lambda~\mathrm{DisCo}^2[R_1(X),R_2(X)],
\end{eqnarray}
where $f_{1,2}, g_{1,2}$ are the respective encoder and decoder functions, $R_i(x) = (f_i(g_i(x))-x)^2$, $\lambda>0$ is a hyperparameter and DisCo is the distance correlation between 0 and 1. DisCo is zero if and only if the arguments are independent. This forces the AEs to learn independent representations of the input data (background). Here, the distance correlation is computed at the level of a batch of examples $x$, which are realisations of the random variable $X$. 

They define counts as $N_{\lessgtr,\lessgtr}(\Vec{c})=\sum_i \mathbb{I}\left[R_1(x_i)\lessgtr c_1\right]\mathbb{I}\left[R_2(x_i)\lessgtr c_2\right]$, where $\Vec{c} = (c_1,c_2)$ are the thresholds and $\mathbb{I}[.]$ is the indicator function that is $1$ when its argument is true or $0$ otherwise. Three of these regions, $N_{<,<}, N_{<,>}, N_{>,<}$, are background dominated, while the fourth, $N_{>,>}$, is the signal sensitive region. The background in the signal region can be predicted from the other three regions with the ABCD method (discussed in the previous section):
\begin{equation}
    N_{>,>}^{predicted}(\vec{c})=\frac{N_{<,>}(\Vec{c}),N_{>,<}(\Vec{c})}{N_{<,<}(\Vec{c})}.
\end{equation}
The paper demonstrates that the decorrelated AE protocol is an effective tool for simulation-free, non-resonant anomaly detection. This method is online compatible and can run in the trigger system. The paper proposes to preserve all events falling within the signal-sensitive region as defined by the two AEs. A random subset of events in the three other regions would be saved for subsequent offline background estimation. 
\medskip

\noindent
{\bf Anomalous jet tagging with VAE:} Ref.~\cite{Cheng:2020dal} uses the VAE to detect anomalous non-QCD jets and shows that unsupervised learning without any guidelines may not give the optimal solution. Simple VAEs based on different anomaly metrics become highly correlated with the jet mass when tagging anti-QCD anomalies, whereas the mass-decorrelated DisCo-VAE show poor discrimination performance in the full test spectrum. The paper, therefore, looks into a simple semi-supervised approach to enhance the performance. The training dataset consists of $600,000$ simulated QCD di-jets (of which $20$\% is reserved for validation set) for the $13$-TeV LHC. The jets are clustered using the anti-$k_T$ algorithm with a cone size of $R = 1.0$ and a selection cut $p_T > 450$ GeV. The first $20$ hardest ($p_T$-ordered) jets are considered as input, with the features ($E, p_x, p_y,  p_z$) being suitably boosted, rotated, and standardised. 

The authors have treated the $\mathbb{KL}$ divergence term in the VAE loss function [Eq.~\eqref{eq:VAEloss}] as a regularisation term and the loss function is subsequently modified to:
\begin{eqnarray}\label{eq:loss_vae}
    L(\theta,\varphi) = -\mathbb{E}_{z \approx q_\theta(z|x_i)}[\log p_\phi(x_i|z)] +\beta ~\mathbb{KL}(q_\theta(z|x_i)||p(z));
\end{eqnarray}
where $\beta$ denotes the relative strength of the latent regularisation. They found that changing the value of $\beta$ in these $\beta$-VAE models affects the competition between fitting the latent distribution and the input space reconstruction. With $\beta$ = 0, the VAE reduces to the deterministic autoencoder. The reconstruction performance drops with increased $\beta$ since there is an extra effort to fit the latent distribution. The authors have tested different values $\beta$ = 0.1, 0.5, 1.0, and 5.0, and observed that $\beta$ = 0.1 gives a better balance between input reconstruction and latent coding. The authors have also looked into different metrics such as the negative log-likelihood (Eq.~\eqref{eq:loss_vae}, with $\beta$ = 1), MSE reconstruction error, $\mathbb{KL}$ divergence in the latent space, and energy mover's distance~\cite{Komiske:2019fks} to compute the anomaly score. They find that the VAE performance in AUC is correlated with the jet mass, with the anomaly metric KL divergence ($\mathbb{KL}(q_\theta(z|x_i)||p(z))$) slightly outperforming the others in most of the BSM cases. This technique brings difficulty in tagging jets with low mass, therefore they next employ the DisCo regularization as a mass-decorrelation baseline but find them to have poor discrimination performance in the full test spectrum. The DisCo regularization term measures the non-linear correlation between the jet mass and the VAE loss, Eq.~\eqref{eq:loss_vae}.

The failure of the simple and mass-decorrelated DisCo-VAE motivated the authors of Ref.~\cite{Cheng:2020dal} to finally introduce outlier-exposed VAE (OE-VAE) as a solution, where some signal events (outliers in this case) were injected into the training process and the model was required to separate the outliers from the training samples. This became a semi-supervised process whereby asking the VAE to separate the outliers from the training set, this approach guides and filters the information learned by the model. The information contained in the jet mass that was earlier used to help discriminate against signals is excluded from the learned representations, by matching the mass distribution.  The same analysis is performed and it is found that the OE-VAEs generally have much higher signal efficiency w.r.t. DisCo-VAEs at the same mass-decorrelation level.

\subsection{Weak supervision}\label{sec:weaksup}
Weakly supervised methods are used when event-level data labels are not available. These methods can be loosely grouped into three categories: Incomplete Supervision, Inexact Supervision, and Inaccurate Supervision. Among these, Incomplete Supervision is the closest to fully supervised methods, as the model learns from a partly labelled data set where only $L$ out of $N$ ($L<N$) samples of the training data are labelled. We will not discuss this further; instead, we will look at examples of the other two. 
\medskip

\noindent
{\bf Learning from labelled proportions:}
Inexact supervision is the scenario where we don't have per-data labels but have truth information for bags of data; i.e., the labels are not available with the precision needed for a supervision task. We can consider the example of Ref.~\cite{Dery:2017fap}, where the authors propose the learning from labelled proportions (LLP) model to separate quark jets from gluon jets at the LHC. Even though it is possible to estimate the fractions of outgoing quark vs gluon jets with our current knowledge of the parton distribution functions and the state-of-the-art simulation tools, it is not possible to accurately tag them at the event level with high efficiency from QCD-jet data. However, Ref.~\cite{Dery:2017fap} shows that the knowledge of the fraction can help us with the task. In the proposed LLP paradigm, the classifiers are trained on multidimensional data with only class proportions instead of individual sample labels. For example, to classify samples from two classes labelled $0$ and $1$, one can construct a function of the following form for training\footnote{Arg min$_x$ is defined as the values of the argument for which the function it is acting on is minimum; i.e., arg min$_xf(x)$ returns the set of $x$ values for which $f(x)$ becomes the minimum.}:
\begin{align}
F= \mbox{arg min}_{f:\mathbb R^n\to\{0,1\}}\ L\left(\frac1N\sum_{i=1}^N f(x_i)-y\right),
\end{align}
where $\lim_{x\to0}L(x)=0$ is a loss function and $y=\sum_it_i/N$ with $t_i$ being the true label of the sample $i$, and $N$ is the batch size.
\begin{figure}[t]
\centering
\includegraphics[scale=0.6]{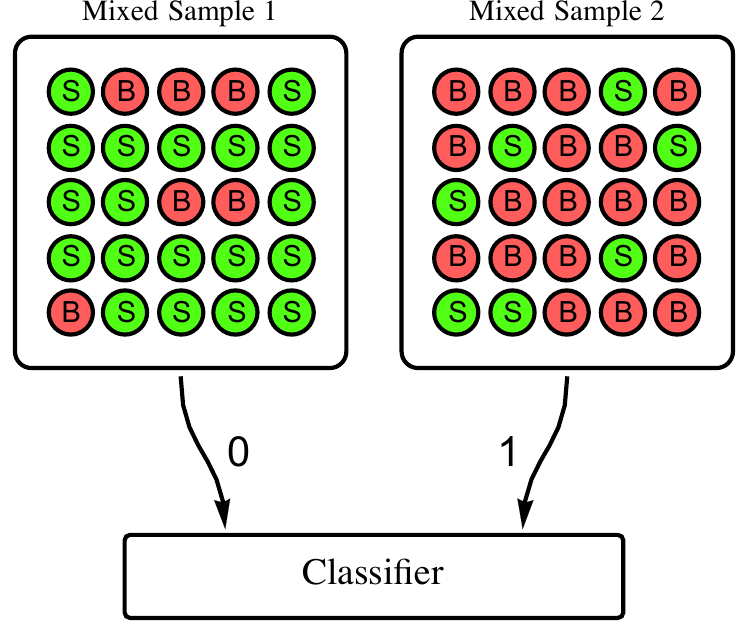}
\caption{\emph{Learning from noisy mixtures using the CWoLa method}~\cite{Metodiev:2017vrx}: Schematic diagram showing the learning process using the Classification without Labels method. A supervised classifier is trained to identify the data from Sample 1 and Sample 2, one of which contains more signal events than the other. The trained classifier is then used to perform the classification between signal and background events.} 
\label{fig:CWoLa}
\end{figure}
\medskip

\noindent
{\bf Classification Without Labels (CWoLa):} In the case of Inaccurate Supervision, the labels are noisy. Classification Without Labels (abbreviated as CWoLa, pronounced ``koala")~\cite{Metodiev:2017vrx} is an example of such learning tasks. CWoLa-based classification models work under a crucial assumption: the signal is localised as an excess or a bump in one of the features, but the background is smooth over the features; i.e., it aims to tag a resonant anomaly, as illustrated in Fig.~\ref{fig:ResAD}. Instead of directly training a classifier on the labelled data, the method creates two noisy mixtures of the signal and background events, say $M_1$ and $M_2$. These mixtures are designed to correspond to the signal and sideband regions, so they are to be different in the number of signal events they contain: one of them (say $M_1$, from the signal region, SR) should have much more signal than the other ($M_2$, samples from the sideband region, SBR). A fully-supervised classifier is then trained over these mixtures to correctly predict the class ($M_1$ or $M_2$) of a given sample. This classifier is then used to perform the actual classification of signal events from that of the backgrounds. If the mixtures are similar, i.e., the fraction of signal in each mixture is $\sim 0.5$, the classifier will perform poorly as the mixtures become indistinguishable. Unlike the LLP paper, the fractions of events are never used in training the classifier. It can be shown that a classifier trained on a mixture of noisy mixture of signals and backgrounds can perform equally well as one trained on pure samples~\cite{doi:10.1142/12294,Metodiev:2017vrx}. Let the signal fraction in $M_1$ and $M_2$ be $f_1$ and $f_2$ respectively. Then,
\begin{align}
    \rho_{M_1}(\vec{x}) &= f_1\:\rho_S(\vec{x}) + (1-f_1)\:\rho_B(\vec{x}), \nonumber \\
    \rho_{M_2}(\vec{x}) &= f_2\:\rho_S(\vec{x}) + (1-f_2)\:\rho_B(\vec{x}).
\end{align}
The optimal classifier to distinguish samples from $M_1$ and $M_2$ is
\begin{equation}
    L_{M1/M2} = \frac{\rho_{M1}(\vec{x})}{\rho_{M2}(\vec{x})} = \frac{f_1\:\rho_S(\vec{x}) + (1-f_1)\:\rho_B(\vec{x})}{f_2\:\rho_S(\vec{x}) + (1-f_2)\:\rho_B(\vec{x})} = \frac{f_1\:L_{S/B} + (1 - f_1)}{f_2\:L_{S/B} + (1 - f_2)},
\end{equation} 
which is monotonically related to the optimal classifier to distinguish samples drawn from the pure distribution of $S$ and $B$, $L_{S/B}(\vec{x}) = \rho_S(\vec{x})/\rho_B(\vec{x})$, provided $f_1 > f_2$.

\begin{figure}[t]
\centering
\includegraphics[scale=0.46]{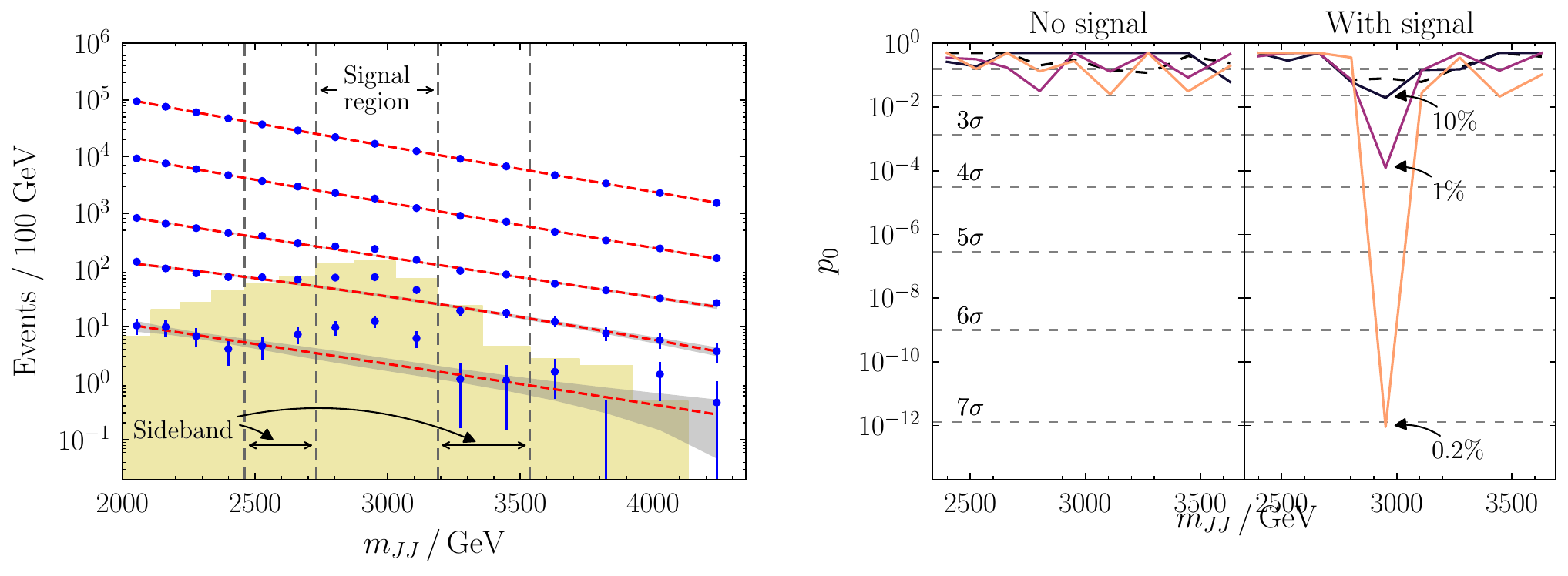}
\caption{\emph{CWoLa hunting in the dijet final state}~\cite{Collins:2019jip}---Left Panel: Background plus signal sample information where the signal and sideband regions are tuned for a $3$ TeV resonance. Blue dotted lines show the effect of increasing the classifier threshold on the dijet background distribution from the ATLAS data; the lowest one has the highest threshold of about $0.2\%$, whereas the topmost one has no cut applied on it at all. Right Panel: When the background-only sample is considered (first plot), no bump is observed. In the signal injected sample (second plot), a sharp bump can be seen at various classifier thresholds, the most prominent one showing up $0.2\%$ classifier threshold.}
\label{fig:CWoLaHunt}
\end{figure}
\medskip

\noindent
{\bf  CWoLa hunting:}
Hunting for resonant anomalies using the CWoLa can also use the information in the auxiliary features rather than the one where we expect the anomaly to be localised. To see how one performs bump hunting using CWoLa~\cite{Collins:2019jip}, let $m_{res}$ be the feature where the signal is localised and $Y$ ($\in \mathcal{R}^n$) be an auxiliary feature uncorrelated to $m_{res}$. CWoLa hunting, as the method is called, uses the information in the auxiliary feature to narrow down the regions with signal-like events. In Ref.~\cite{Collins:2019jip}, an ensemble of fully connected NNs are trained on $Y$ to differentiate samples from the SR and the SBR using five-fold cross-validation (the training data is split into five sets, and four of them are used for training and validation one at a time while the other is kept aside as the test set) and the average of the best-performing model for each $k$-fold validation is selected as the ensemble classifier. The performance metric chosen for this classification task is the true positive rate (SR events) for a given false positive rate (SBR events) of $\sim 1\%$. An ensemble classifier is used to select events passing a threshold (a percentage of most signal-like events) on the test set; these are the signal-like events. These events are binned in $m_{res}$ and a smooth fit for the background is done by masking out the SR. 

For an illustration, let us consider the example given in Ref.~\cite{Collins:2019jip}. The benchmark signal considered there is the production of $3$-TeV $W'$ bosons in the fully hadronic channel. The heavy $W'$ bosons decay to a $W$ boson and a new scalar $X$ of mass about $400$ GeV, which further decays to a pair of $W$ bosons (since $2m_W \ll M_X \ll M_{W'}$): $p p \to W' \to W X,\;X \to W^{+} W^{-}$. The final-state signature is taken as two fatjets, where one of them is two-pronged ($W$ boson) and the other is four-pronged (the new scalar $X$), and QCD-dijets events are considered as the background. The CWoLa hunting method is evaluated on two datasets, one without any signal events and the other a mixture of signal and QCD dijet events. The feature of interest is $m_{JJ}$, the invariant mass of the two-fatjet system.
The auxiliary variable, which has to be uncorrelated with $m_{JJ}$, is created from the mass of the fatjet and substructure variables: for each fatjet $J_i$, $Y_i = \left( m_J, \sqrt{\tau^{\beta=2}_{1}}/\tau^{\beta=1}_{1}, \tau^{\beta=1}_{21}, \tau^{\beta=1}_{32}, \tau^{\beta=1}_{43}, n_{trk} \right)$, where $n_{trk}$ is the number of tracks associated with the fatjet and $\tau_{\alpha\delta}$'s are the $N$-subjettiness ratios~\cite{Thaler:2010tr} for $\alpha$-prongness check. Thus, the auxiliary variable becomes the $12$-features long, $Y\equiv (Y_1, Y_2)$. 
The SR and the SBR for the $m_{JJ}\simeq 3$ TeV hypothesis are chosen as in the left panel of Fig.~\ref{fig:CWoLaHunt}. As seen in the right panel of Fig.~\ref{fig:CWoLaHunt}, a clear bump is visible in the case of the signal-injected sample for different classifier thresholds, whereas no such peak is seen in the other mixture.

The authors also compare CWoLa hunting with a fully supervised tagger for the $WX$ final states using ROC curves (Reciever Operating Characteristic curve showing the background rejection rate against signal efficiency) and show the performance of the CWoLa-based classification improves with increasing signal events in the signal region. 

The Tag n' Train method proposed in Ref.~\cite{Amram:2020ykb} is similar to CWoLa hunting. It assumes that the data has two distinct objects (like jets or some other final-state objects) that are uncorrelated with each other. First, a weak classifier is trained to identify signal-like and background-like events based on one of the objects, say $O1$. Signal-rich and background-rich events tagged by the weak classifier serve as the input data to train another classifier, which is trained on features of $O2$ to distinguish between the signal-rich and background-rich samples. For example, while using the dijet anomaly dataset (mentioned above) for evaluating performance, the authors pick the first massive jet as $O1$ and the second one as $O2$; they use AEs for weak classification and CNNs as the second classifier as they use jet images as input data.
\medskip

\noindent
{\bf ANOmaly detection with Density Estimation (ANODE):}
CWoLa hunting shows that auxiliary features, uncorrelated to the feature where the anomaly is localised, can play a critical role in narrowing down the region of interest.
Ref.~\cite{Nachman:2020lpy} uses a more robust background estimation technique based on normalising flows to 
extend the sensitivity of the resonant anomaly (i.e., an unexpected resonance) searches. The method---called `ANODE' (short for ANOmaly detection with Density Estimation)---learns the densities, $\rho_{\text data}(Y|m)$ and 
$\rho_{B}(Y|m)$ for $m\in$ SR, in the feature space $Y$ and then the likelihood ratio, 
\begin{equation}
    \mathcal L^R(Y|m) = \frac{\rho_{\text data}(Y|m)}{\rho_{\text B}(Y|m)}\label{eq:ANODE1}
\end{equation}
is used for classification tasks. The Neyman-Pearson lemma tells us this is the optimal classifier for an ideal scenario where $\rho_{\text data}(Y|m) = \alpha\:\rho_{B}(Y|m) + (1-\alpha)\:\rho_{S}(Y|m)$, where $0\leq\alpha\leq 1$. In the absence of signal, $\mathcal L^R(Y|m) = 1$. Thus, $\mathcal L^R_{\text data}(Y|m)$ is greater than $1$ in the SR and close to $1$ in the SBR. The density, $\rho_{\text data}(Y|m)$, is estimated in the SR and $p_{B}(Y|m)$ is estimated in the SBR using masked autoregressive flow (MAF) and then interpolated into the SR. 

\begin{figure}[t]
\centering
\includegraphics[scale=0.28]{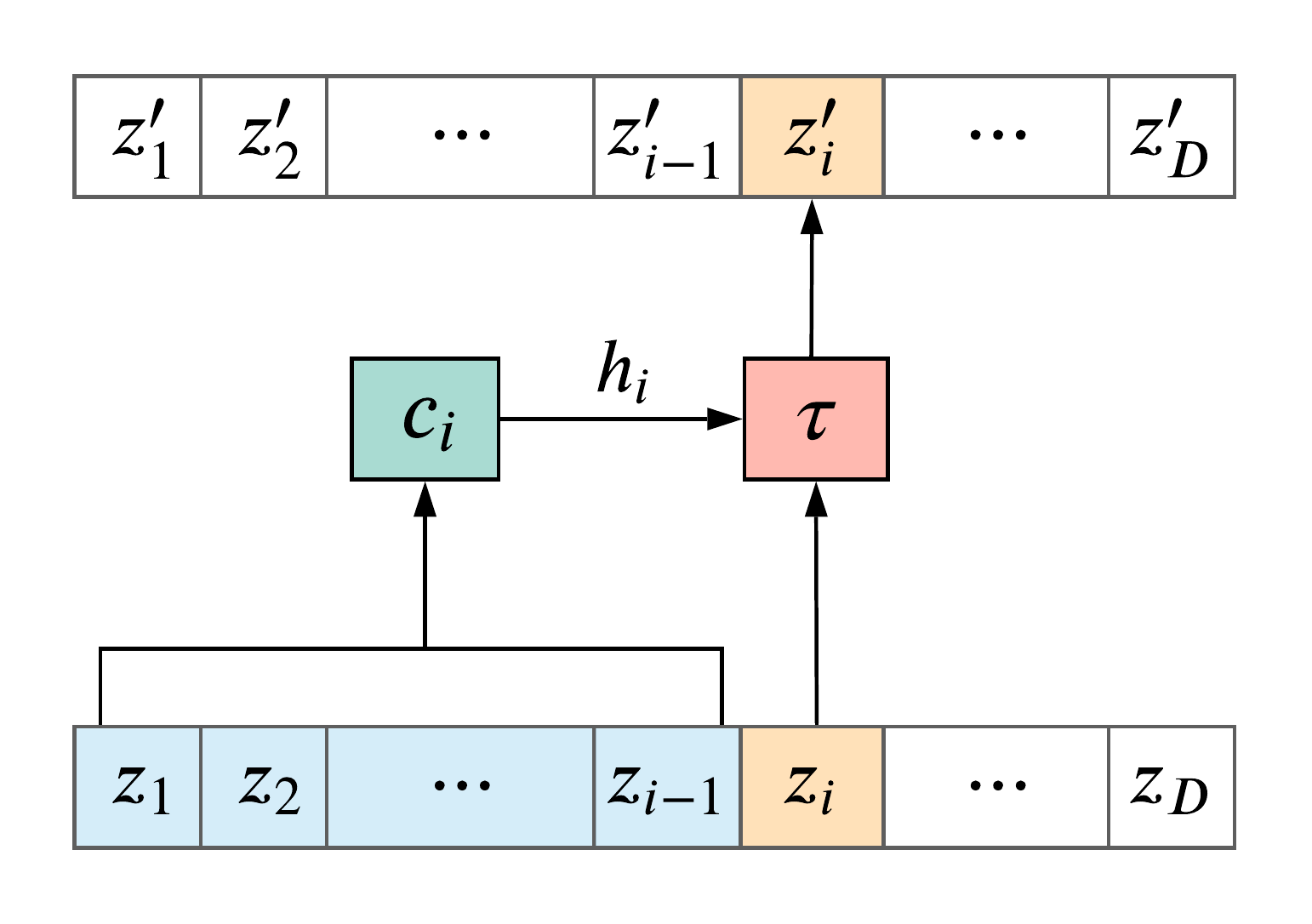}
\caption{Autoregressive flow schematic taken from Ref.~\cite{NF-review}.}
\label{fig:AutRegFlo}
\end{figure}

We briefly described how to model a general normalising flow in Eq.~\eqref{eq:NF}. An autoregressive flow is modelled as
\begin{equation}
    z' = \tau(z_i, h_i), \text{ with } h_i = c_i(z_{< i}),
\end{equation}
where, $\tau$ is a transformer and $c_i$ is the $i$-th conditioner. The transformer $\tau$ is a strictly monotonic function of $z_i$ and the $i$-th conditioner can only take values of $z$ with indices less than $i$; for $z, z' \in \mathcal{R}^D$, there are $D$ conditioners. This requirement on the conditioners gives the flow model its name. A schematic for the working of an autoregressive flow is shown in Fig.~\ref{fig:AutRegFlo}. Masked conditioners are modelled as feedforward DNNs that take in the $z$ vector and output the $(h_1, h_2, ..., h_D)$ in one pass. To implement the autoregressive structure, connections between the $h_i$ and $z_{\geq i}$ need to be removed. Weight matrices of the feed-forward DNN can multiplied with a binary matrix (where each element is either $0$ or $1$) to remove the undesired connections or to ``mask'' them out. 

\begin{figure*}[h!]
\captionsetup[subfigure]{labelformat=empty}
\centering
\subfloat[\quad\quad(a)]{\includegraphics[width=0.46\textwidth]{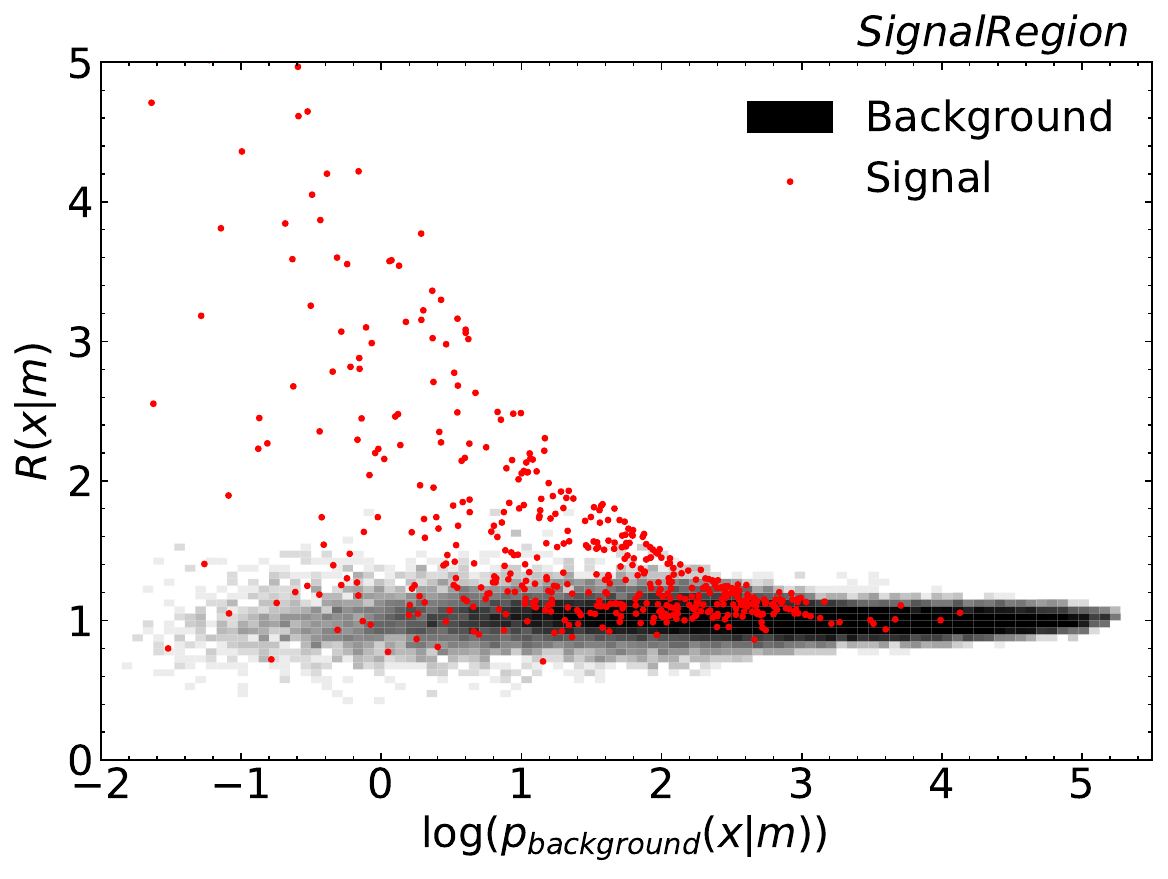}\label{fig:ANODEa}}\hspace{1cm}
\subfloat[\quad\quad(b)]{\includegraphics[width=0.46\textwidth]{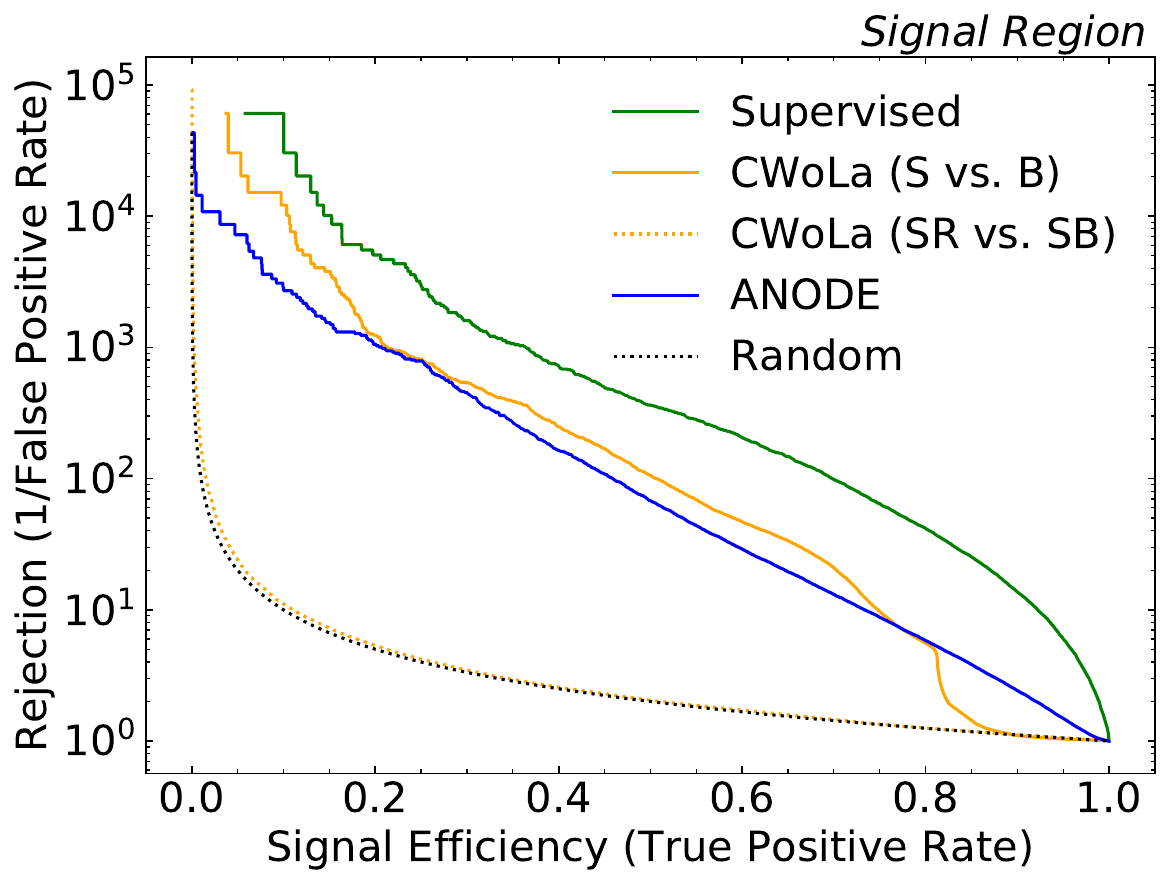}\label{fig:ANODEb}}
\caption{\emph{Anomaly Detection using Density Estimation (ANODE)}~\cite{Nachman:2020lpy}: Here, $R(x|m)\equiv\mathcal L^R(x|m)$ defined in Eq.~\eqref{eq:ANODE1} and $p_{background}(x|m)\equiv\rho_B(x|m)$ in our notation. (a) The effectiveness of the density estimation can be seen here. Background distribution, as one would expect, is mainly concentrated along the $\mathcal L^R(x|m) = 1$; The signal distribution has a long tail at higher values of $\mathcal L^R(x|m)$ in the SR. (b) ROC curve showing the performance of ANODE compared with a fully-supervised classifier and the CWoLa method-based classifier.}\label{fig:ANODE}
\end{figure*}
Ref.~\cite{Nachman:2020lpy} uses a masked autoencoder implementation~\cite{MADE} for both SR and SBR density estimators and use a dijet search to showcase the performance of ANODE, reproduced in  Fig.~\ref{fig:ANODE}. The effectiveness of the density estimation in the signal region on a test set is shown in Fig.~\ref{fig:ANODEa}. As expected, the background distribution is mainly concentrated on $\mathcal L^R(x|m) = 1$. It shows a spread at the lower end of $\log \rho_{B}(x|m)$, as the density estimation becomes less accurate because of the lower background density. The signal distribution has a long tail at higher values of $\mathcal L^R(x|m)$ as expected in the SR. Fig.~\ref{fig:ANODEb} shows the ROC curve of ANODE contrasted with a fully supervised classifier and CWoLa hunting; ANODE is comparable with CWoLa hunting, but the latter outperforms ANODE at lower signal efficiencies. For the particular dataset, the authors show that ANODE can increase the significance score by a factor of $7$. An update to the ANODE method called Residual ANODE or R-ANODE~\cite{Das:2023bcj} fits the signal distribution in the SR using normalising flows while using a fixed background template. 
\medskip

\noindent
{\bf Classifying Anomalies THrough Outer Density Estimation (CATHODE):}
Ref.~\cite{Hallin:2021wme} introduces a learning method CATHODE (short for Classifying Anomalies THrough Outer Density Estimation) that combines the ideas of CWoLa hunting and ANODE. Like ANODE, a MAF is trained to estimate the background distribution in the SBR, which is then interpolated into the SR. The MAF is constructed with an affine transformer and is defined as 
\begin{equation}
\tau(z_i, h_i) = \alpha_i\:z_i + \beta_i, \text{ where } h_i = \left\{ \alpha_i, \beta_i \right\},    
\end{equation} 
which is invertible for $\alpha_i \neq 0$. However, instead of constructing likelihood ratios, here one generates samples from the interpolated region. Like CWoLa hunting, a classifier is trained to distinguish $\rho_{\text{data}}(x|m)$ and $\rho_{B}(x|m)$, where the latter is obtained by generating using the density estimator. CATHODE outperforms both ANODE CWoLa hunting and maintains nearly optimal performance even in challenging cases. CATHODE outperforms ANODE because it does not have to directly learn $\rho_{\text{data}}$ in the SR or the sharp increases in $\rho_{\text{data}}$ where the signal is localised. It also outperforms CWoLa hunting for two reasons:  (1) CWoLa hunting is limited to actual data from the SBR, however, with CATHODE, one can oversample the SBR to learn a better classifier and (2) CWoLa requires the auxiliary features to be independent of $m_{\text{res}}$, while CATHODE incorporates information from these features through a neural density estimation. 
\medskip

\noindent
{\bf Constructing Unobserved Regions by Transforming Adjacent Intervals (CURTAINs):}
In another technique called CURTAINs~\cite{Raine:2022hht}---short for Constructing
Unobserved Regions by Transforming Adjacent Intervals---a conditional invertible neural network (cINN, an example can be seen in Fig.~\ref{fig:cinn}) is trained with an optimal transport loss to learn the background from the SBR. The cINNs are trained to transform data from the lower sideband of $m_{res}$ ($m_{JJ}$ in Ref.~\cite{Raine:2022hht}) to the higher sideband in the forward pass and in the opposite direction in the inverse pass. These are conditioned on $m^{low}_{JJ}$ and $m^{high}_{JJ}$ (belonging to the respective sidebands) using the conditioning function $f(m^{low}_{JJ},m^{high}_{JJ}) = m^{high}_{JJ} - m^{low}_{JJ}$. In the forward pass, $m^{low}_{JJ}$ is the true value of the input and $m^{high}_{JJ}$ becomes the target value. 
The authors pick the same dataset as described in the CWoLa hunting example before; the target values are sampled from a fit to QCD dijet background invariant mass distribution.
The loss function is taken as the Sinkhorn divergence, which measures the distance between the original and the transformed distributions. After every epoch of training, intra-sideband training is performed where each of the sidebands is divided into two halves and the network is trained to make transformations between those. The authors show that CURTAINs method performs similarly to CATHODE in anomaly detection tasks using only the sideband region information and much less training data. A Normalising Flow-based update to CURTAINs can be found in Ref.~\cite{Sengupta:2023xqy}. 

\medskip 

Some hybrid techniques take input from simulations to aid the unlabeled classification task. Simulation-assisted likelihood-free anomaly detection (or SALAD for short)~\cite{Andreassen:2020nkr} uses a reweighting function to transform simulation into observed data parametrised on the auxiliary features. First, a classifier is trained to distinguish between simulated and observed data in the SB. This classifier is used to build a weighting function, $w(Y|m)$, which reweights the simulated events in the SR where $w$ is interpolated by a neural network into the SR. A second classifier is trained to distinguish the reweighted simulation from unlabeled data. The second classifier is monotonically related to the optimal classifier, and events can be selected by putting a high threshold on this classifier output. Another method, Flow-enhanced transportation for anomaly detection (FETA)~\cite{Golling:2022nkl} used normalising flows to transform simulation to observed data in the sideband regions. This transformation is used to create a simulation-informed background template in the SR.

\begin{figure}[t!]
\centering
\includegraphics[width=0.6\textwidth]{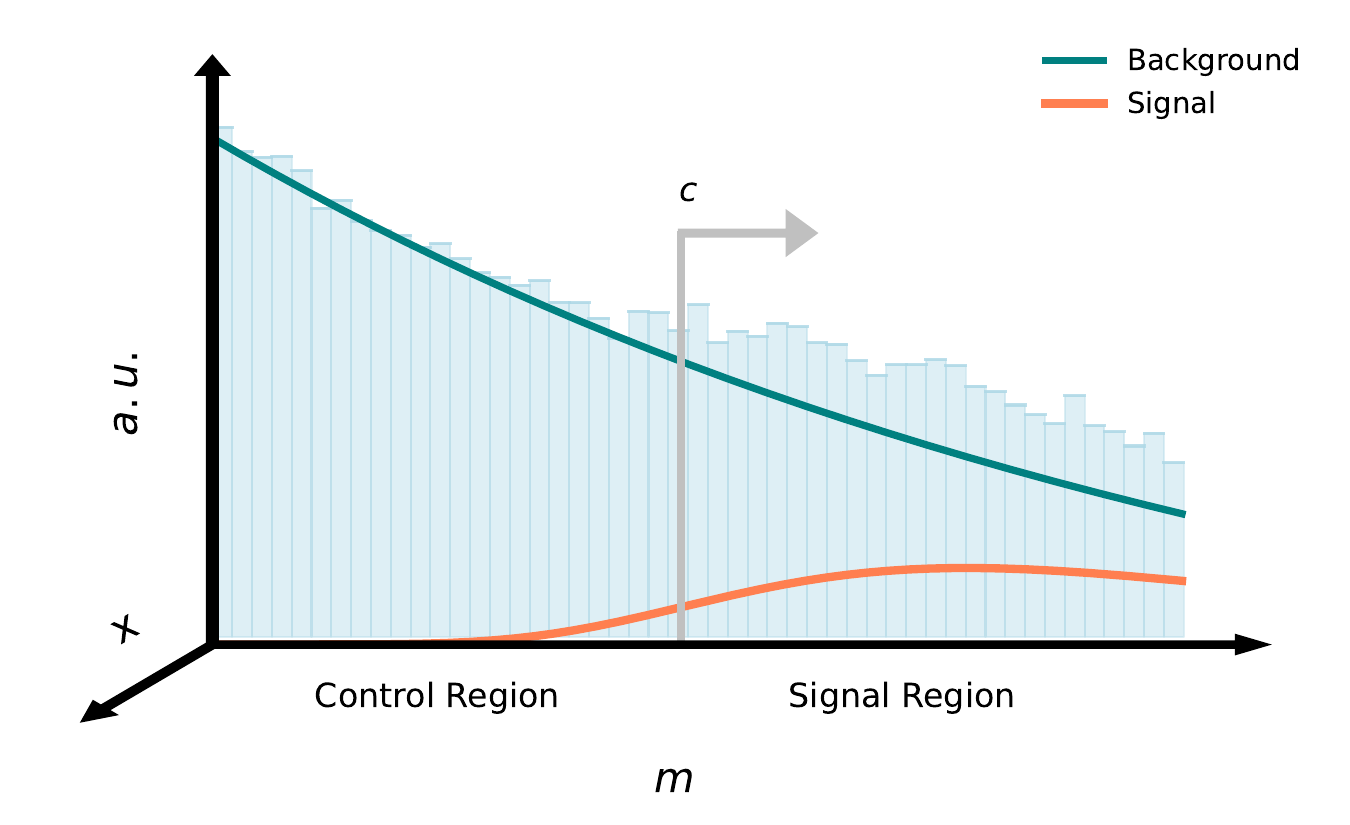}
\caption{Non-resonant anomaly detection setup from Ref.~\cite{Bai:2023yyy} showing signal and control regions. Unlike in resonant anomaly detection, the background distribution learned in the CR needs to be extrapolated into the SR.}
\label{fig:nonResAD}
\end{figure}

So far, we have seen how to tag resonant anomalies using weakly supervised methods. Ref.~\cite{Bai:2023yyy} looks at anomalous signals that are non-resonant by combining the best parts of both simulation-based methods (like FETA, SALAD, etc. mentioned above) and the sideband-based methods to estimate backgrounds. The SR is modelled in a $n$-dimensional space with thresholds, i.e. $m_i > c_i$, where $m \in \mathcal{R}^n$ and $c_i \in \mathcal{R}$ are cutoff values (Fig.~\ref{fig:nonResAD}). First, $\rho_{\text{data}} (x|m_{res})$ is estimated in the CRs (one-sided, unlike the SBRs in the case of resonant anomaly tagging) conditioned on $m$ for extrapolation into the SR. The distribution $\rho_{bg}(m)$ is estimated in the SR by reweighing $\rho_{\text{simulation}}(m)$ in the CR to match the data and then extrapolating into the SR (similar to SALAD). A binary classifier trained to distinguish simulated data from the observed data gives the weight function $w(m)$. To estimate $\rho_{\text{data}} (x|m)$ in the CR, the authors use a generative model inspired by the CATHODE method, a flow-based model like in FETA, and a simple reweighing process as in SALAD. They pick a dataset based on hidden valley models which have a strongly interacting dark sector, where massive quarks (charged under the dark sector) are produced through a heavy mediator and undergo showering to form dark hadrons. Some of these hadrons decay to SM final states, whereas the rest remain invisible, leading to semi-visible jet signatures at the collider. They report significant improvements by the three methods and find that these methods can potentially boost $1$-$2\sigma$ signals to the thresholds of discovery.

\subsection{Topic modelling}
\begin{figure*}[t]
\captionsetup[subfigure]{labelformat=empty}
\centering
\subfloat[\quad\quad(a)]{\includegraphics[width=0.45\textwidth, height=3.7cm]{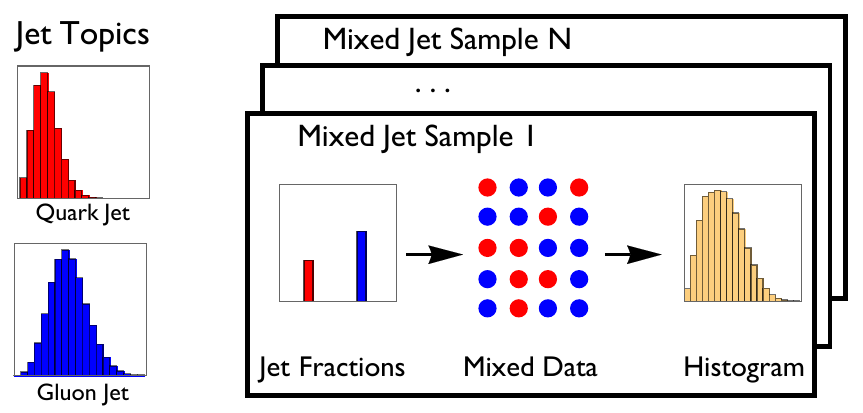}\label{fig:jettopicsA}}\hspace{1cm}
\subfloat[\quad\quad(b)]{\includegraphics[width=0.45\textwidth, height=3.7cm]{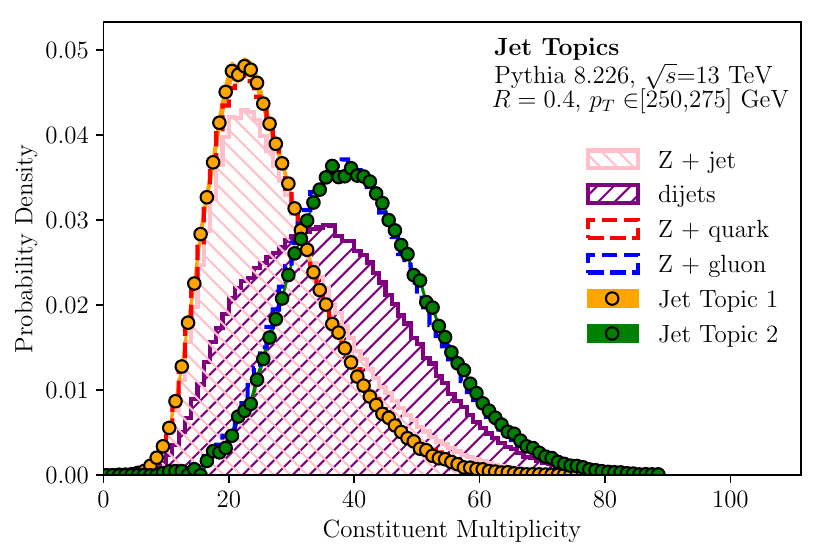}\label{fig:jettopicsB}}
\caption{\emph{Jet Topics}~\cite{Metodiev:2018ftz}: (a) Illustration of how an observable can interpreted in the context of topic modelling. Each jet comes from either a quark or gluon sampled with a particular quark fraction. Each observable is a mixture of two universal distributions; the existence of anchor bins (pure regions of phase space for each topic) is necessary for learning jet topic distributions. (b) The topic distributions learned by the topic model show good agreement with the truth-level (\textsc{Pythia}-labelled) distributions.}
\label{fig:JetTopics}
\end{figure*}

\noindent Probabilistic topic models are used to annotate large collections of text with underlying patterns (for a short review, see Ref.~\cite{10.1145/2133806.2133826}). These algorithms can discover dominant themes in the text and connections between them without any prior annotation. A theme or a topic is defined as a distribution of words in the vocabulary. Documents, the collection of words, are taken as an unstructured bag of words. Each document is an unknown mixture of topics and a collection of documents is called a corpus. Topic modelling has interesting applications to high-energy physics data. The process of generating word counts can be mapped to how jet observables are generated in a straightforward manner~\cite{Metodiev:2018ftz}. Table~\ref{tab:topicmodels} and Fig.~\ref{fig:jettopicsA} show how the mapping is done and the process of generating observables or jet histograms, respectively. 
\begin{table}[t!]
\begin{centering}
\begin{tabular}{r@{\hspace{2em}}l}
\hline
\hline
 {\bf Topic Model} & {\bf Jet Distributions}\\
\hline
Word & Histogram bin\\
Vocabulary & Jet observable(s) \\
Anchor word & Pure phase-space region (\emph{anchor bin}) \\
Topic & Type of jet (\emph{jet topic}) \\
Document & Histogram of jet observable(s) \\
Corpus & Collection of histograms \\
\hline
\hline
\end{tabular}
\caption{\emph{Jet Topics}~\cite{Metodiev:2018ftz}: The mapping between topic modelling in the context of large text documents and the collider event data. \label{tab:topicmodels}}    
\end{centering}
\end{table}

We can model the event data recorded at the LHC as statistical mixtures $(M_a)$ from $K$ different types of jets. For any observable, the distribution is generated from a combination of underlying jet distributions, i.e.,
\begin{equation}
    \rho_{M_a}(x) = \sum_{k=1}^K f^{(a)}_k \rho_k (x),
\end{equation}
where $f^{(a)}_k$ is the fraction of $k$-type jets in mixture $a$ with $\sum_k f^{(a)}_k = 1$ and $\int dx \rho_k (x) = 1$ for all $k$. The goal is to obtain the fractions, $f^{(a)}_k$, and the probability distributions, $\rho_k (x)$, using topic modelling. Following Ref.~\cite{Metodiev:2018ftz}, we can consider two mixtures of quark and gluon jets: mixture 1 contains events generated from the $Z+\text{jets}$ process mostly containing quark jets, whereas mixture 2 contains events generated from QCD dijet background which are dominated by gluon jets. Ref.~\cite{Metodiev:2018ftz} separates the two mixtures maximally by subtracting one distribution from the other. If $\kappa (M_1|M_2)$ be the maximum amount of subtraction possible such that $\rho_{M_1} - \kappa \rho_{M_2} \geq 1$, the jet topic for $T_1$ is then defined as the normalised maximal subtractions $M_2$ from $M_1$,
\begin{equation}
    \rho_{T_1} = \frac{\rho_{M_1}(x) - \kappa(M_1|M_2) \rho_{M_2}(x)}{1- \kappa(M_1|M_2)}.
\end{equation}
Similarly, $\rho_{T_2}$ can be obtained from  $\kappa(M_i|M_j) = \min_x \frac{\rho_{M_i}(x)}{\rho_{M_j}(x)}$. 

As we saw in Section~\ref{sec:weaksup}, CWoLa-based taggers trained on such mixtures could separate quark jets from gluon jets well, but they do not give any information about the underlying distributions $\rho_k (x)$. One can match these topic distributions $\rho_{\mathcal T_i}(x)$ to the actual underlying quark or gluon distributions, $\rho_{g}(x)$ and $\rho_{q}(x)$, if certain requirements are fulfilled by the samples. The main criterion is that the mixtures should have different purities, i.e., $f^{(a)}_q (x)$ for each mixture should be different. Another condition is the presence of anchor bins or mutual irreducibility of the distributions---an underlying distribution $\rho_k (x)$ cannot be a mixture of the rest of the distributions plus some other distribution. To illustrate the effectiveness of the algorithm, Ref.~\cite{Metodiev:2018ftz} generates the mixtures using some selection cuts: central jets ($|\eta|<2$) are selected, one for the $Z+\text{jets}$ events and two for the QCD dijet events if they have transverse momentum $p_T \in \left[250,275\right]$ 
GeV. At the truth level (\textsc{Pythia} labelled), these mixtures have $f_{q}^{(1)} = 0.88$ and $f_{q}^{(2)} = 0.37$. Fig.~\ref{fig:jettopicsB} shows the constituent multiplicity of the jets in each of these mixtures along with the \textsc{Pythia}-labelled distributions. The learned jet topics show good agreement with the truth-level distributions.

Another way to approach the problem of modelling observables is to treat them as a result of a generative process arising from some hidden variables. In Generative Probabilistic Modelling, a generative process defines a joint probability distribution of the observables and the hidden variables. The main goal of such models is to find the conditional probability distribution of the hidden variables given the observables, called the posterior distribution. Ref.~\cite{Dillon:2019cqt} uses a probabilistic topic model called Latent Dirichlet Association (LDA) to interpret jet substructures with a generative Bayesian model and, using the learned underlying distributions, to develop event classifiers. It assumes that some of the parton-shower information can be recovered by looking at the clustering history of a jet-clustering algorithm, particularly Cambridge-Aachen. Given a set of topics, an observable is populated from a mixture of the latent distributions or topics. The likelihood of populating a bin $i$ of the observable given a topic ($\mathcal T$) is described by the multinomial distribution $\rho(i|\mathcal T, \beta)$. The likelihood of a given topic contributing to an event is given by $\rho(\mathcal T|\omega)$, where $\omega$-s are drawn from a probability distribution $\rho(\omega|\alpha)$ which models the topic proportions in a parametric manner on the hyperparameter $\alpha$. Topics $(\mathcal T)$ are parametrised by $\beta$, and the topic proportions $\omega$ are parameterised by $\alpha$---the hidden variables of the model are then $\beta$ and $\omega$. The likelihood of generating the full jet observable $J$, with $n$ bins, is then given by
\begin{equation}
    \rho(\:J\:|\:\alpha, \beta) = \int_{\omega} \rho(\omega|\alpha) \prod_{i=1}^{n} \left( \sum_{\mathcal T} \rho(\mathcal T|\omega) p (i|\mathcal T, \beta) \right). 
    \label{eq:bayesGen}
\end{equation}
The latent variables can be found by inverting Eq.~\eqref{eq:bayesGen} and finding the best fit for a given set of events. In Ref.~\cite{Dillon:2019cqt}, the authors optimise the LDA model using Approximate Variational Inference or Gibbs Sampling. They find that the learned structure from a two-theme LDA model can be used to build unsupervised jet taggers or event classifiers that efficiently discriminate signal and background in previously unseen data. The details of the implementation of the LDA model have been elaborated further in Ref.~\cite{Dillon:2020quc}.

In a similar manner, Ref.~\cite{Alvarez:2022qoz} explores the idea of data-driven unsupervised top taggers free of simulation biases using Bayesian inference. Top taggers can isolate top jets from QCD jets; generally, they are built (trained) upon simulated data. The tagging is done by estimating the top jet fraction of a mixed sample set via Bayesian inference, where one tries to infer the distribution from which the data is assumed to be sampled. The idea can be understood as follows. For a tagging observable $x$, one can define a likelihood function for its $N$ independent measurements ($X\equiv\{x_1,x_2,\ldots,x_N\}$) as~\cite{Alvarez:2022qoz}
\begin{align}
\rho(X) = \prod_{i=1}^N\sum_{k=\{0,1\}} \pi_k\; \rho(x_i|k),\label{eq:Bayes1}
\end{align}
where $\pi_k$ is the probability of sampling a jet of class $k$ ($=1$ for top and $0$ for QCD, say). The probabilities, $\rho(x_i|k)$ depend on some parameters ($\zeta$) which are assumed to be random variables. From some prior distribution of these parameters $\rho(\zeta)$, Bayes’ theorem gives their posterior PDF:
\begin{align}
\rho(\zeta|X) \propto \rho(X|\zeta)\; \rho(\zeta),
\end{align}
where $\rho(X|\zeta)$ is read from Eq.~\eqref{eq:Bayes1}. Ref.~\cite{Alvarez:2022qoz} uses Stochastic Variational Inference (SVI)~\cite{hoffman2013stochastic} to obtain the set of model parameters maximizing the posterior probability (maximum a posteriori, MAP) and conclude that their unsupervised taggers (using different observables) show promise but perform somewhat poorer than the existing supervised taggers trained on MC data.

\subsection{Self-supervised learning methods}

In self-supervised learning methods, one creates a set of pseudo-labels to train the network. In contrast to density estimation-based approaches or generative models, these work by training a model on a discriminative loss using pseudo-labels. Reconstruction-based approaches, such as auto-encoders, have their latent spaces highly aligned to the underlying density of the data; however, these density approaches have some limitations. For instance, even perfect density estimation cannot guarantee anomaly detection ~\cite{Cheng:2022gma}. Moreover, the scores are not invariant under simple transformations of the phase space~\cite{Dillon:2021gag}. Here, self-supervised learning methods provide an avenue to incorporate invariances and physics inductive biases through pseudo-labels to increase the sensitivity to anomaly detection.  

\begin{figure}[t]
    \centering
    \includegraphics[width=0.6\textwidth]{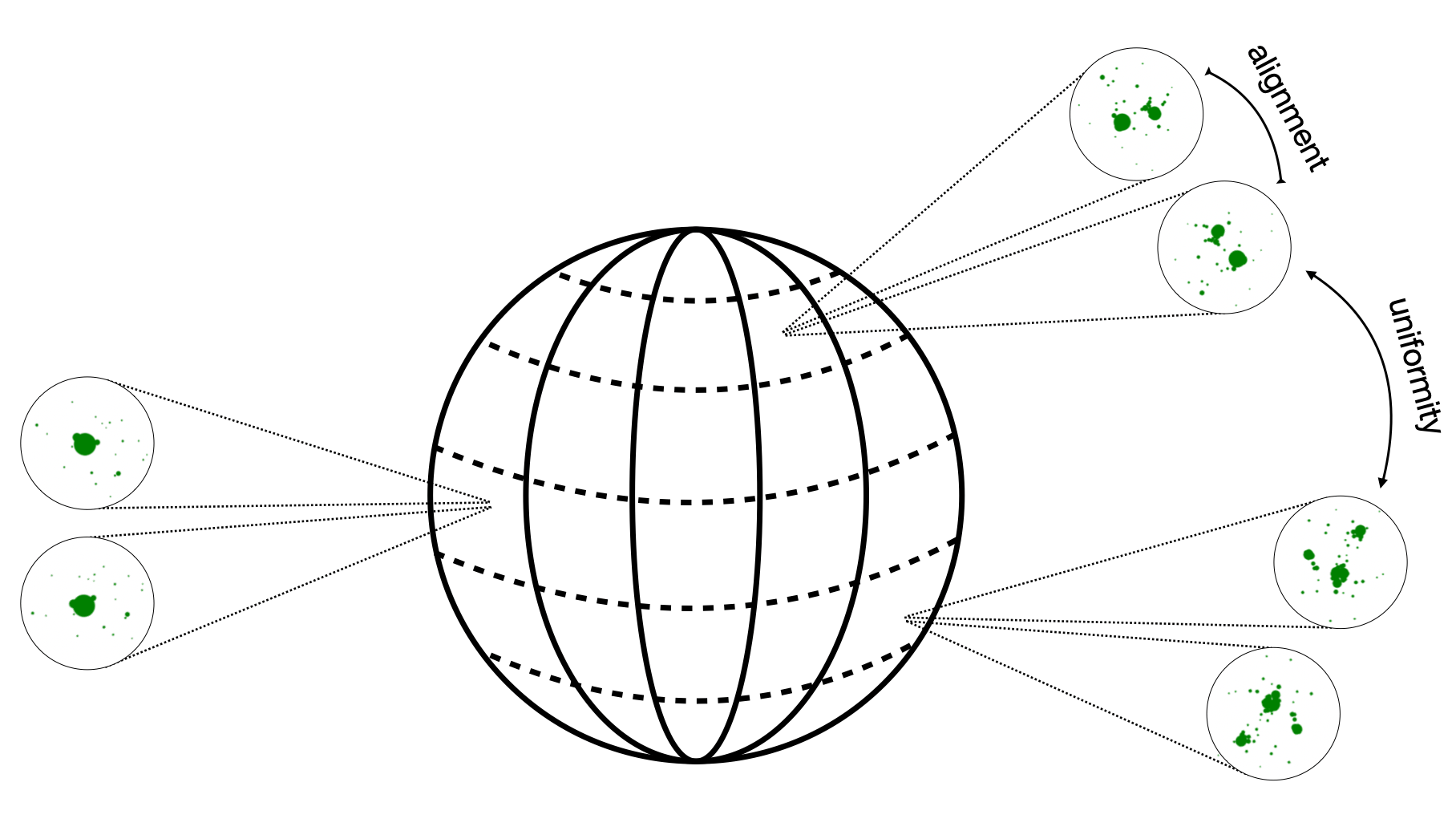}
    \caption{\emph{JetCLR}~\cite{Dillon:2021gag}: Alignment and Uniformity in the learned embeddings on $\mathbb{S}^{n-1}$ hypersphere with a contrastive loss. Alignment enforces that positive pairs have similar embeddings and, therefore, occupy the same location in the hypersphere. Uniformity ensures that the learned embeddings occupy the entire hypersphere to maximize the information learned~\cite{wang2020understanding}.}
    \label{fig:contrastive-loss-img}
\end{figure}

Most self-supervision applications use a different kind of discriminative loss than the cross-entropy loss. Instead of predicting the probabilities for each pseudolabel, models predict an embedding in the $\mathbb{S}^{n - 1}$ (hypersphere in $\mathbb{R}^n$) space. These embeddings are then directly compared using cosine similarity to contrast between embeddings of different pseudo-labels. One such kind of loss, commonly found in Computer Vision literature, is the NT-Xent (Normalised Temperature-scaled Cross Entropy) loss~\cite{10.5555/3524938.3525087}, defined as:  
\begin{equation} \label{eq:contrastive-loss}
    L_{ij} = -\log \left(\frac{e^{(\text{sim}(z_i, z_j)/\tau)}}{\sum_{k=1}^{2N} \mathbbm{I}[k \neq i]e^{(\text{sim}(z_i, z_k)/\tau)}}\right),  
\end{equation}
where $\mathbbm{I}[k \neq i] \in \{0, 1\}$ is the indicator function defined earlier (i.e., it is $1$ iff $k \neq i$) and $\tau$ denotes a temperature parameter. The total loss is computed across all positive pairs, both $(i, j)$ and $(j, i)$ in a mini-batch. Without the normalisation constraint, the softmax function (the argument of the log expression) can be made arbitrarily sharp by simply scaling the features (since dot products are proportional to the scale of features). The loss function in Eq.~\eqref{eq:contrastive-loss} can be visualised in terms of alignment and uniformity on the hypersphere. If two data samples form a positive pair, they should have similar embeddings and, therefore, should be invariant to noise factors. Maximizing the numerator of the fraction within the $\log$ is equivalent to maximizing the alignment between the positive pairs. On the other hand, the denominator of the fraction ensures that the embeddings are spread uniformly around the sphere, in a manner that negative pairs have minimal alignment with the positive ones.
Fig.~\ref{fig:contrastive-loss-img} shows an illustration of learned representations of similar jets aligned on the surface of the hypersphere.

One of the first works, \texttt{JetCLR}~\cite{Dillon:2021gag}, aims to learn robust representations for QCD vs Top jet tagging. The pseudolabels are created such that the positive pairs are formed by a jet present in the dataset and an augmented version of that sample. The negative samples are all the other jets in the (mini-batch) dataset (these include both top jets and QCD jets). The augmentations are based on developing invariances to (approximate) symmetries (such as rotational symmetry in the $\eta-\phi$ plane) and theory-inspired augmentations (such as QCD splitting). To encode permutation invariance, the authors use a transformer-based architecture to learn the representations. This strategy of pseudo-labels is generally used to pretrain networks to learn good embeddings (representations) for jets present in the dataset. This design is inspired by a seminal work in computer vision \texttt{SimCLR}~\cite{10.5555/3524938.3525087} which applies a similar strategy to learn representations for images in the ImageNet dataset. \texttt{JetCLR} provides pretty good results, and a linear classifier trained on the learned representation obtains an AUC score of $0.98$ on the dataset. Furthermore, the learned representation shows signs of (approximate) invariances to rotational symmetries as well as soft-collinear augmentations. This shows that even without the actual labels, the network is able to learn a representation space where top jets are linearly separable from QCD jets.

Ref.~\cite{Dillon:2022tmm} extends \texttt{JetCLR} to learn robust event representations. Here, pseudolabels are constructed to form positive pairs between an event and an augmented version of that event. The augmentations are applied both at the jet and event levels. At the jet level, (1) jets are (independently) rotated around the jet axis in the $\eta-\phi$ plane to encode (approximate) rotational symmetries, (2) jet constituents are randomly shifted using a Gaussian noise to simulate the smearing from detector effects, and (3) jets are split collinearly. At the event level, global $\eta$ and $\phi$ shiftings are performed to make the learned representations invariant under these transformations. Similar to \texttt{JetCLR}, a transformer-encoder architecture is used to enforce permutation invariance to event constituents. The training dataset is based on the LHC $2020$ Olympics R\&D dataset. After the contrastive pretraining, the authors find that a linear classifier trained on the learned representation for the events outperforms a linear classifier trained directly on the (particle) event space (AUC: $0.926$ vs $0.883$); complex architectures trained on the learned representation outperform the linear classifier as expected. The authors also test the viability of such an unsupervised learned representation for CWoLa (weak-supervision paradigm) and find that a CWoLa weakly supervised classifier performance of a small latent space is comparable to (but cannot match) that of the full (particle) event space with better performance with larger latent spaces.

\texttt{AnomalyCLR}~\cite{Dillon:2023zac} adapts ideas from \texttt{EventCLR} to construct expressive representations for use in anomaly detection tasks. \texttt{AnomalyCLR} differs from the previous methods in two ways: (1) It introduces a new kind of augmentation, anomaly augmentation, to mimic unphysical potential anomalies, which we want the representation to be highly discriminative towards; and (2) an associated anomaly-pair pseudolabel which matches an event in the dataset to an anomaly-augmented version of itself. To incorporate the new pseudolabel, the authors modify the original contrastive loss:
\begin{equation}
    L_{\text{AnomCLR}} = -\log\left( \frac{e^{\left[s(z_i, z_i') - s(z_i, z_i^*)\right] / \tau}}{\sum_{j \neq i \in \text{batch}} \left[ e^{s(z_i, z_j)/\tau} + e^{s(z_i, z_j')/\tau} \right]} \right)\label{eq:anomaly-clr-eq1},
\end{equation}
where $s(z_i, z_i')$ is the (cosine) similarity between the latent representation $z$ of event $i$ and a physically augmented version itself, $s(z_i, z_i^*)$ is the similarity between the latent representation of event $i$ and an anomaly-augmented version itself; the denominator term does not change here. The denominator term in Eq.~\eqref{eq:anomaly-clr-eq1} is useful for learning representations that are useful for discriminating each sample from each other (uniformity). However, for anomaly detection, this property is not necessary and the anomaly-pair should provide the representations with sufficient discriminative power. Therefore, the authors also introduce a modified loss that does away with the denominator:
\begin{equation}
    L_{\text{AnomCLR}}^+ = -\log e^{\left[s(z_i, z_i') - s(z_i, z_i^*)\right] / \tau} = \frac{1}{\tau}(s(z_i, z_i') - s(z_i, z_i^*)). \label{eq:anomaly-clr-eq2}
\end{equation}
Eq.~\eqref{eq:anomaly-clr-eq2} results in a less computationally expensive loss since we do not need to calculate the $N^2_{\text{batch}}$ term for the denominator. For anomaly augmentations, the authors introduce some simple scenarios: (1) Multiplicity-shifts: for each event, a random number of electrons, muons, and jets are added to the event with each object's physical properties also randomly selected within a range. (2) Multiplicity shifts, keeping MET and total $p_T$ constant: this is similar to the previous augmentation, however, the multiplicity is changed by splitting the existing objects and smearing the $\eta-\phi$. (3) $p_T$ and MET shifts: the $p_T$'s in the event, or the MET or both are shifted by a random factor. An autoencoder is trained on the learned representations to use the reconstruction error as the anomaly score. The authors find that $L_{\text{AnomCLR}}^+$ outperforms autoencoders trained on $L_{\text{AnomCLR}}$, standard \texttt{CLR} loss and the raw particle-level input. 

\texttt{DarkCLR}~\cite{Favaro:2023xdl} extends \texttt{AnomalyCLR} to identify anomalous semi-visible jets~\cite{semivisble_jets_prl, Cohen:2017pzm, Pierce:2017taw, Beauchesne:2017yhh, Bernreuther:2019pfb, Bernreuther:2020vhm, Batz:2023zef} at LHC. Semi-visible jets show up in models that incorporate the dark sector (from the general class of Hiddden Valley Models~\cite{Strassler:2006im, MORRISSEY20121, Knapen:2021eip}) and generally contain fewer constituents compared to usual jets. Therefore, the anomalous augmentation, in this case, involves dropping constituents from the jets. The authors also use a normalized autoencoder (NAE) to calculate the anomaly score. The method results in better performance compared to \texttt{AnomalyCLR} for low small values of $\epsilon_B^{-1}$.

In a different line of work, Ref.~\cite{Cheng:2022gma} explores the possibility of using a pretrained network, trained for SM jet classification, for anomalous anti-QCD jet tagging. While this is more of an Out-of-Distribution detection strategy, the motivation for using a pretrained classifier is similar to training a self-supervised method, which is to learn/utilize a robust representation invariant to physical augmentations as well as sensitive to anomalous events. The authors find that both the softmax probabilities and the penultimate latent vectors of the pretrained network can be used to estimate the model uncertainty on a new event. By doing so, the authors are able to effectively show that a multi-class jet SM tagger can be used to reduce copious amounts of the QCD background (anti-QCD tagging). 

\section{Detector unfolding and simulation}\label{sec:det_sim}

There are two aspects to testing the prospects of a NP model at the LHC. First, one needs precise simulations\footnote{For example, one can use MC tools like \texttt{MadGraph5}~\cite{Alwall:2014hca,Frederix:2018nkq} and \texttt{Pythia}~\cite{Bierlich:2022pfr} to produce events and study them within a detector environment with \texttt{Delphes}~\cite{deFavereau:2013fsa} or \texttt{Geant4}~\cite{GEANT4:2002zbu,Allison:2016lfl}, etc.} to test the viability of the model. Full simulations of the particle showers in calorimeters can be computationally expensive, sometimes taking minutes per event even on a high-performance platform. As the LHC enters the high-luminosity mode, simulations will become even more computationally expensive and the need for alternative approaches for fast simulation will increase. The other is the inverse process of mapping the observed data to hard scatterings accurately. This process is called unfolding, and generally, it is highly challenging as it has to take real detector effects, such as acceptance, smearing, threshold effects, etc., into account. 
Some matrix-based methods have been developed to perform unfolding of 1D histograms~\cite{Prosper:1306523,Cowan:2002in}. However, these methods scale exponentially with the joint unfolding of multiple variables making them practically impossible. They also require binning which can lead to further problems like statistical uncertainties and figuring out the (near-)optimal binning. Deep generative models, especially generative adversarial networks (GANs) and their variations, offer promising solutions to these issues. We present a brief overview of how they work below. (For detailed reviews on GANs, see Refs.~\cite{DBLP:journals/corr/abs-2001-06937,DBLP:journals/corr/abs-2110-01442}.)
\medskip

\noindent
{\bf Understanding GANs:} In generative modelling, the main objective is to learn the underlying distribution of the dataset and generate new data from the distribution. GANs are a way to make a generative model with two neural networks (generator and discriminator) competing with each other. The generator takes random noise as input and produces synthetic data (ideally resembling real data) by learning to map random noise from a simple distribution (e.g., a normal distribution) to complex data distributions. The discriminator trains itself to classify the input data as real or generated. The generator (discriminator) is defined by a differentiable function $G$ ($D$) that takes $z$ ($x$) as input and uses $\theta^{(G)}$ ($\theta^{(D)}$) as parameters. Both networks have their respective cost functions defined in terms of both their parameters. The generator (discriminator) tries to minimize $L^{(G)}(\theta^{(D)}, \theta^{(G)})$ ($L^{(D)}(\theta^{(D)}, \theta^{(G)})$), and must do so while only controlling $\theta(G)$ ($\theta(D)$). A straightforward optimisation can not optimise these loss functions as each network's cost depends on the other network’s parameters, over which it has no control. The solution is obtained through a game theoretic approach of looking for the local differential Nash equilibrium, a local minimum of $L^{(G)}$($L^{(D)}$) with respect to $\theta^{(G)}$($\theta^{(D)})$.
The discriminator is trained to minimise a standard binary cross-entropy:
\begin{equation}\label{eq:loss_ganD}
    L^{(D)}(\theta^{(D)}, \theta^{(G)}) = -\frac{1}{2}\mathbb{E}_{x\sim p_{data}} \log D(x) -\frac{1}{2}\mathbb{E}_{z} \log(1-D(G(z))).
\end{equation}
Here, the first term is the discriminator output for real data $x$ and the second term captures the discriminator output for fake data $G(z)$. The generator cost function varies across different GAN models, with the simplest version being the zero-sum game with 
\begin{equation}
    L^{(G)}(\theta^{(D)}, \theta^{(G)}) = -  L^{(D)}(\theta^{(D)}, \theta^{(G)}).
\end{equation}
In this zero-sum game where the generator tries to generate data that is indistinguishable from real data, and the discriminator tries to correctly distinguish between real and generated data, the value function $V(\theta(D), \theta(G))$ captures the adversarial nature of the training process and represents the expected gain for the generator and discriminator. 
\begin{equation}
     V(\theta^{(D)}, \theta^{(G)}) = -  L^{(D)}(\theta^{(D)}, \theta^{(G)}).
\end{equation}
Zero-sum games are also called minimax games because their solution involves minimization in an outer loop and maximization in an inner loop:
\begin{equation}\label{eq:jsd}
    \theta^{(G)*} = \arg \min_{\theta^{(G)}} \max_{\theta^{(D)}}V(\theta^{(D)}, \theta^{(G)}).
\end{equation} 
The learning in this game resembles minimizing the Jensen-Shannon divergence (JSD)(see Appendix \ref{sec:JSD} for more details) between the data and the model distribution, and the game converges to its equilibrium if both neural networks can be updated directly in parameter space.

\subsection{Unfolding detector effects}

Recent developments in ML-based methods have made some progress in solving the unfolding problem. These methods are broadly of two kinds: (1) density estimation-based approaches, where a generative model is trained to estimate the densities of the measurements as well as generate samples from the unfolded truth level data given the detector measurements; and (2) classifier based approaches, which learn the likelihood ratio to reweight the simulated parton distributions to generate the truth level distributions.

\medskip
\noindent
{\bf Density estimation-based approaches}:
Ref.~\cite{Datta:2018mwd} trains a generative adversarial network (GAN) to reconstruct samples from the unfolded distribution. The GAN is trained to learn an inverse function from the detector data to the parton-level data. The dataset is created such that there is a training pair data of generation level distribution and the reconstructed distribution. This allows the training to occur in a semi-supervised fashion with two losses: (1) a binary cross entropy-based adversarial loss due to the discriminator, and (2) a MSE loss to check the quality of the reconstruction. The method provides a proof-of-concept demonstration that GANs could indeed be used for unfolding providing competitive performance to the standard Iterative Bayesian Unfolding method at much faster speeds.
\begin{figure}[t]
     \centering
\definecolor{Gcolor}{HTML}{2c7fb8}
\definecolor{Dcolor}{HTML}{f03b20}
\tikzstyle{generator} = [thick, rectangle, rounded corners, minimum width=1.5cm, minimum height=1cm,text centered, draw=Gcolor]
\tikzstyle{discriminator} = [thick, rectangle, rounded corners, minimum width=1.5cm, minimum height=1cm,text centered, draw=Dcolor]
\tikzstyle{mmd} = [thick, rectangle, rounded corners, minimum width=1.5cm, minimum height=1cm,text centered, draw=black]
\tikzstyle{io} = [thick,circle, trapezium left angle=70, trapezium right angle=110, minimum width=1.2cm, minimum height=1cm, text centered, draw=black]
\tikzstyle{cond} = [thick, rectangle, dotted, rounded corners, minimum width=10.0cm, minimum height=2cm,text centered, draw=gray!50!black]
\tikzstyle{iodotted} = [thick, circle, trapezium left angle=70, trapezium right angle=110, minimum width=1.2cm, minimum height=1cm, text centered, draw=black, dotted]
\tikzstyle{process} = [thick, rectangle, minimum width=1cm, minimum height=1cm, text centered, draw=black]
\tikzstyle{xG} = [thick,rectangle, minimum width=2.2cm, minimum height=3cm, text depth= 2.2cm, draw=black]
\tikzstyle{s0} = [thick,rectangle, minimum width=2cm, minimum height=3cm, text centered]
\tikzstyle{s1} = [thick, dotted, rectangle, minimum width=1.6cm, minimum height=1.1cm, text centered, draw=black]
\tikzstyle{decision} = [thick,rectangle, minimum width=1cm, minimum height=1cm, text centered, draw=black]
\tikzstyle{dots} = [circle, minimum size=2pt, inner sep=0pt,outer sep=0pt, draw=Dcolor, fill = Dcolor]
\tikzstyle{arrow} = [thick,->,>=stealth]
\begin{tikzpicture}[node distance=2cm]
\node (generator) [generator] {$G$};
\node (random) [io, left of=generator, xshift=-0.2cm, yshift=0cm] {$\{ r \}$};
\draw [arrow, color=black] (random) -- (generator);
\node (xG) [io, right of = generator, xshift=1.0cm, yshift=0cm] {$\{x_G\}$};
\node (discriminator) [discriminator, right of = xG, xshift=1.0cm, yshift=0cm] {$D$};

\node (cond) [cond, above of = generator, xshift=1.5cm, yshift=0.5cm] {};
\node (condi) [above of = xG, xshift=1.9cm, yshift=1.2cm, color=gray!50!black] {Condition};

\node (xd) [io, above of = generator, xshift=0.cm, yshift=0.5cm] {$\{x_d\}$};
\node (xp) [io, below of = xG, xshift=0.5cm, yshift=0cm] {$\{x_p\}$};

\node (detector) [process, left of=xd, xshift=-0.2cm, yshift=0cm] {detector};
\node (parton) [process, left of=xp, xshift=-0.2cm, yshift=0cm] {parton};

\coordinate[ above of= discriminator, xshift=-0.1cm, yshift=0.5cm] (in1);
\draw [thick, color=black] (xd) -- (in1);
\draw [arrow, color=black] (in1) -- ([xshift=-0.1cm] discriminator.north);

\draw [arrow, color=black] (detector) -- (xd);
\draw [arrow, color=black] (parton) -- (xp);
\draw [arrow, color=black] (xd) -- (generator);
\draw [arrow, color=black] (xp) -- (discriminator);
\draw [arrow, color=Gcolor] (generator) -- (xG);
\draw [arrow, color=Gcolor] (xG) -- (discriminator);

\node (dloss) [process, right of=discriminator, xshift=0.5cm, yshift=0cm] {$L_{D}$};
\node (gloss) [process, below of=dloss, xshift=0.0cm, yshift=0cm] {$L_{G}$};
\node (mmd) [mmd, below of=discriminator, xshift=0.0cm, yshift=0cm] {MMD};

\draw [arrow, color=Gcolor] (discriminator) -- (gloss);
\draw [arrow, color=Dcolor] (discriminator) -- (dloss);

\coordinate[ above of = dloss, xshift=0cm, yshift=-1cm] (d1);
\coordinate[ above of = discriminator, xshift=0.1cm, yshift=-1cm] (d2);
\draw[thick, dashed, color=Dcolor] (dloss) -- (d1);
\draw[thick, dashed, color=Dcolor] (d1) -- (d2);
\draw[arrow, dashed, color=Dcolor] (d2) -- ([xshift=0.1cm] discriminator.north);

\draw[arrow, color=Gcolor] (xG) --  (mmd);
\draw[arrow, color=black] (xp) --  (mmd);
\draw[arrow, color=Gcolor] (mmd) --  (gloss);

\coordinate[ below of = gloss, xshift=0cm, yshift=1.0cm] (out1);
\coordinate[ below of = generator, xshift=0cm, yshift=-1.0cm] (out2);
\draw[thick, dashed, color=Gcolor] (gloss) --  (out1);
\draw[thick, dashed, color=Gcolor] (out1) --  (out2);
\draw[arrow, dashed, color=Gcolor] (out2) --  (generator);
\end{tikzpicture}
\caption{\texttt{FCGAN} setup. The setup illustrates a conditional GAN. The generator $G$ receives a random noise vector $\{r\}$ as input and the detector response $\{x_d\}$. The discriminator $D$ receives either the generated unfolded sample $\{x_G\}$ or the true parton distribution $\{x_p\}$ along with the detector response $\{x_d\}$. The discriminator is trained using the loss $L_D$ to classify true and generated samples appropriately. On the other hand, the generator is trained using both the loss based on the discriminator output $L_G$ and a MMD loss to check for the quality of the unfolded sample. Image taken from Ref.~\cite{Diefenbacher:2023wec}}.
\label{fig:fcgan}
\end{figure}
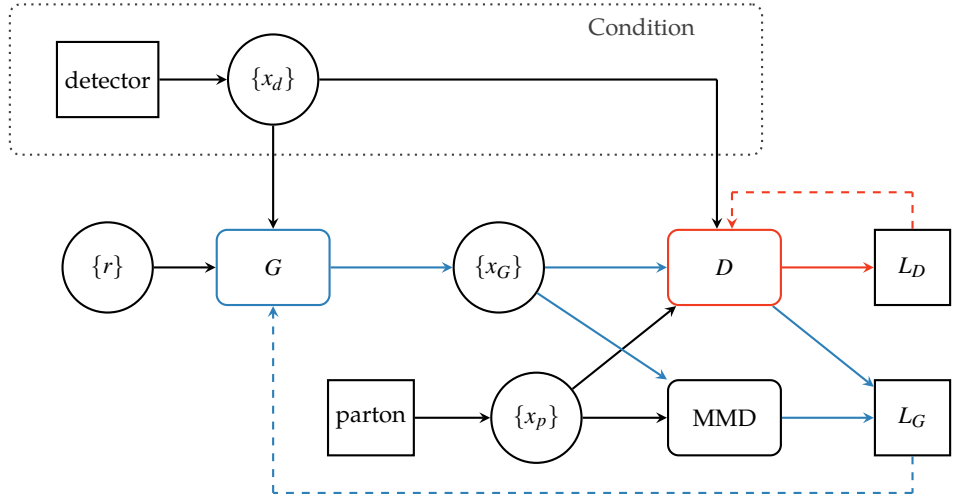

A follow-up work on unfolding done in Ref.~\cite{Diefenbacher:2023wec} also trains a GAN to perform unfolding. They test between two GAN frameworks: a naive one and a fully conditional GAN. For the naive GAN, the generator input is the detector response (measurement) and is trained to output the parton distribution. The discriminator is trained to identify between the generated distribution and the true distribution. One can note that, in this scenario, there is no need for an event-by-event matched training pair between the detector response and the true distribution. Additionally, to improve the performance of the generator and provide extra training signal, the authors calculate the Maximum Mean Discrepancy (MMD) distance between a batch of generated samples and the true parton distributions. This naive GAN can unfold events effectively as long as the training and test datasets are statistically identical. In general, one cannot assume that the training and test datasets are arbitrarily similar, so to resolve this, the authors introduce a fully conditional GAN (FCGAN). The idea behind a conditional setup is not to learn a deterministic link between the input and the output (as is the case in supervised learning), but rather to learn to sample from a structured distribution $\rho(x|y)$. In FCGAN, the training is modified such that the generator is provided with a random vector $r$ along with the detector response as inputs to generate the parton distribution. The discriminator is also modified such that it also has access to the detector response when classifying between real and generated samples. In this case one does require an event-to-event matched training pairs of detector response and parton distribution. The authors do not modify the MMD loss to be conditional since there is no efficient implementation of the same. Fig.~\ref{fig:fcgan} illustrates the model setup and the training losses applied. In general, FCGAN outperforms the naive GAN, even in cases where the training and the test datasets are not identical. However, it fails to reconstruct the invariant mass distribution, which the authors hypothesise is due to the lack of a fully conditional MMD loss. Using FCGAN, the authors also test the applicability of a GAN trained on the SM for unfolding BSM events. They find that, usually, the model is able to reconstruct the parton distributions for the BSM process, though not as well as in the previous cases. 

\begin{figure}
    \centering
\definecolor{Gcolor}{HTML}{2c7fb8}
\definecolor{Dcolor}{HTML}{f03b20}
\tikzstyle{cINN} = [thick, rectangle, rounded corners, minimum width=2.5cm, minimum height=3.5cm,text centered, draw=Gcolor]
\tikzstyle{preprocessor} = [thick, rectangle, rounded corners, minimum width=1.5cm, minimum height=1cm,text centered, draw=Dcolor]
\tikzstyle{mmd} = [thick, rectangle, rounded corners, minimum width=1.5cm, minimum height=1cm,text centered, draw=black]
\tikzstyle{io} = [thick,circle, minimum width=1.2cm, minimum height=1cm, text centered, draw=black]
\tikzstyle{cond} = [thick, rectangle, dotted, rounded corners, minimum width=10.0cm, minimum height=2cm,text centered, draw=gray!50!black]
\tikzstyle{iodotted} = [thick, circle, minimum width=1.2cm, minimum height=1cm, text centered, draw=black, dotted]
\tikzstyle{process} = [thick, rectangle, minimum width=1cm, minimum height=1cm, text centered, draw=black]
\tikzstyle{xG} = [thick,rectangle, minimum width=2.2cm, minimum height=3cm, text depth= 2.2cm, draw=black]
\tikzstyle{s0} = [thick,rectangle, minimum width=2cm, minimum height=3cm, text centered]
\tikzstyle{s1} = [thick, dotted, rectangle, minimum width=1.6cm, minimum height=1.1cm, text centered, draw=black]
\tikzstyle{decision} = [thick,rectangle, minimum width=1cm, minimum height=1cm, text centered, draw=black]
\tikzstyle{dots} = [circle, minimum size=2pt, inner sep=0pt,outer sep=0pt, draw=Dcolor, fill = Dcolor]
\tikzstyle{arrow} = [thick,->,>=stealth]
\begin{tikzpicture}[node distance=2cm]
\node (cINN) [cINN] {cINN};
\node (xG) [io, left of = cINN, xshift=-0.5cm, yshift=1cm] {$\{\tilde{x}_p\}$};
\node (rG) [io, right of = cINN, xshift=0.5cm, yshift=-1cm] {$\{ \tilde{r} \}$};
\node (xp) [io, left of = cINN, xshift=-0.5cm, yshift=-1cm] {$\{x_p\}$};
\node (parton) [process, below of=xp, xshift=0cm, yshift=0.25cm] {parton};
\node (mmd) [mmd, left of=xp, xshift=0.0cm, yshift=1cm] {$L_\text{MMD}$};

\node (random) [io, right of=cINN, xshift=0.5cm, yshift=1cm] {$\{ r \}$};
\node (gauss) [mmd, right of=rG, xshift=0.0cm, yshift=1cm] {$L$};
\node (cond) [cond, above of = xG, xshift=1.5cm, yshift=0.7cm] {};
\node (condi) [above of = xG, xshift=5.5cm, yshift=1.3cm, color=gray!50!black] {condition};
\node (preprocessor) [preprocessor, above of=cINN, xshift=0cm, yshift=1.5cm] {subnet};
\node (xd) [io, left of = preprocessor, xshift=-0.5cm, yshift=0cm] {$\{x_d\}$};
\node (detector) [process, left of=xd, xshift=-0.2cm, yshift=0cm] {detector};
\coordinate[ right of = rG, xshift=0cm, yshift=0cm] (Gin1);
\coordinate[ right of = random, xshift=0cm, yshift=0cm] (Gin2);
\coordinate[ left of = xG, xshift=0cm, yshift=0cm] (MMDin1);
\coordinate[ left of = xp, xshift=0cm, yshift=0cm] (MMDin2);
\draw [arrow, color=black] ([yshift=0em]parton.north) -- ([yshift=0em]xp.south);
\draw [thick, color=Gcolor] ([yshift=0em]xp.west) -- ([yshift=0em]MMDin2);
\draw [arrow, color=Gcolor] ([yshift=0em]MMDin2) -- ([yshift=0em]mmd.south);
\draw [thick, color=Gcolor] ([yshift=0em]xG.west) -- ([yshift=0em]MMDin1);
\draw [arrow, color=Gcolor] ([yshift=0em]MMDin1) -- ([yshift=0em]mmd.north);
\draw [thick, color=Gcolor] ([yshift=0em]rG.east) -- ([yshift=0em]Gin1);
\draw [arrow, color=Gcolor] ([yshift=0em]Gin1) -- ([yshift=0em]gauss.south);
\draw [arrow, color=black] ([yshift=0em]detector.east) -- ([yshift=0em]xd.west);
\draw [arrow, color=Dcolor] ([yshift=0em]xd.east) -- ([yshift=0em]preprocessor.west);
\draw [arrow, color=Dcolor] ([yshift=0em, xshift=1mm]preprocessor.south) -- node[scale=0.8, anchor=center, right, color=Dcolor]{$f(x_d)$} ([yshift=0em,xshift=1mm]cINN.north);
\draw [arrow, dashed, color=black] ([yshift=0em, xshift=-1mm]cINN.north) -- ([yshift=0em, xshift=-1mm]preprocessor.south);
\draw[arrow, thick, color=Gcolor] (random.west) -- node[scale=0.8, sloped, anchor=center, above, color=Gcolor]{$\bar{g}(r, f(x_d))$} (xG.east);
\draw[arrow, thick, color=Gcolor] (xp.east) -- node[scale=0.8, sloped, anchor=center, above, color=Gcolor]{$g(x_p, f(x_d))$} (rG.west);
\draw[arrow, thick, dashed, color=black] (mmd) -- (cINN.west);
\draw[arrow, thick, dashed, color=black] (gauss) -- (cINN.east);
\end{tikzpicture}
    \caption{cINN Setup: The inputs are the random numbers $\{r\}$ while $\{x_{d,p}\}$ denote the detector level and the parton level data. The latent dimension loss is denoted by $L$ is the negative log-likelihood loss to maximize the (posterior) probability $p(\theta | x_p, x_d)$ for a network with parameters $\theta$. A tilde indicates the INN generation. $L_{\text{MMD}}$ indicates the MMD loss between the unfolded distribution $\{\tilde{x}_p\}$ and the truth level parton distribution $\{x_p\}$. Image taken from Ref.~\cite{Bellagente:2020piv}.}
    \label{fig:cinn}
\end{figure}
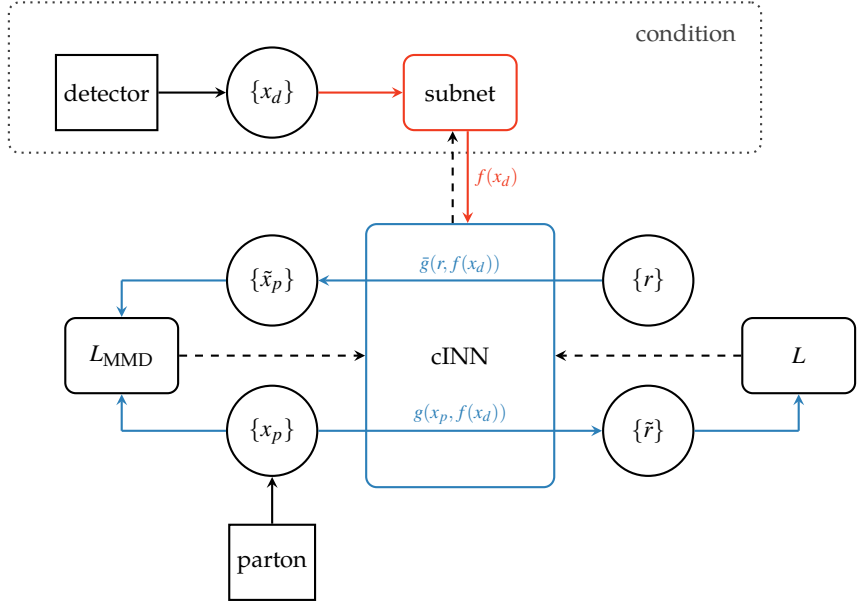

In a follow-up work~\cite{Bellagente:2020piv}, authors instead utilize a cINN to model the transformation from the detector response to the parton distribution and vice versa. Invertible neural networks (INNs) are a class of neural networks that are designed such that the learned functions are invertible. They allow for information-preserving transformations and are effective at tackling probabilistic modelling and generative tasks. Specifically, unlike traditional neural networks that tackle ambiguous inverse problems, INNs focus on learning the forward process while implicitly capturing the reverse process. The motivation for using an INN is that the extra training signal provided by testing the parton to detector simulation would improve performance. The authors first test a naive implementation of INNs using a MSE loss to check for the reconstruction on both sides and the MMD loss to improve reconstruction. Similar to the naive GAN, the naive INN fails to capture the hard cuts imposed by the jet definition, especially in the tails of the distributions. To mitigate these, the authors introduce a conditional INN (cINN) that works by linking the parton-level data ($x_p$) to random noise ($r$) conditioned on the detector-level data ($x_d$). Trained on a given process, this network can generate the particle-level distribution given the detector-level events and an unfolding model. While this two-way training does not represent the inversion of a detector, it stabilises the training by requiring the input/reconstructed noise to be Gaussian. A small subnet first processes the detector-level data to provide the conditioning, i.e., $x_d \rightarrow f(x_d)$. This subnet is trained alongside the cINN and does not need to be reversed or adapted. Fig.~\ref{fig:cinn} illustrates the conditional INN setup used and the losses applied. The cINN model outperforms the simple INN model in representing hard cuts on the tails of the input distribution. Furthermore, the cINN model appears to be more calibrated than FCGAN. This fully conditional setup outperforms both the naive INN setup as well as a noise-extended INN setup for all the test cases. Finally, the authors test the performance of cINN on processes with initial-state radiations which changes the number of final jets at the detector level. Despite the variable number of jets (from $2$ to $4$), the model is effectively able to unfold the quantities with excellent performance individually for events with 2 jets, 3 jets, and 4 jets.

Other unfolding strategies that are based on modern generative modelling methods have also been developed~\cite{NEURIPS2023_cd830afc,NIPS2015_8d55a249}. Ref.~\cite{NEURIPS2023_cd830afc} utilises a latent diffusion model (LDM), resulting in unfolded distributions with the least earth mover's distance~\cite{Komiske:2019fks} to parton distributions over a wider range of values.

\begin{figure}[t]
    \centering
    \includegraphics[width=0.6\textwidth]{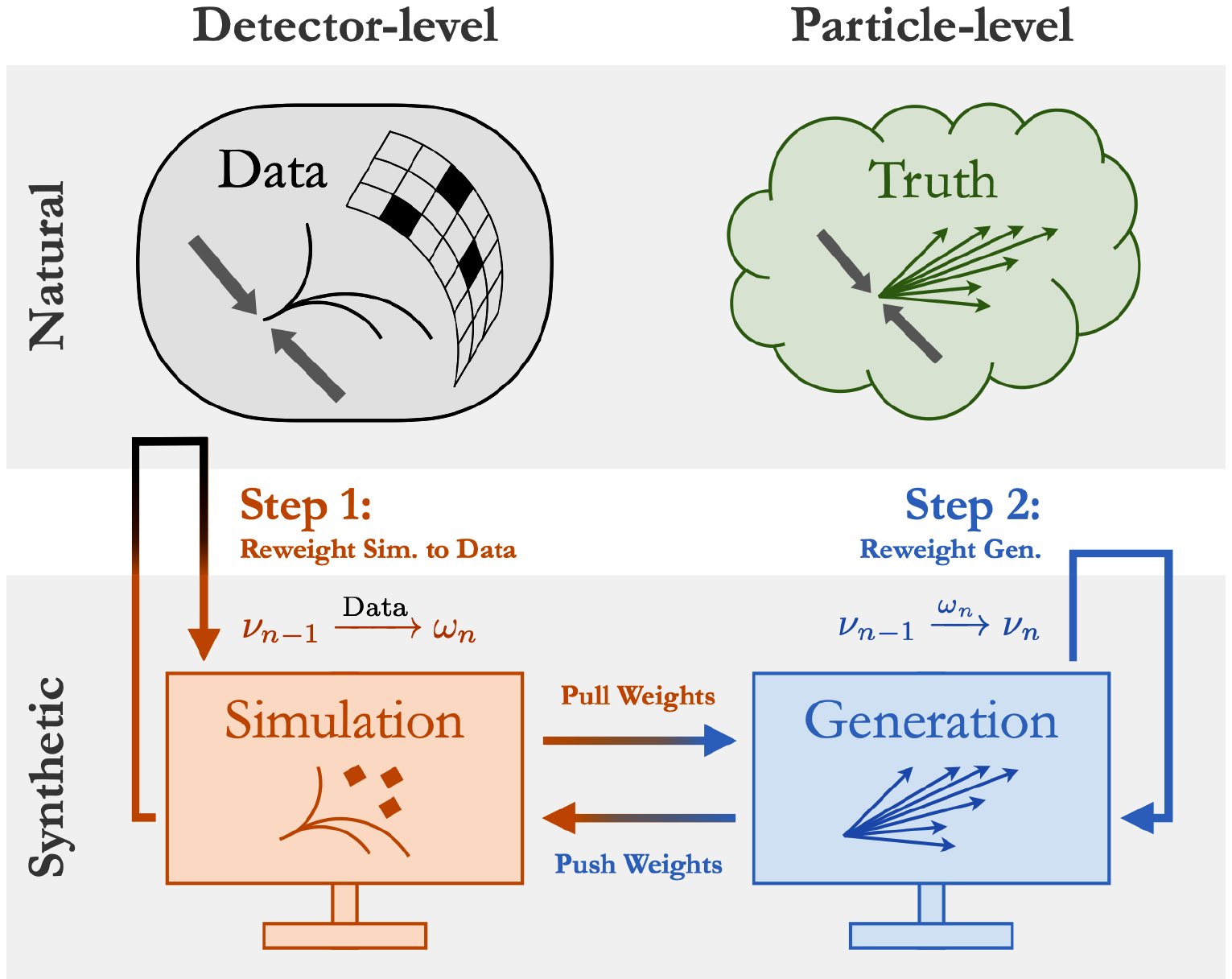}
    \caption{\texttt{Omnifold} schematic: The setup has two steps: (1) pulling weights based on the likelihood ratio between the simulated detector response and the actual detector measurement to reweigh the simulated parton-level MC samples, and (2) pushing weights of the generated samples to readjust the simulated detector-level response so that step (1) can be performed again. The process continues iteratively until convergence, and the final truth-(parton-)level distributions are obtained by reweighing the simulated parton-level distributions. Image taken from Ref.~\cite{Andreassen:2019cjw}}
    \label{fig:omnifold}
\end{figure}
\medskip
\noindent
{\bf Classifier-based approaches}:
Among the recent classifier-based unfolding approaches, \texttt{\texttt{Omnifold}}~\cite{Andreassen:2019cjw,Andreassen:2021zzk} has emerged as a compelling framework due to its ability to unfold all observables simultaneously. \texttt{Omnifold} generalizes the binned iterative Bayesian unfolding (also known as Richardson-Lucy deconvolution) to the unbinned full-phase space. The authors achieve this by iteratively reweighting the parton and detector-level MC simulations, using ML to capitalise on all available information. Essentially, \texttt{Omnifold} can be understood as an expectation maximization (EM) procedure~\cite{Diefenbacher:2023wec} that proceeds as:
\begin{itemize}
    \item \textbf{E step:} $\omega_{i+1}(x)\: = \:\rho_{\text{data}}(x)/\tilde{\rho}_{\text{sim.}}(x)$, where $\displaystyle\tilde{\rho}_{\text{sim.}}(x) \equiv \int \text{d}z\; \rho_{\text{sim.}}(x, z) \nu_{i}(z)$,

    \item \textbf{M step:} $\nu_{i+1}(z)\: = \:\bar{\rho}_{\text{sim.}}(z) / \rho_{\text{sim.}}(z)$, where $\displaystyle\bar{\rho}_{\text{sim.}}(z) \equiv \int \text{d}x\; \rho_{\text{sim.}}(x, z) \omega_{w+1}(x)$.
\end{itemize}
In practice, both expectation (\textbf{E step}) and maximization (\textbf{M step}) are achieved by training a classifier and interpreting its score as the likelihood ratios. After $n$ iterations, the unfolded distribution is
\begin{equation*}
    \rho_{\text{unfolded}}^{(n)} = \nu_{n}(z) \rho_{\text{sim.}}(z).
\end{equation*}

The \texttt{Omnifold} method takes into account the correlations between observables as well. The authors evaluate the method on a realistic jet-substructure example. The results show that the \texttt{OmniFold} method is able to achieve state-of-the-art performance in terms of both efficiency and accuracy. The NN in \texttt{OmniFold} is trained on a simulated dataset that includes both the true underlying distribution and the detector effects. It is then used to reweight the simulated dataset to match the observed data. 
We note that the density estimation-based strategies may also be viewed as modelling the expectation step of the EM algorithm, while the classifier is used to estimate the weights for reweighing (M Step). Therefore, the two approaches can complement each other. An issue with the standard generative models-based methods is that the accuracy of the unfolding method depends heavily on (and therefore limited by) the ability of the simulated training samples to model the actual data being unfolded. In Ref.~\cite{Backes:2022vmn}, the authors propose to use the cINN model within an iterative framework (IcINN) for unfolding that adjusts for deviations between simulated training samples and data. The algorithm uses a classifier-based approach to reweight events in the simulated dataset to match the detector measurements as follows.
\begin{enumerate}
    \item[(1)] The first two steps are identical to the cINN setup: one first trains the cINN on the simulated data and applies it to the measured data, i.e., samples $z$ in the latent space under the condition of a measured event $y$ to obtain an unfolded distribution $f_{u, i}(x)$ (starting with $i=0$).
    \item[(2)] In the third step, a classifier is trained to learn the ratio between the phase-space densities of the unfolded distribution and the truth-level parton (prior) distribution. One reweighs the simulation on parton level distribution to match the unfolded distribution $f_{u, i}(x)$. Since each parton-level distribution is linked to a detector measurement, the event weights can be transferred from the parton level to the detector level.
    \item[(3)] The process is then repeated --- training-unfolding-reweighing with the new reweighed simulations until convergence.
\end{enumerate}
By using the classifier in the third step, we are able to include more information about the measured data in the simulation and, therefore improve our unfolding. 
The similarities between this approach and \texttt{Omnifold} are evident when we realise that both methods follow the EM algorithm:
\begin{itemize}
    \item \textbf{E step:} $\rho_i(z|x)\: \propto \:\rho_{\text{sim.}}(x|z)\nu_{i}(z)$, 
    
    \item \textbf{M step:} $\nu_{i+1}(z)\: = \:\hat{\rho}_{\text{sim.}}(z) / \rho_{\text{sim.}}(z)$, where $\displaystyle\hat{\rho}_{\text{sim.}}(z) \equiv \int \text{d}x\; \rho_{i}(z|x) \rho_{\text{data}}(x)$.
\end{itemize}
The expectation step is trained using a normalising-flow model (cINN), while the maximisation step uses a standard classifier to reweigh the MC events.
Compared to the standard cINN, the iterative process of IcINN is able to adjust to test-time distributions different from the training dataset. Furthermore, the method works effectively at unfolding comparable to the performance of reweighing MC simulated samples.

\texttt{SBUnfold}~\cite{Diefenbacher:2023wec} presents a strategy to perform the expectation step of the EM algorithm by transforming the MC samples to detector-level measurements using Schr\"odinger bridges. \texttt{SBUnfold} shows improved performance in some cases compared to the cINN setup.

None of the unbinned unfolding strategies mentioned above allows for simultaneously constraining (profiling) of the nuisance parameters. To resolve this issue, Ref.~\cite{Chan:2023tbf} introduces unbinned profiled unfolding (UPU), a ML-based method that results in an unbinned differential cross-section and can profile nuisance parameters. Specifically, UPU uses the binned maximum likelihood at the detector level as the metric to optimise the unfolding reweighing function $\omega_0(t)$, which takes unbinned parton level simulations as input. The function $\omega_0(t)$ and the nuisance parameters ($\theta$) are optimised simultaneously. This also requires learning a conditional likelihood ratio $\omega_1(t, r | \theta)$ which reweighs the detector-level spectra based on the profiled values of nuisance parameters and is taken as input for the optimisation of $\omega_0(t)$ and $\theta$. By profiling the nuisance parameter, the authors show that the model is able to better unfold and reconstruct the truth level distribution with the example of Higgs cross-section measurement.

Other kinds of methods, such as neural empirical Bayes~\cite{pmlr-v130-vandegar21a}, have been developed that take advantage of developments in neural density estimators to perform unfolding by learning $\rho(x|y)$ and $\rho(x)$ to maximise $\rho(y)$ using the Bayes theorem. The learned likelihoods can be used to reweigh the simulations to obtain parton-level distributions. Another method mitigates the problem of unfolding by learning optimized observables that can be calculated directly from the detector measurements~\cite{Arratia:2022wny} and can be unfolded easily using traditional unfolding strategies.  

\subsection{Detector simulation}

Generative models like GANs have been explored for detector simulations due to their ability to learn and replicate complex data distributions~\cite{deOliveira:2017pjk,Paganini:2017hrr,Hashemi:2023ruu,Hashemi:2023ple}. In this section, we look at some prominent applications of these models in detail.

\medskip

\noindent
{\bf CaloGAN}: Ref.~\cite{Paganini:2017dwg} introduced a new GAN-based method for simulating the 3D shapes of particle showers in electromagnetic calorimeters while retaining the spatiotemporal relations among layers.  Electromagnetic calorimeters are used in high-energy physics experiments to measure the energy and momentum of electrons, photons, and other high-energy particles. Particle showers are created when these particles interact with the material of the calorimeter. The shape of the shower can be used to infer the particle's energy and momentum. The calorimeter considered in Ref.~\cite{Paganini:2017dwg} is a simplified version of the ATLAS electromagnetic calorimeter, with three layers, of dimensions provided in Table~\ref{tab:decdim}.
\begin{table}[]
    \centering
    \begin{tabular}{c|c|c|c} \hline
    Layer &$z$ length &$\eta$ length &$\phi$ length \\
    &[mm] &[mm] &[mm] \\ \hline
    0 &90 &5 &160 \\
    1 &347 &40 &40 \\
    2 &43 &80 &40 \\ \hline
    \end{tabular}
    \caption{Dimension of a calorimeter cell. The $z$ direction is the direction of particle propagation, the $\eta$ direction would be along the $pp$ beam axis in a full experiment, and $\phi$ is perpendicular to $z$ and $\eta$~\cite{Paganini:2017dwg}.}
    \label{tab:decdim}
\end{table}
The training data is the \textsc{Geant4} generated calorimeter data corresponding to $e^+, \pi^+$, and $\gamma$’s being perpendicularly incident on the centre of the calorimeter with energies uniformly sampled from 1–100 GeV. A 3-dimensional particle energy is visualized as a series of three 2D images in $\eta - \phi$ space, where the pixel intensity represents the sum of the energies of all particles incident to that cell. The first layer can be represented as a $3 \times 96$ image, the middle layer as a $12 \times 12$ image, and the last layer as a $12 \times 6$ image.

\begin{figure*}
    \centering
    \includegraphics[width=0.9\textwidth]{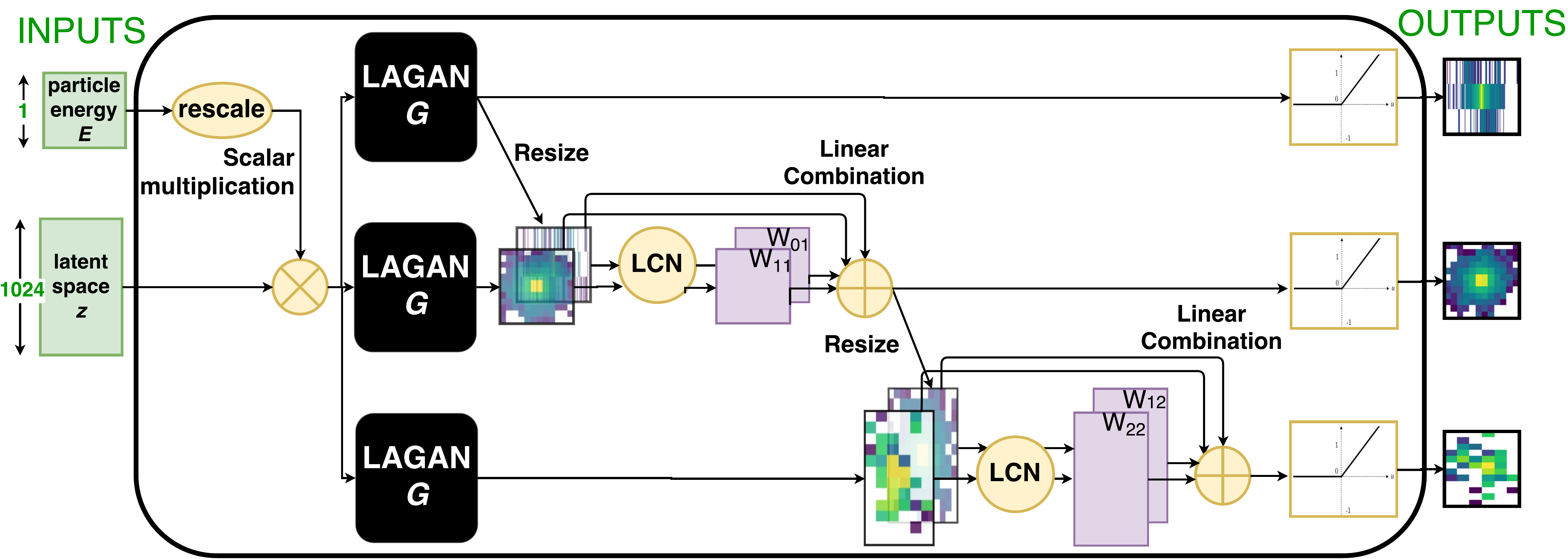}
    \caption{Composite Generator, illustrating three streams with attentional layer-to-layer dependence, Ref.~\cite{Paganini:2017dwg}.}
    \label{fig:G}
\end{figure*}

\begin{figure*}
    \centering
    \includegraphics[width=0.9\textwidth]{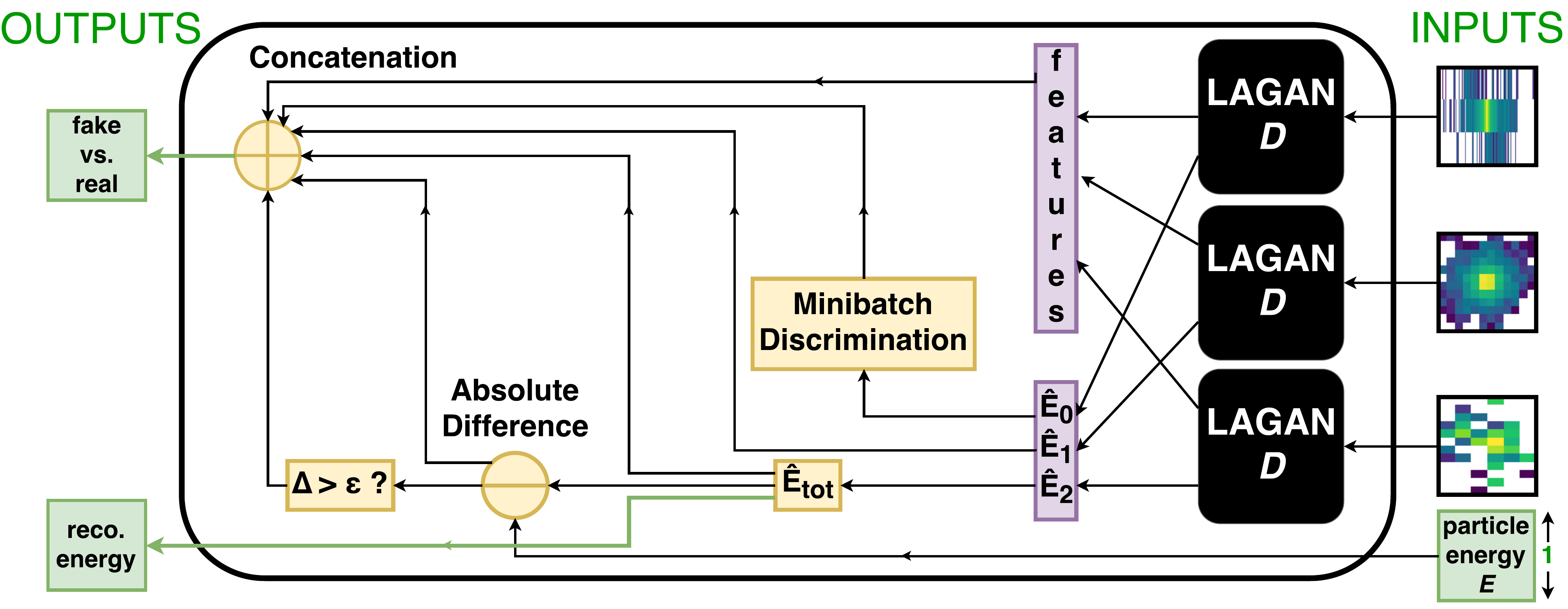}
    \caption{Composite Discriminator, depicting additional domain-specific expressions included in the final feature space, Ref.~\cite{Paganini:2017dwg}.}
    \label{fig:D}
\end{figure*}

\textsc{CaloGAN} consists of two neural networks: a generator (G) and a discriminator (D). It follows the DCGAN (Deep Convolutional GAN) architecture, with the inclusion of LAGAN (Locally Aware Global Anomaly Network)\footnote{The key idea in LAGANs is the integration of location information into the GAN architecture, enabling the generation of images with improved spatial coherence and context-awareness. It concatenates spatial coordinates or spatial embeddings with the input at various layers of both the generator and discriminator networks. Spatial coordinates could be absolute pixel positions or relative positions within the image. Spatial embeddings are learned representations of location information that can capture spatial relationships within the image. By incorporating location information, the generator can learn to produce images with better spatial coherence, such as smooth transitions between objects or accurate placement of features.} guidelines, as shown in Figs.~\ref{fig:G} and~\ref{fig:D}. The generator takes a random noise vector (latent space) as input and maps the latent input to three grayscale image outputs with different numbers of pixels, as per the three calorimeter layers. The mismatch of dimensionality and granularity among the three longitudinal segmentations of the detector, demands separate streams of operations with suitably sized kernels. The generator architecture, therefore, has multiple independent processing paths, each resembling a LAGAN architecture,  with a trainable attention mechanism focusing on the relationships between these parallel LAGAN streams. The LAGAN submodules consist of a 2D convolutional unit, for spatial feature extraction followed by two locally-connected units which focus on specific regions within the data, leveraging the spatial features extracted by the previous layer. The attention mechanism allows dependence among layers, helping in utilizing information from the previous layers to the generation of the subsequent layer readout. The discriminator D accepts the three images as inputs, along with the energy of the incoming particle, $E$. The images are augmented with a sub-differentiable version of the sparsity percentage, and minibatch discrimination\footnote{Minibatch discrimination is used in the discriminator of GAN to increase diversity in generated samples and improve training stability. It addresses the problem of mode collapse, where the generator produces a limited variety of samples that are easily classified as fake by the discriminator.} is used on both the generated images as well as the augmented ones with the output sparsity. The inputs are mapped to a binary output that classifies showers into real and fake and a continuous output that calculates the total energy deposited in the three layers and then compares it with the requested energy E.

\begin{figure*}
  \centering
    \includegraphics[width=0.25\textwidth]{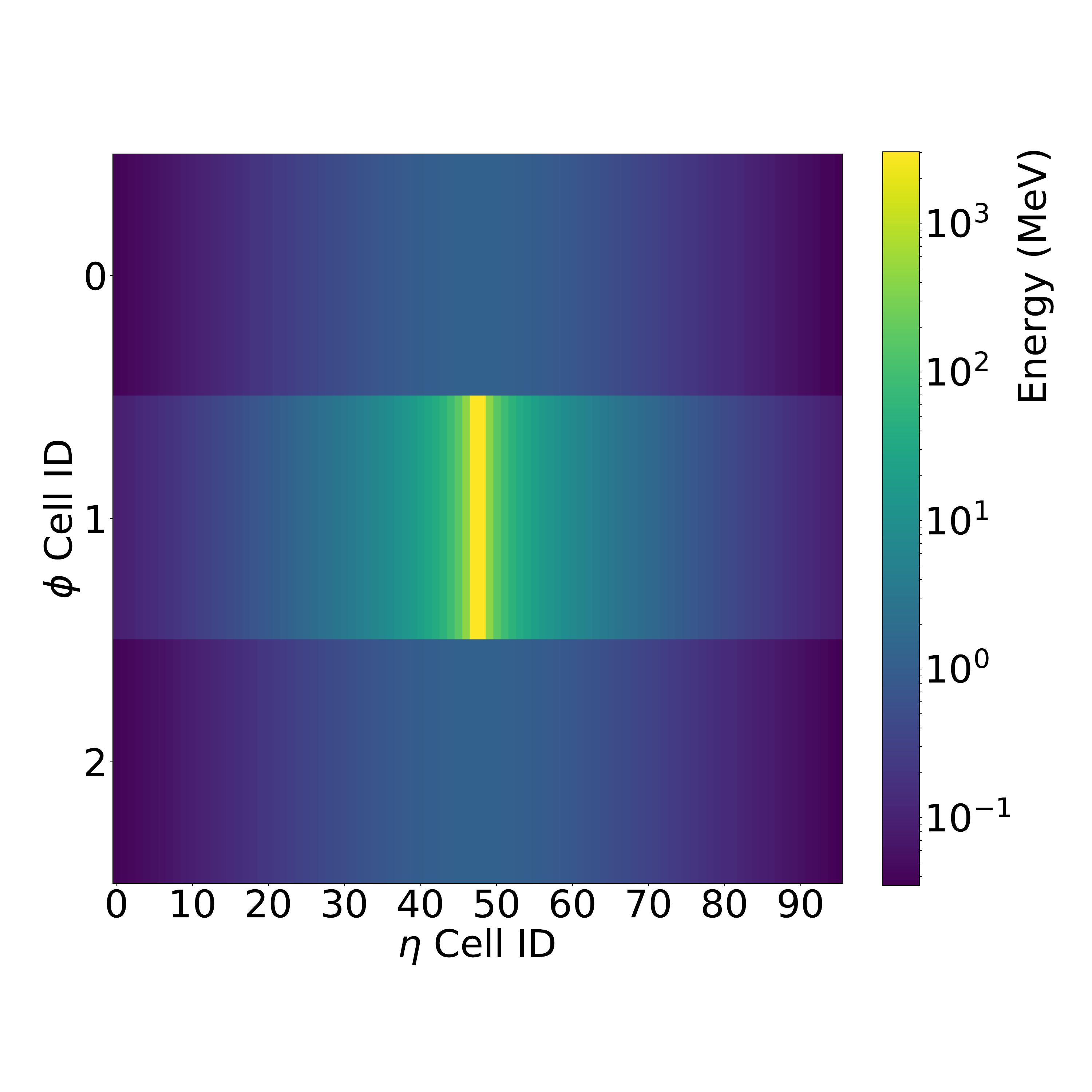}
    \includegraphics[width=0.25\textwidth]{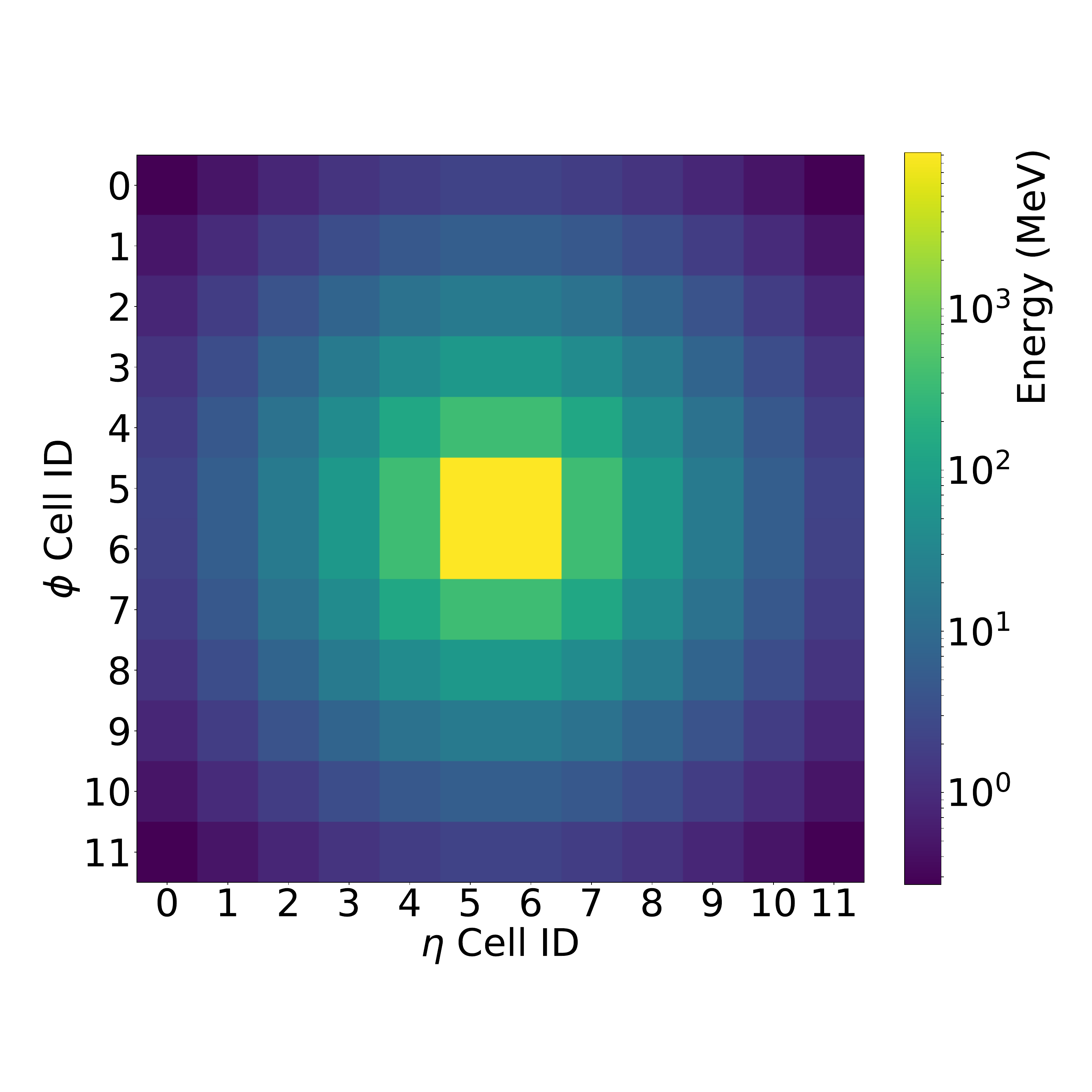}
    \includegraphics[width=0.25\textwidth]{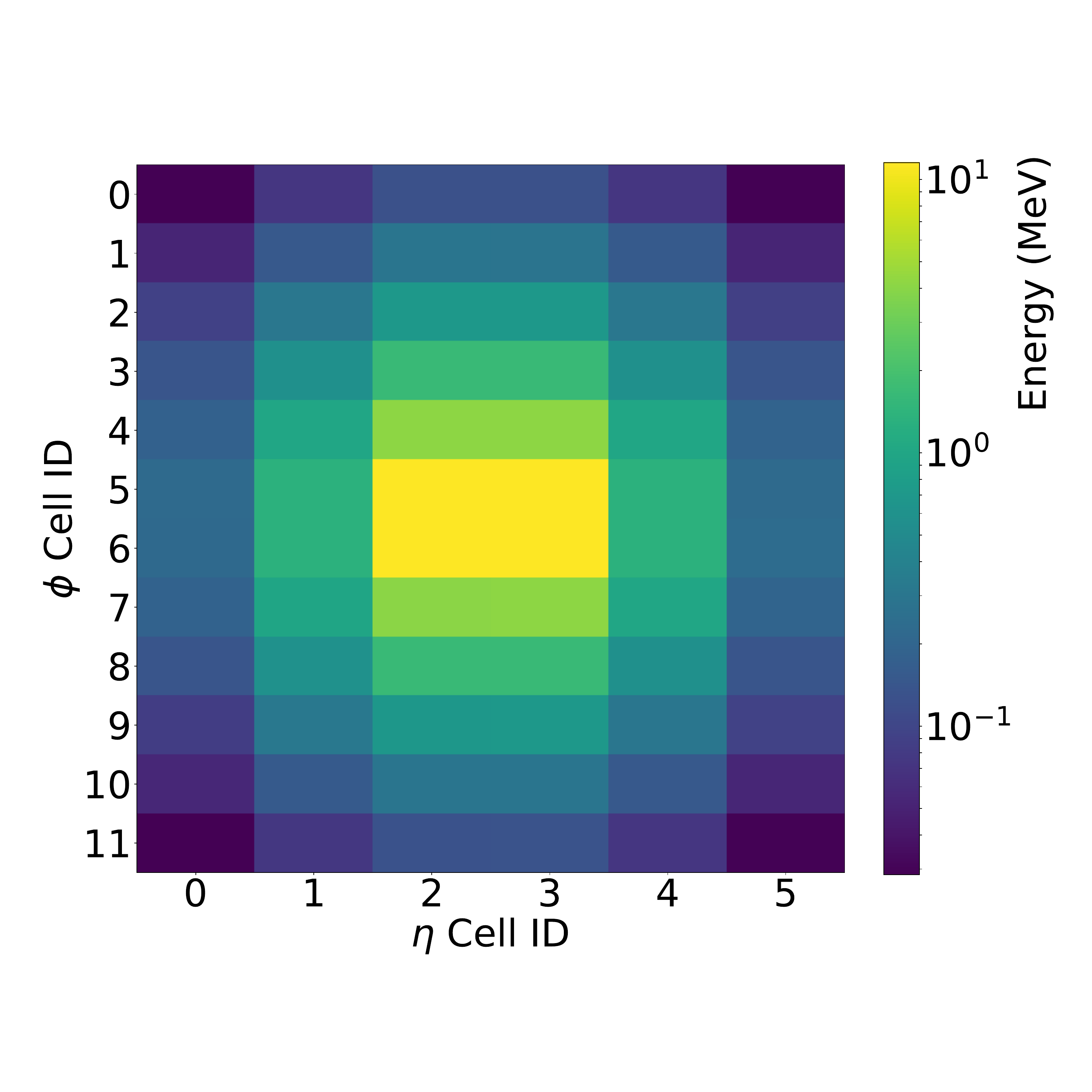}\\
    \includegraphics[width=0.25\textwidth]{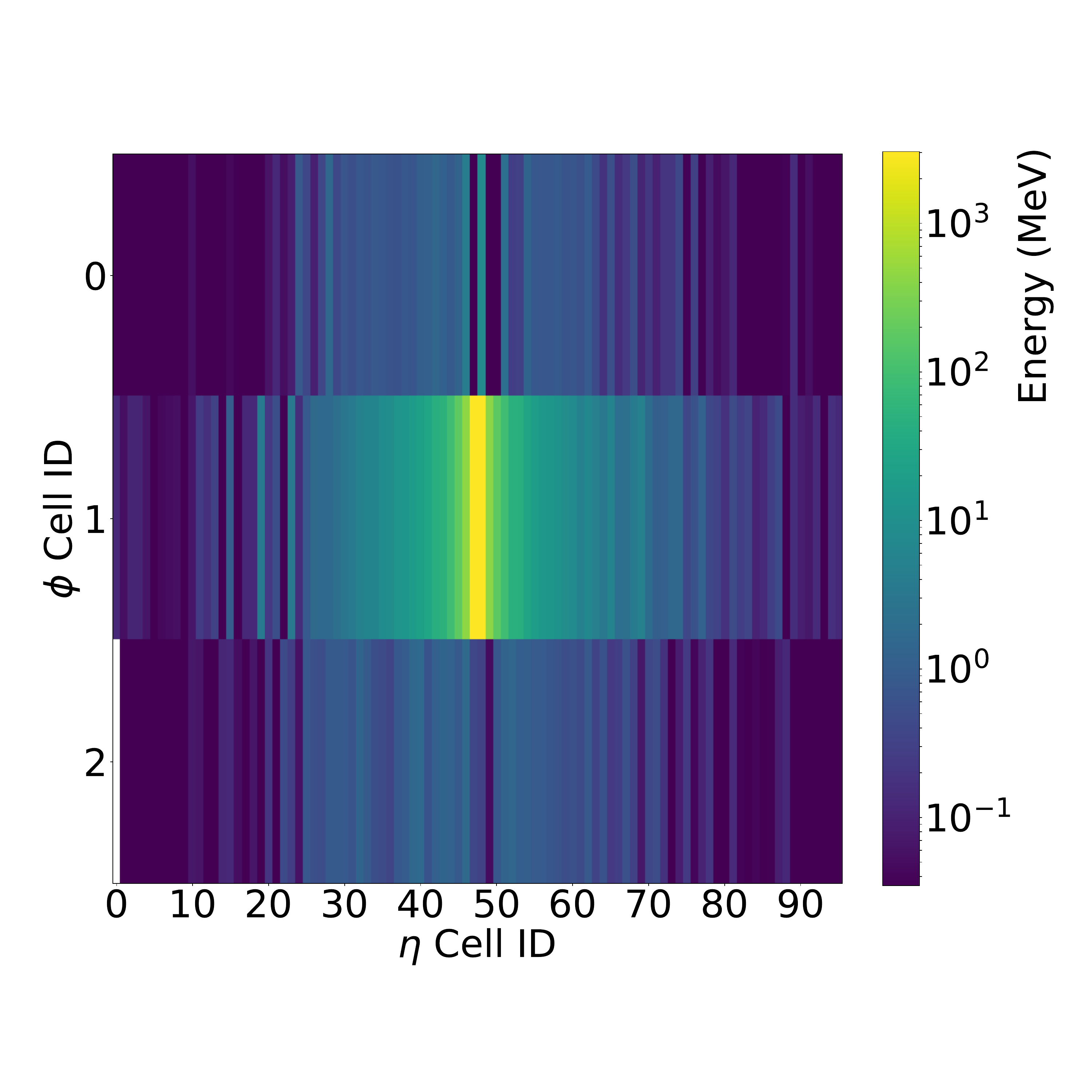}
    \includegraphics[width=0.25\textwidth]{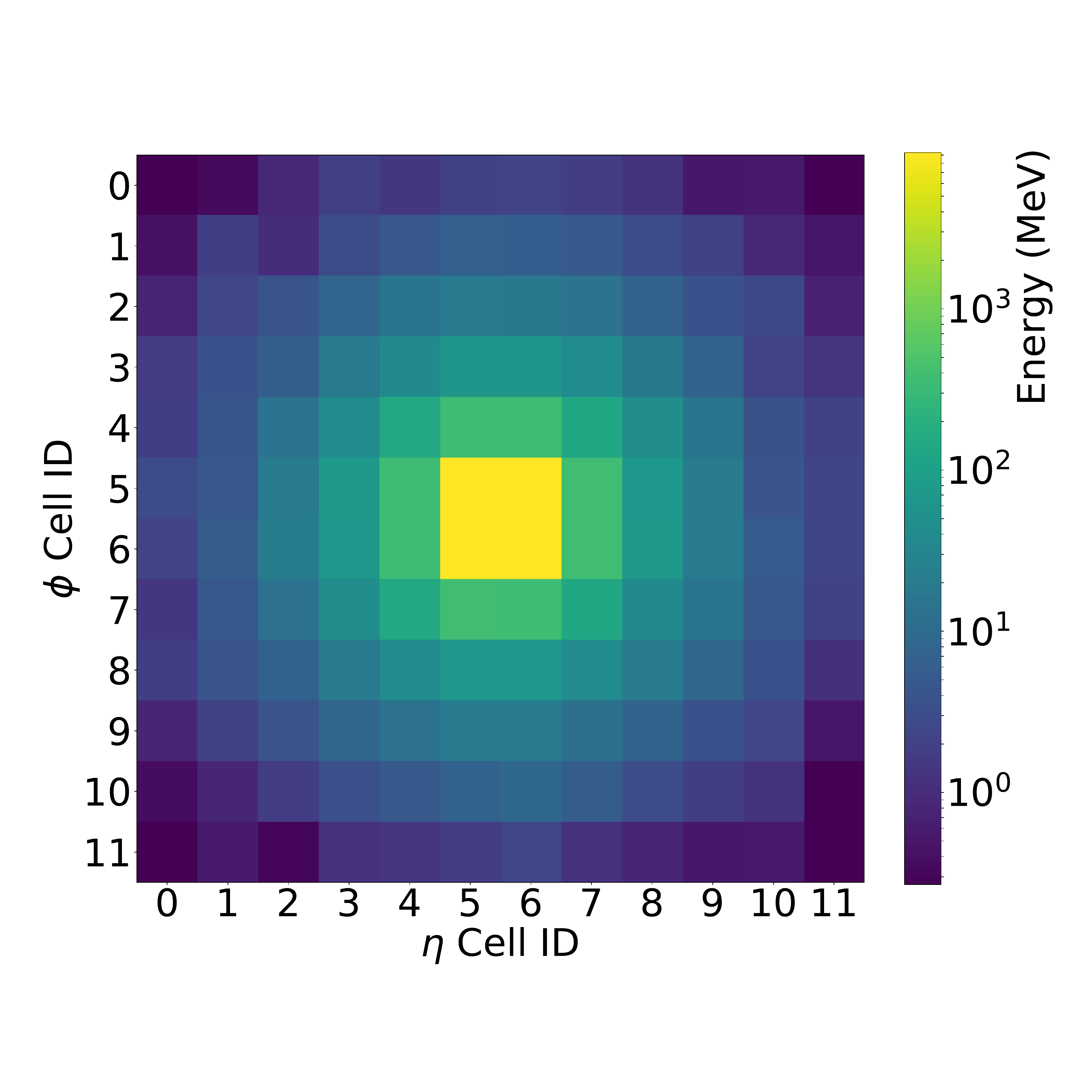}
    \includegraphics[width=0.25\textwidth]{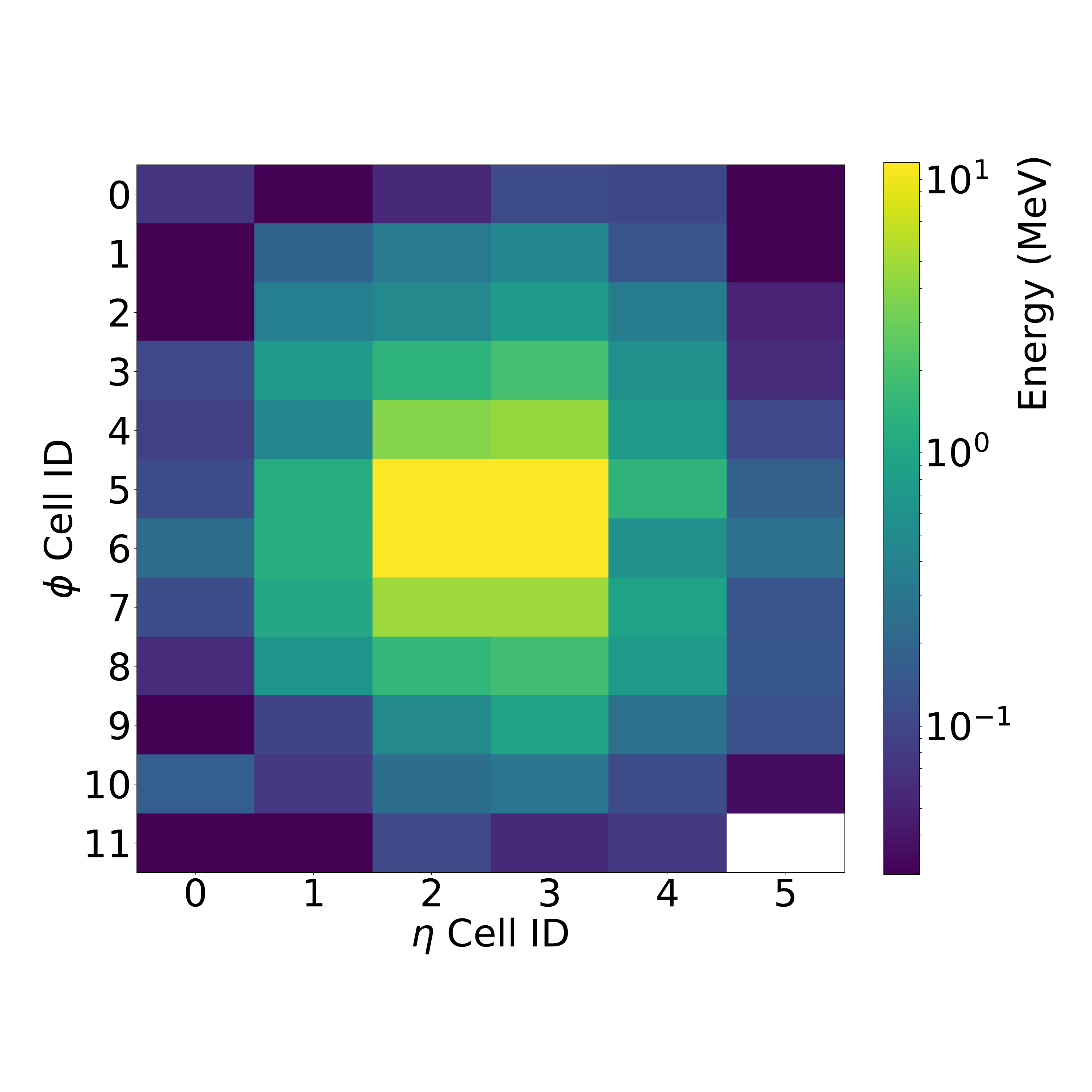}\\
    \caption{The average calorimeter deposition per voxel is examined, with the network learning a complete picture of the physical processing governing the cascades of $e^+$. The energy of the underlying process is uniformly distributed between 1 GeV and 100 GeV. The average $e^+$ \textsc{Geant4} shower (top), and average $e^+$ \textsc{CaloGAN} shower (bottom), are shown with a progressive increase of calorimeter depth (left to right). See Ref.\cite{Paganini:2017dwg} for more details.}
    \label{fig:eplus_avg}
\end{figure*}

The \textsc{CaloGAN} model was trained and was able to learn a complete picture of the underlying physical processes governing the cascades of $e^+, \gamma$, and $\pi^+$  with uniform energy between 1 GeV and 100 GeV. The average $e^+$ \textsc{Geant4} shower (top), and \textsc{CaloGAN} shower (bottom), with progressive calorimeter depth (left to right), is shown in Fig.~\ref{fig:eplus_avg}.
\textsc{CaloGAN} image quality was tested from the transferability of classification performance from GAN-generated samples to \textsc{Geant4}-generated samples and a good agreement was found. 

The \textsc{CaloGAN} model has the potential to significantly reduce the computational cost of simulating particle showers in electromagnetic calorimeters. Once trained, the simulator can be used at a cost of as little as 12 microseconds per event---a five-order-of-magnitude speedup compared to the full simulation. 

The non-convex min-max loss function associated with GAN sometimes can not converge to an optimum leading to "mode collapse", with the GAN learning to generate only a subset of the data. Moreover, the ability of GANs to truly reproduce the underlying distribution of the data is not clear mathematically as GANs can only learn the likelihood implicitly. Other state-of-the-art GAN architectures, such as WGAN-GP~\cite{Arjovsky:2017fjr},~\cite{Gulrajani:2017kpy} (the generator minimizes the Wasserstein distance in WGAN), Bounded Information Bottleneck Autoencoder (BIB-AE)\footnote{The BIB-AE architecture unifies architectures from GANs and VAEs with the principles of Information Bottleneck method. The primary aim is to balance the compression and the retention of relevant information in a way that is beneficial for representation learning tasks. The Encoder Network maps the input calorimeter images to some latent representation, with the decoder generating the calorimeter images from the latent space. The information bottleneck here refers to the principle that the latent representation is optimized by the model while also maximizing the mutual information between the input and output.}~\cite{Buhmann:2021lxj}, and GANs using the Deep neural networks using the Classification for Tuning and Reweighting protocol (DCTRGAN)\footnote{The key idea here is to leverage classification-based tuning and reweighting strategies to enhance the quality and stability of the generated outputs from GANs. Classification-based tuning uses a classifier network trained alongside the GAN to guide the tuning of the GAN components, ensuring that the generated data aligns with desired class distributions. The reweighting part consists of adjusting the importance of different data points or features during training to address class imbalances or to emphasize certain characteristics in the generated data. For example, if certain classes are underrepresented, their corresponding data points can be given higher weights. The loss function for both the generator and discriminator can be adjusted using these weights, ensuring that the GAN learns to generate balanced and representative data.},~\cite{Diefenbacher:2020rna} have been proposed to help with training stability and mode collapse. The loss function in WGAN, comparable to the JSD, Eq.~\eqref{eq:jsd}, is defined as:
\begin{equation}\label{eq:WGan}
   \min_{\theta^{(G)}} \max_{\theta^{(D)}} L_{WGAN} = \min_{\theta^{(G)}} \max_{\theta^{(D)}} \lbrace \mathbb{E}_{x\sim p_{data}} D(x) -\mathbb{E}_{z \sim p(z)} D(G(z))\rbrace.
\end{equation} 
The JSD can lead to vanishing gradients and instability during training especially when the distributions have disjoint supports. WGANs address this issue by using Wasserstein distance as their distance metric, which encourages smoothening convergence. Wasserstein distances are a measure of dissimilarity between two distributions, mathematically measuring the minimum "cost" of transporting one probability distribution into another. Since the transport distance increases with the separations between the two distributions, there is no gradient collapse when the distributions are non-overlapping.

\medskip
\noindent
{\bf CaloFLOW:} Another popular approach was introduced in Ref.~\cite{Krause:2021ilc} with the introduction of \textsc{CaloFLOW}, a fast detector simulation framework based on normalizing flows. Normalizing flows (defined in Eq.~\ref{eq:NF}) offers a unique and powerful approach to generative modelling, blending high-quality data generation with the crucial ability to understand and reason about the underlying probabilities. They excel in tractability because they maintain a clear mathematical connection (bijective transformation) between the simple initial distribution (latent space $z$) and the complex final one (data space $x$). Normalizing Flows use invertible transformations so that the probability density can be computed and the generator is optimized using the log-likelihood. In simple words, these models map the data $x$ to latent variables $z$ through a sequence of invertible transformations $f = f_1$ o $f_2$ o $\cdots$ o $f_n$, such that $z = f(x)$ or $x = g(z) = f^{-1}x$. The ability to estimate the likelihood of samples is a key advantage of normalizing flows. The latent space $z$ is modelled with a simple base distribution, typically chosen to be a standard normal (Gaussian) distribution. Once the model is trained, it can generate new samples by transforming samples from a simple distribution (e.g., Gaussian noise) through the learned sequence of invertible transformations. 
 
The authors have used a combination of Masked Autoencoders for Distribution Estimation (MADE) and Neural Spline Flows to maximize expressive power (see Appendix \ref{sec:MADE} for more details). They have used the same calorimeter configuration as \textsc{CaloGAN} and trained two separate normalizing flows, a smaller one to learn the distribution of energies deposited in the three layers of the calorimeter, ensuring energy conservation and another one to learn the shower shapes in each layer. They present a detailed comparison of \textsc{CaloFlow} results with \textsc{CaloGAN} and \textsc{Geant4}. The comparison is on the average shower shapes images of the 100K events, nearest neighbour images (selecting elements of the \textsc{Geant4} set at random and looking for their nearest neighbours in the set of generated \textsc{CaloFlow}/\textsc{CaloGEN} samples) and histograms of relevant physical quantities to better assess the quality of the generated shower shapes and voxel level information. \textsc{CaloGAN} in some regions of the phase space shows a sign of mode collapse, with the results from \textsc{CaloFLOW} very close to their \textsc{GEANT4} counterpart.

The metric earlier used in \textsc{CaloGAN} to test the image generation was to distinguish between different categories of data (e.g. $e^+$ vs. $\pi^+$) and to see if there is any difference in classifier performance when real data and generated data are interchanged. The authors here have used a new quantitative metric, where a binary classifier is trained to distinguish \textsc{Geant4} and generated images, demonstrating the high fidelity of \textsc{CaloFlow}-generated images. If the generated and true probability densities are equal, and the classifier is optimal, then according to the Neyman-Pearson lemma it will be no better than random guessing.

\begin{figure}[t]
\centering
\includegraphics[width=0.6\textwidth]{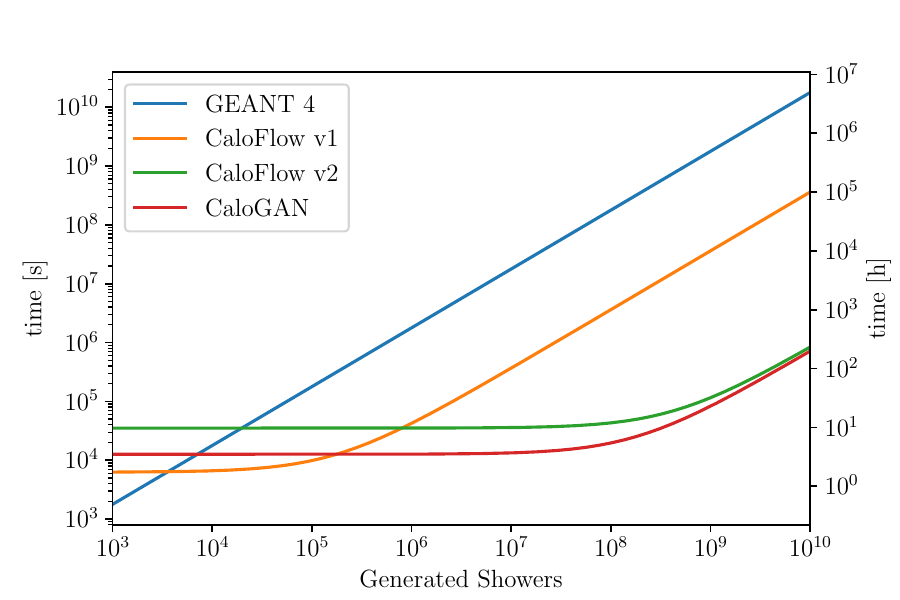}
\caption{Comparison of shower generation times, using the fastest \textsc{CaloGAN} numbers for comparison. \textsc{CaloFlow v2} leads to a faster shower generation while still preserving the high fidelity of \textsc{CaloFlow},~Ref.~\cite{Krause:2021wez}.}
\label{fig:timing}
\end{figure}

Normalizing flows as we saw, learns a series of invertible transformations that map a simple distribution (e.g., Gaussian) to the target data distribution. The model is trained to maximize the likelihood of the observed data under these transformations. Vanilla NFs have difficulty scaling to higher-dimensional datasets, especially when used for the study of ultra-fine calorimeters~\cite{ATLAS:2021pzo}. It has been observed that for two of the three datasets in the recent Fast Calorimeter Simulation Challenge 2022~\cite{Krause:2022jna}, the NF training time is prohibitive and would require significant research and development~\cite{Buckley:2023rez, Diefenbacher:2023vsw}. \textsc{CaloFlow}, though $\sim 50$ times faster than $\textsc{GEANT4}$, is found to be considerably slower than the GAN-based alternative. The Masked Autoregressive Flow (MAF) architecture used in \textsc{CaloFlow} though fast in density estimation, is slower in sampling. The Inverse Autoregressive Flow (IAF), on the other hand, leads to a fast sampling and a slow density estimation. Implementing IAF in \textsc{CaloFlow} is not straightforward as the high dimensionality collider data cannot be trained using the negative log-likelihood objective, due to memory and time limitations (see Ref.~\cite{Krause:2021wez} for more details). The authors in Ref.~\cite{Krause:2021wez} have used the Probability Density Distillation method, alternatively known as the teacher-student training for \textsc{CaloFlow v2}. The MAF (teacher) leads to a fast density estimation, by quickly mapping the target data points to the latent space, $x \rightarrow z$, while the IAF (student) trains on the output of the trained MAF model, leading the IAF inverse to the mapping of the latent space back to the target data space, $z \rightarrow x'$. The teacher-student training leads to \textsc{CaloFlow v2} preserving the same high fidelity of \textsc{CaloFlow}, but also significantly faster comparable to the previous version as shown in Fig.~\ref{fig:timing}.

\medskip
\noindent
{\bf Score-based Generative Models (CaloSCORE)}: An alternative class of generative models is proposed in Ref.~\cite{Mikuni:2022xry} known as score-based models. These models operate by estimating the gradient of the data distribution's log-likelihood w.r.t. the model parameters. 
The training of the score-based generative model uses a diffusion process that involves iteratively transforming a simple distribution to match the target data distribution. The diffusion process is modelled using a stochastic differential equation, which describes how the probability density evolves. The training process is divided into discrete time steps, where the model updates the samples by applying a transformation that introduces noise. This transformation gradually deforms the simple distribution towards the target distribution. At each time step, the model estimates the gradient of the log-likelihood of the data distribution w.r.t the model parameters. The score is used to update the model parameters through gradient-based optimization methods, such as stochastic gradient descent (SGD). Importantly, the transformations applied during the diffusion process are designed to be reversible. This ensures that, given a sample from the target distribution, it is possible to trace back the diffusion process to obtain a sample from the simple distribution. The objective during training is to make the transformed samples at each time step indistinguishable from samples drawn from the true data distribution. This is achieved by minimizing a suitable divergence or distance measure between the transformed distribution and the true data distribution. By iteratively applying the diffusion process and updating the model parameters based on the estimated gradients, the score-based generative model learns to generate realistic samples that match the true data distribution.

A detailed evaluation of the quality of generated samples using different \textsc{CaloScore} SDE implementations and a WGAN-GP model for fast calorimeter simulation is presented. Various distributions are used to assess the agreement between the \textsc{Geant4} and generated samples, focusing on parameters such as total energy deposited, calorimeter hits, energy deposition distribution, and angular distribution of the calorimeter shower. They obtain good agreement between the \textsc{CaloScore} implementations and \textsc{Geant4} predictions, with some deviations observed in specific scenarios. The WGAN-GP model shows higher deviations compared to \textsc{CaloScore}. \textsc{CaloScore} includes a scalable and stable training schedule based on the minimization of the convex score-matching loss and exact likelihood estimation. The major challenge to be addressed in \textsc{CaloScore} is the generation time, currently requiring hundreds of model evaluations to solve the reverse SDE. The potential for accelerating the generation process is noted and taken care of in Ref.~\cite{Mikuni:2023tqg}, where the \textsc{CaloScore} architecture is updated through the use of the attention layer which produces higher-quality samples. The additional use of progressive distillation leads to an iterative reduction in the number of time steps required during sampling. This technique further shows that single-shot diffusion models can be achieved for fast and high-fidelity simulation in collider physics. Overall, \textsc{CaloScore} offers a new generative paradigm with the potential for producing realistic calorimeter showers in collider physics while addressing some limitations of existing generative models.

\medskip
\noindent
{\bf ATLAS calorimeter simulation using GANs:} The experimentalists at LHC have also established that a fast calorimeter simulation is essential to produce the number of simulated events needed to match the High Luminosity LHC. ATLAS first provided the AtlFastII tool~\cite{ATLAS:2010arf} based on the FastCaloSim package~\cite{ATLAS:2010bfa}. This tool provided a parametrized simulation of the particle energy response and the energy distribution in the ATLAS calorimeter system hence reducing the calorimeter simulation time to a few seconds per event. The parametric simulations simulate the energy of a particle shower as a single step based on an underlying parametrization instead of simulating how every particle propagates and interacts inside the calorimeter volume. This tool has difficulty in modelling boosted objects and jets of particles reconstructed with large-radius clustering algorithms. Moreover, some observables in electromagnetic showers are not well reproduced. ATLAS has also proposed FastCaloGAN~\cite{FaucciGiannelli:2742369}, based on GANs that can perform a full calorimeter simulation for three particle types (electrons, photons, and pions) for a wide range of energies. The FastCaloGAN tool is based on the conditional WGAN-GP algorithm, doing the conditioning on the true momentum of the input samples since WGAN has been demonstrated to provide good performance and stability of the training. A good level of agreement was observed on reconstructed variables for both single-particle showers and jets with the most notable difference coming from the total energy of $e/\gamma$ showers. The $e/\gamma$ showers were well reproduced by FastCaloSimV2. Therefore, the ATLAS experiment designed the AtlFastIII tool~\cite{ATLAS:2021pzo}, a combination of the above two. It uses the $e/\gamma$ and hadron showers at high and low energies simulated by FastCaloSimV2 and medium energy hadrons by FastCaloGAN and also provides a high-performance fast simulation. It does a smooth interpolation between the two tools and is designed to simulate particle showers to a level of precision such that no sizeable differences from the \textsc{Geant4} simulation can be resolved by the reconstruction algorithms. Additionally, it requires only 20\% as much CPU as \textsc{Geant4} to simulate an event. 

\medskip
\noindent
{\bf GANs at highly granular detectors:} The 3DGAN conditional model is used as an imaging tool for electromagnetic calorimeter (ECAL) simulation of the CLIC detector in Ref.~\cite{Vallecorsa:2019ked}. It is based on three-dimensional convolutions to capture the whole shower development along the three spatial dimensions corresponding to a specific particle energy. The discriminator in this adversarial network, apart from discriminating the real and fake samples, also performs two additional tasks, estimating the incoming particle energy ($E_p$) and the total energy measured by the calorimeter ($E_{cal}$). The authors have successfully generated three-dimensional images of energy showers, according to different energy and incident angle values. A good agreement is found between their 3DGAN and GEANT4 output, for both low-level variables describing the calorimeter energy response, i.e., the single cell energy distributions, and more complicated features such as average profiles of the electromagnetic showers along the three calorimeter axes. Additional works have been done in the context of planned highly granular International Large Detectors (ILD) for future linear colliders, or other detectors of high granularity, see Refs.~\cite{Buhmann:2021caf, Buhmann:2023bwk, Diefenbacher:2023prl, Acosta:2023zik}.

\medskip
\noindent
{\bf Other frameworks:} JUNIPR framework discussed in Ref.~\cite{Andreassen:2018apy}, uses a deep neural network to learn the full differential distribution of the data. It can be trained using the tag-and-probe method or weakly supervised learning to find the different probability functions on separate samples. These learned functions can be compared to discriminate the different samples and find interesting features. JUNIPR once trained on data, can also be used for generation with the events generated from the learned probability distributions. The strength of JUNIPR lies in its interpretable nature. It deconstructs the probability density into contributions from each point in the clustering history, thereby providing insights into the underlying physics.  The authors of Ref.~\cite{Howard:2021pos} propose a new method, One-shot Topological Unsupervised Simulation (OTUS) for simulating high-energy particle collisions from unlabeled data. Their method is based on a probabilistic autoencoder, where the autoencoder is trained on a dataset of particle collision events, and the latent space of the autoencoder is then identified with the space of theoretical models of particle collisions. This allows the decoder network of the autoencoder to be used as a fast, predictive simulator of particle collisions.
The authors have demonstrated the effectiveness of OTUS on two-particle physics examples, $Z$-boson and top-quark decays, and show that OTUS can accurately simulate the kinematic distributions of particles produced in these decays, even though it was not trained on any data that explicitly contained information about these distributions.

\section{Conclusions and outlook}\label{sec:conclusions}

Particle physics in the era of high-luminosity LHC and other data-intensive experiments has led us to think of new ways to analyse and interpret experimental data. This, in turn, has paved the way for the natural synergy between particle physics and modern ML techniques. Supervised learning has been widely used for classifying signals from backgrounds at the colliders. However, all supervised methods work on labelled data, which rely heavily on our ability to generate synthetic datasets for training. We can not use supervised techniques on real data which has no labels like synthetic data. People have approached this issue in different ways. For example, depending on how much a priori information one has on the real data, one can use unsupervised, weakly supervised, or semi-supervised techniques. Alternatively, one can also use ML techniques to determine how well the simulations agree with data and then try to improve the simulations. Currently, all these approaches are actively being studied and applied in particle physics (see, e.g., Refs.~\cite{Capozi:2019xsi,Romao:2024gjx}).

Unsupervised clustering methods identify patterns and group objects in high-dimensional data from particle detectors---we use them to reconstruct and classify jets. Anomaly detection algorithms can identify rare or unexpected events in the data that may indicate the presence of NP or detector malfunctions. AEs and VAEs are commonly used for anomaly detection. A comparison of the performance of AEs to other shallow ML models for different anomaly detection tasks is presented in Ref.~\cite{CrispimRomao:2020ucc}. AE models require a better understanding of the learned (latent) space. Some studies look at this aspect and other challenges associated with AEs~\cite{Fraser:2021lxm}. Many weak-supervision techniques---particularly the CWoLa-based ones---assume that the anomaly is localised in a subspace of the entire feature space. As shown in Ref.~\cite{Metodiev:2017vrx}, there is an asymptotic guarantee of optimality when the classifier is trained on mixed samples of signal and backgrounds, provided the samples have different signal fractions. However, in practice, the performance of such classifiers depends on several factors, particularly the signal fractions in SRs. There are studies that look into the correlations between these methods~\cite{Golling:2023yjq} (in the context of resonant anomaly detection)---a combination of these methods looks the most promising for anomaly detection. Comparisons between fully and partially unsupervised methods show that they are complementary and can be used in combination with each other~\cite{Collins:2021nxn}. ATLAS~\cite{ATLAS:2020iwa} and CMS~\cite{Zipper:2023ybp} have started exploring these strategies for NP searches. By combining unsupervised learning with domain-specific knowledge and expertise, physicists can leverage the strengths of both approaches to extract meaningful insights from experimental data. The use of generative modelling in simulating particle collisions and detector unfolding is rising. Density estimation-based models like normalising flows (that can efficiently map a simple distribution to a target one) are being used for detector simulation. There are also classifier-based tools like \texttt{Omnifold}, which learn the mapping from simulation to observed data and then invert the mapping to unfold detector effects.

In this review, we mainly focused on the interesting ideas in a small subset of the burgeoning field of data-driven physics models. These techniques look promising and their complete scope is yet to be explored. With the advent of noisy intermediate-scale quantum computers, the efficiency of the data-driven methods may take a higher leap in the near future.\footnote{Quantum algorithms for ML~\cite{Guan:2020bdl,Hammad:2023wme,Blance:2020nhl,Terashi:2020wfi,Gianelle:2022unu}, partiularly in anomaly detection~\cite{Alvi:2022fkk,Wozniak:2023xbe,Schuhmacher:2023pro,Ngairangbam:2021yma}, is already being explored.} Moreover, beyond the colliders, data-driven models, or ML methods in general, can play crucial roles in other domains like neutrino physics, string theory, lattice QCD, etc. (see Ref.~\cite{Schwartz2021Modern} and the references therein).

\appendix
\refstepcounter{section}
\section*{Appendix \thesection\quad Kullback-Liebler divergence}\label{sec:KLD}

Kullback-Liebler ($\mathbb{KL}$) divergence~\cite{10.1214/aoms/1177729694} is a measure of how a probability distribution diverges from an expected probability distribution. In simple words, we can think it of as a measure that shows how different the two distributions are from each other. The $\mathbb{KL}$ divergence can be used for both discrete and continuous distributions. Given two discrete probability distributions, $\rho$ (actual distribution) and $\pi$ (reference distribution), the $\mathbb{KL}$ divergence is calculated as the sum over all possible outcomes:
\begin{equation}
\mathbb{KL}(\rho||\pi) = \sum_{i=1}^N \rho_i \log \left(\frac{\rho_i}{\pi_i}\right),
\end{equation}
where $\rho_i$ and $\pi_i$ are the probabilities for the outcome $i$. If the distributions are continuous, it is calculated as the integral over the domain:
\begin{equation}
\mathbb{KL}(\rho||\pi) = \int \rho(x) \log \left(\frac{\rho(x)}{\pi(x)}\right) dx.
\end{equation}
If the two distributions $\rho$ and $\pi$ perfectly match, $\mathbb{KL}(\rho||\pi)=0$. Therefore, the lower the $\mathbb{KL}$ divergence value, the better the match between the original and the reference distribution. However, the $\mathbb{KL}$ divergence is not a distance metric, as it is not symmetric in $\rho$ and $\pi$. It will result in a different number if we swap the actual and reference distributions. 

$\mathbb{KL}$ divergence in VAE is used to measure the information loss when approximating the latent space distribution with a prior distribution, usually a normal distribution. In the case of two multivariate $J$-dimensional Gaussian distributions, the $\mathbb{KL}$ divergence is given by:
\begin{equation}
\mathbb{KL}(\rho||\pi) = \frac{1}{2}\left[(\mu_2-\mu_1)^T \Sigma_2^{-1}
(\mu_2-\mu_1) + \mathrm{Tr}(\Sigma_2^{-1}\Sigma_1)+\log \frac{|\Sigma_2|}{|\Sigma_1|}-J\right],
\end{equation}
where the probability distributions of $\rho$ and $\pi$ are defined by their mean vectors $\mu_{1,2}$ and covariance matrices $\sum_{1,2}$. 
\begin{equation}
\rho = \Phi(\mu_1,\Sigma_1), \pi = \Phi(\mu_2,\Sigma_2).
\end{equation}
If the distribution $\pi$ is a normal distribution with $\mu_2$ = 0 and $\sum_2$ = $\mathbb{I}$, it reduces to
\begin{equation}
\mathbb{KL}(\rho||\pi) = \frac{1}{2}\left[\mu_1^T \mu_1 + \mathrm{Tr}(\Sigma_1)-J-\log|\Sigma_1|\right].   
\end{equation}
From the above, we get Eq.~\eqref{eq:KLloss} if we substitute $\mu_1 = \mu$, $\Sigma_1 = \mathrm{diag}(\sigma^2)$ (the encoder outputs diagonal covariance matrices).

\refstepcounter{section}\section*{Appendix \thesection\quad Jensen-Shannon divergence}\label{sec:JSD}

The Jensen-Shannon divergence~\cite{MENENDEZ1997307,lin1991divergence} is another measure of similarity between two probability distributions. In the context of GANs, the JSD is often used to quantify the difference between the distribution of real data and the distribution of generated data. Let $\rho$ represent the distribution of real data and $\pi$ represent the distribution of generated data from a GAN. The Jenson-Shannon Divergence between $\rho$ and $\pi$ is defined as:
\begin{equation}
    \mathrm{JSD}(\rho,\pi) = \frac{1}{2}\left(\mathbb{KL}(\rho||M)+\mathbb{KL}(\pi||M)\right),
\end{equation}
where $\mathbb{KL}(\rho||M)$ is the Kullback-Liebler divergence between $\rho$ and the distribution $M = (\rho+\pi)/2$. Similar definition holds for $\mathbb{KL}(\pi||M)$. The JSD is essentially a symmetric version of the $\mathbb{KL}$ divergence. It measures how much information is lost by representing $\rho$ and $\pi$ with the average distribution, $M$. During the training of GANs, the generator tries to minimize the JSD, as a smaller JSD indicates that the generated distribution is becoming more similar to the real data distribution.

\refstepcounter{section}\section*{Appendix \thesection\quad MADE and neural spline flow}\label{sec:MADE}

MADE (Masked Autoencoder for Distribution Estimation) uses masked neural networks to enforce an autoregressive structure by allowing each variable in the input to depend only on the variables that precede it. The masks restrict the connections between layers in the neural network. These masks ensure that each feature in the input is only connected to features in subsequent layers that have been observed previously, preserving the autoregressive property. Autoregressive models like MADE are well-suited for density estimation tasks because they model the conditional probability of each feature given the previous features. By doing so, MADE can efficiently model complex, high-dimensional distributions. 

Neural Spline Flows use piecewise polynomial transformations, specifically spline functions, to model the data distribution. Spline functions offer a flexible way to capture complex shapes in the data distribution. The piecewise nature allows the model to focus on different regions of the input space, capturing local variations effectively. Neural Spline Flows have been shown to provide expressive modelling capabilities, especially in scenarios where the data distribution exhibits non-trivial structures.

\vspace{1mm}
MADE and Neural Spline Flows combine to maximize the expressive power of the normalizing flow model.  Training such a model involves jointly optimizing the parameters of the MADE network and the parameters defining the spline functions. The optimization process typically involves maximizing the likelihood of the training data under the model.

\bibliographystyle{JHEPCust}
\bibliography{review_v3}{}

\end{document}